\documentclass[sigconf]{acmart}
\pdfoutput=1

\makeatletter
\define@key{Gin}{artifact}[true]{}
\makeatother

\usepackage{paralist}
\usepackage{grumble}
\usepackage{tikz}
\usepackage{acronym}
\usepackage{enumitem}
\usepackage{multicol,multirow}
\usepackage{xspace}
\usepackage{colortbl}
\usepackage{xcolor}
\usepackage{realboxes}
\usepackage{siunitx}
\usepackage{cleveref}
\usepackage{tabularx}
\usepackage{booktabs}
\usepackage{listings}
\usepackage{lstautogobble}
\usepackage{subcaption}
\usepackage{wrapfig}
\usepackage{makecell}
\usepackage{svg}
\usepackage[most]{tcolorbox}
\usepackage{fontawesome5}
\usepackage{arydshln}
\usepackage{pifont}
\usepackage[percent]{overpic}
\usepackage{mdframed}

\begin{document}
\makeatletter
\@clubpenalty=0
\makeatother
\clubpenalty=0
\widowpenalty=0
\displaywidowpenalty=0
\brokenpenalty=0

\newcommand*\Let[2]{\State #1 $\gets$ #2}

\newcounter{sarrow}
\newcommand\xleadsto[1]{%
\stepcounter{sarrow}%
\mathrel{\begin{tikzpicture}[baseline= {( $ (current bounding box.south) + (0,-0.5ex) $ )}]
\node[inner sep=.5ex] (\thesarrow) {$\scriptstyle #1$};
\path[draw,<-,decorate,
  decoration={zigzag,amplitude=0.7pt,segment length=1.2mm,pre=lineto,pre length=4pt}] 
    (\thesarrow.south east) -- (\thesarrow.south west);
\end{tikzpicture}}%
}

\theoremstyle{plain}

\newtheorem{property}{Property}

\theoremstyle{definition}

\newtheorem{parameter}{Parameter}
\newtheorem{assumption}{Assumption}
\newtheorem{notation}{Notation}
\newtheorem{remark}{Remark}

\newcommand{\myparagraph}[1]{\noindent{\bf {#1}.}}

\renewcommand{\ttdefault}{cmtt}

\newcommand{\sevcik}{\v{S}ev\v{c}\'{\i}k\xspace}

\colorlet{colorDEPS}{violet}
\colorlet{colorPO}{darkgray!80!black}
\colorlet{colorRF}{green!60!black}
\colorlet{colorJF}{blue}
\colorlet{colorEW}{brown}
\colorlet{colorSI}{violet}
\colorlet{colorCA}{brown}
\colorlet{colorMO}{orange}
\colorlet{colorCO}{orange}
\colorlet{colorCA}{blue}
\colorlet{colorFR}{purple}
\colorlet{colorECO}{red!80!black}
\colorlet{colorSW}{green!40!black}
\colorlet{colorHB}{blue}
\colorlet{colorWB}{blue}
\colorlet{colorPPO}{magenta}
\colorlet{colorBOB}{magenta}
\colorlet{colorPB}{olive}
\colorlet{colorRMW}{olive!70!black}
\colorlet{colorRSEQ}{green!40!black}
\colorlet{colorSC}{violet}
\colorlet{colorPSC}{violet}
\colorlet{colorREL}{olive}
\colorlet{colorCONFLICT}{olive}
\colorlet{colorRACE}{olive}
\colorlet{colorRMW}{brown}
\colorlet{colorSYN}{green!40!black}
\definecolor{shadeblue}{RGB}{190,255,186}

\tikzset{
   every path/.style={>=stealth},
   po/.style={->,color=colorPO,thin,shorten >=-0.5mm,shorten <=-0.5mm},
   sw/.style={->,color=colorSW,shorten >=-0.5mm,shorten <=-0.5mm},
   rf/.style={->,color=colorRF,thin,shorten >=-0.5mm,shorten <=-0.5mm},
   jf/.style={->,color=colorJF,thin,shorten >=-0.5mm,shorten <=-0.5mm},
   ew/.style={<->,color=colorEW,thin,shorten >=-0.5mm,shorten <=-0.5mm},
   fr/.style={->,color=colorFR,thin,shorten >=-0.5mm,shorten <=-0.5mm},
   hb/.style={->,color=colorHB,thick,shorten >=-0.5mm,shorten <=-0.5mm},
   co/.style={->,color=colorCO,thin,very thick,shorten >=-0.5mm,shorten <=-0.5mm},
   no/.style={->,dotted,thick,shorten >=-0.5mm,shorten <=-0.5mm},
   deps/.style={->,color=colorDEPS,dotted,thick,shorten >=-0.5mm,shorten <=-0.5mm},
   ppo/.style={->,color=colorPPO,thick,shorten >=-0.5mm,shorten <=-0.5mm},
   rmw/.style={->,color=colorRMW,thick,shorten >=-0.5mm,shorten <=-0.5mm},
   rseq/.style={->,color=colorRSEQ,thick,dotted,shorten >=-0.5mm,shorten <=-0.5mm},
   xrf/.style={->,color=colorRF,shorten >=-0.5mm,shorten <=-0.5mm},
   si/.style={<->,color=colorSI,thin,shorten >=-0.5mm,shorten <=-0.5mm},
   bob/.style={->,color=colorBOB,thin,shorten >=-0.5mm,shorten <=-0.5mm},
   unknown/.style={<->,color=gray,dashed,shorten >=-0.5mm,shorten <=-0.5mm},
}
\newcommand{\edgeUnknown}{\mathsf{\color{gray}?}}

\newcommand{\step}[1]{\textcolor{dark-red}{\$\ensuremath{#1}}}
\newcommand{\cstep}[1]{\textcolor{dark-red}{\$\ensuremath{C_{#1}}}}

\newcommand{\vle}{\preceq}

\newcommand{\setof}[1]{\{#1\}}
\newcommand{\tup}[1]{\langle#1\rangle}
\newcommand{\Step}[3]{#1 \rightarrow_{#2} #3}
\newcommand{\Steps}[3]{#1 \mathbin{{\rightarrow_{#2}}{}^*} #3}
\newcommand{\BasicStep}[3]{{#1} \xRightarrow{#2} {#3}}
\newcommand{\GStep}[2]{{#1} \longrightarrow {#2}}

\newcommand{\AddJF}{\mathsf{AddJF}}
\newcommand{\racons}{\textsc{ra}}
\newcommand{\MOna}{\textsc{na}}
\newcommand{\MOrlx}{\textsc{rlx}}
\newcommand{\MOacq}{\textsc{acq}}
\newcommand{\MOcon}{\textsc{con}}
\newcommand{\MOacqrel}{\textsc{acq-rel}}
\newcommand{\MOrel}{\textsc{rel}}
\newcommand{\MOsc}{\textsc{sc}}
\newcommand{\MOunord}{\textsc{unord}}
\newcommand{\MOmono}{\textsc{mono}}
\newcommand{\behav}{\mathsf{Behav}}
\newcommand{\rw}{\textsc{rw}}
\newcommand{\labell}{\mathit{lab}}
\newcommand{\irreflexive}{\mathsf{irreflexive}}
\newcommand{\rlx}{\mathsf{Rlx}}
\newcommand{\mrk}{\mathsf{mark}}
\newcommand{\mrkd}{\mathsf{marked}}
\newcommand{\mrke}{\mathsf{mrke}}
\newcommand\lht{\mathsf{ht}}
\newcommand\ew{\mathsf{\color{colorEW}ew}}
\newcommand\cyc{\mathsf{cyc}}
\newcommand\rcyc{\mathsf{rcyc}}
\newcommand\robust{\mathsf{robust}}

\newcommand\mrobust{\mathsf{mrobust}}

\newcommand\getmf{\mathsf{getG}}

\newcommand{\MOat}{\textsc{a}}

\newcommand\oncyc{\mathsf{OnCyc}}

\newcommand\tidset{\mathsf{TIDSet}}

\newcommand\init{\mathsf{init}}
\newcommand\Ginit{{G_\init}}

\newcommand\fadd{\mathsf{fadd}}

\newcommand\existsw{\mathsf{existsW}}

\newcommand\ex{\mathsf{X}}

\newcommand\rmax{\mathsf{Rmax}}

\newcommand\rloc[1]{{{#1}|_\lloc}}
\newcommand\rnloc[1]{{#1}|_{\neq\lloc}}

\newcommand\rlocop[2]{{#1}\! \! \mid^{#2}_\lloc}
\newcommand\rnlocop[2]{{#1}\! \! \mid^{#2}_{\neq\lloc}}

\newcommand\wf{\mathsf{wf}}
\newcommand\xwf{\mathsf{xwf}}

\newcommand\prmfree[1]{{#1}\! \! \mid_{\promises=\emptyset}}

\newcommand\weakest{\textsc{weakest}\xspace}
\newcommand\weakestmo{\textsc{weak\-est\-mo}\xspace}
\newcommand\wmo{\textsc{wmo}}

\newcommand\weakestmocpp{\textsc{weakestmo}\text{-}\textsc{C11}}
\newcommand\weakestmollvm{\textsc{weakestmo}\text{-}\textsc{llvm}}
\newcommand\weakestmodef{\textsc{weakestmo}\text{-}\textsc{def}}

\newcommand\weakestmoncf{\textsc{wmo-ncf}}
\newcommand\weakestmoncfra{\textsc{wmo-ncf-ra}}
\newcommand\weakestmoncfsc{\textsc{wmo-ncf-sc}}

\newcommand\weaken{\textsc{weaken}}

\newcommand\frace{\mathsf{f}\text{-}\mathsf{race}}

\newcommand\slocs{\mathsf{slocs}}
\newcommand\baserf{\mathsf{baseRF}}

\newcommand\concurrent{\mathsf{concurrent}}

\newcommand\start{\mathsf{start}}

\newcommand\xconcurrent{\mathsf{xconcurrent}}

\newcommand\fromms{\mathsf{fromMS}}
\newcommand\toms{\mathsf{toMS}}

\newcommand\whb{\mathsf{whb}}

\newcommand\seq{\mathsf{seq}}

\newcommand\xy{\mathsf{xy}}

\newcommand{\dobcc}{\mathsf{dobcc}}
\newcommand{\ndobcc}{\mathsf{ndobcc}}

\newcommand\fulfill{\mathsf{fulfill}}

\newcommand\mapto{\mathsf{mapsTo}}
\newcommand\wtomsg{\mathbb{W}}
\newcommand\rmsg{\mathsf{rmsg}}
\newcommand\wmsg{\mathsf{wmsg}}
\newcommand\ptoe{\mathsf{P2E}}
\newcommand\stepptoe{\mathsf{T}}
\newcommand\stateptoe{\mathbb{S}}
\newcommand\simurel{\mathbb{S}}
\newcommand\gstep{\mathsf{Gstep}}
\newcommand\TS{TS}
\newcommand\promises{\mathsf{P}}
\newcommand\promise{\mathsf{promise}}
\newcommand\MS{\mathsf{MS}}
\newcommand\TSmap{\mathcal{T\!S}}
\newcommand\M{\mathsf{M}}

\newcommand\srcm{\mathsf{M_{s}}}
\newcommand\tgtm{\mathsf{M_{t}}}

\newcommand\E{\mathsf{E}}
\newcommand\TM{\mathsf{TM}}

\newcommand\wmap{\mathbb{W}}

\newcommand\timestamp{\mathsf{ts}}

\newcommand\np{\mathsf{np}}

\newcommand\udval{\mathsf{udval}}

\newcommand\promisem{\textsc{promise}}

\newcommand\scal{\mathcal{S}}

\newcommand\ecommit{\mathsf{commit}}

\newcommand\rcommit{\mathcal{C}}

\newcommand\starch{\mathsf{st}}
\newcommand\ldarch{\mathsf{ld}}

\newcommand\mov{\texttt{MOV}}
\newcommand\movld{\texttt{RMOV}}
\newcommand\movst{\texttt{WMOV}}
\newcommand{\lfence}{\texttt{LFENCE}}
\newcommand{\sfence}{\mathsf{SFENCE}}
\newcommand{\mfence}{\texttt{MFENCE}}
\newcommand\cmpxchg{\texttt{CMPXCHG}}
\newcommand\lock{\texttt{LOCK}}
\newcommand\irmw{\texttt{RMW}}

\newcommand\ldrex {\texttt{LDREX}}
\newcommand\strex {\texttt{STREX}}

\newcommand\ldr{\texttt{LDR}}
\newcommand\str{\texttt{STR}}
\newcommand\ldar{\ldr_\Alab}
\newcommand\stlr{\str_\Llab}

\newcommand{\ldxr}{\texttt{LX}}
\newcommand{\stxr}{\texttt{SX}}
\newcommand{\ldaxr}{\ldxr{}_\Alab{}}
\newcommand{\stlxr}{\stxr{}_\Llab{}}

\newcommand\ildaxr {\texttt{ldaxr}}
\newcommand\istlxr {\texttt{stlxr}}
\newcommand\icasal {\texttt{casal}}

\newcommand\idmbfull{\texttt{DMBFF}}
\newcommand\idmbld{\texttt{DMBLD}}
\newcommand\idmbst{\texttt{DMBST}}

\newcommand\idmb {\texttt{DMB}}
\newcommand\isb {\texttt{ISB}}

\newcommand\cbisync {\texttt{cbisync}}
\newcommand\cbisb {\texttt{CBISB}}

\newcommand\prsv{\mathsf{preserve}}

\newcommand\sstate{\mathsf{state}}

\newcommand\future{\mathsf{future}}
\newcommand\silent{\mathsf{silent}}

\newcommand\wrpair{\mathsf{WR}}
\newcommand\rrpair{\mathsf{RR}}
\newcommand\rwpair{\mathsf{RW}}
\newcommand\wwpair{\mathsf{WW}}

\newcommand\pair{\mathsf{pair}}

\newcommand\pairs{\mathsf{pairs}}
\newcommand\mpairs{\mathsf{MM}}
\newcommand\nlocmpairs{\mathsf{nMM}}

\newcommand\ord{\mathsf{ord}}
\newcommand\ghb{\mathsf{ghb}}

\newcommand\nlocpairs{\mathsf{nlocpairs}}
\newcommand\slocpairs{\mathsf{slocpairs}}

\newcommand\ordered{\mathsf{Ordered}}

\newcommand\allordered{\mathsf{allOrdered}}

\newcommand\fenced{\mathsf{fenced}}

\newcommand\cfgpath{\mathsf{CFGPath}}

\newcommand\allpairfenced{\mathsf{allPairFenced}}

\newcommand\pairpath{\mathsf{pairpath}}

\newcommand\mustalias{\mathsf{mA}}

\newcommand\paircyc{\mathsf{paircyc}}

\newcommand\cfg{\mathsf{G}}

\newcommand\subseq{\mathsf{subseq}}

\newcommand\dealloc{\mathsf{dealloc}}
\newcommand\alloc{\mathsf{alloc}}

\newcommand\mbbp{\mathbb{P}}
\newcommand\mbbe{\mathbb{E}}
\newcommand\mbbc{\mathbb{C}}

\newcommand\visited{\mathbb{S}}
\newcommand\exs{\mathbb{S}}

\newcommand\srcp{\mbbp_{\!\sf src}}
\newcommand\tgtp{\mbbp_{\!\sf tgt}}

\newcommand\srcps{\mbbp_{\sf s}}
\newcommand\tgtpt{\mbbp_{\sf t}}

\newcommand\srcx{\ex_{\sf s}}
\newcommand\tgtx{\ex_{\sf t}}

\newcommand\xp{\mbbp_{\!\sf x86}}
\newcommand\armp{\mbbp_{\!\sf ARMv8}}
\newcommand\oarmp{\mbbp_{\!\sf ARMv7}}
\newcommand\oarmpmca{\mbbp_{\!\sf ARMv7\text{-}mca}}
\newcommand\btirp{\mbbp_{\!\sf BTIR}}

\newcommand\xwr{\mathsf{wr}}

\newcommand\markr{\mathsf{markR}}
\newcommand\markw{\mathsf{markW}}
\newcommand\marke{\mathsf{markE}}

\newcommand\maxmo{\mathsf{max}\text{-}\mo}

\newcommand\maxr{\mathsf{maxR}}

\newcommand\scrf{\mathsf{Srf}}
\newcommand\scmo{\mathsf{Smo}}

\newcommand\AddMO{\mathsf{AddMO}}
\newcommand\AddEW{\mathsf{AddEW}}

\newcommand\maxmsg{\mathsf{maxmsg}}

\newcommand\orig{\mathsf{\color{blue!40!black}\mbbp_{cmap}}}
\newcommand\naive{\mathsf{\color{blue!40!black}\mbbp_{naive}}}
\newcommand\atam{\mathsf{\color{blue!40!black}\mbbp_{ATAM}}}
\newcommand\fd{\mathsf{\color{blue!40!black}\mbbp_{opt}}}

\newcommand{\MOnarel}{\textsc{\{na,rel\}}}
\newcommand{\MOnaacq}{\textsc{\{na,acq\}}}

\newcommand{\TODO}[1]{\textcolor{red}{\textbf{[TODO: {#1}]}}}
\newcommand{\denot}[1]{[\![{#1}]\!]}
\newcommand{\RAsem}[1]{\denot{#1}_\racons}
\newcommand{\SCsem}[1]{\denot{#1}_\SC}
\newcommand{\LLVMsem}[1]{\denot{#1}_\LLVM}
\newcommand{\POWsem}[1]{\denot{#1}_\POWER}

\newcommand{\powerset}[1]{\mathcal{P}({#1})}
\newcommand{\defeq}{\stackrel{\mathsf{def}}{=}}
\newcommand{\limp}{\Rightarrow}
\newcommand{\lequiv}{\Leftrightarrow}

\newcommand\inarr[1]{\begin{array}{@{}l@{}}#1\end{array}}
\newcommand\inarrC[1]{\begin{array}{@{}c@{}}#1\end{array}}
\newcommand\inpar[1]{\left(\begin{array}{@{}l@{}}#1\end{array}\right)}
\newcommand\inparII[2]{\begin{array}{@{}l@{~~}||@{~~}l@{}}\inarr{#1}&\inarr{#2}\end{array}}
\newcommand\inparIII[3]{\begin{array}{@{}l@{~~}||@{~~}l@{~~}||@{~~}l@{}}\inarr{#1}&\inarr{#2}&\inarr{#3}\end{array}}
\newcommand\inparIV[4]{\begin{array}{@{}l@{~~}||@{~~}l@{~~}||@{~~}l@{~~}||@{~~}l@{}}
\inarr{#1}&\inarr{#2}&\inarr{#3}&\inarr{#4}\end{array}}
\newcommand\inparV[5]{\begin{array}{@{}l@{~~}||@{~~}l@{~~}||@{~~}l@{~~}||@{~~}l@{~~}||@{~~}l@{}}
\inarr{#1}&\inarr{#2}&\inarr{#3}&\inarr{#4}&\inarr{#5}\end{array}}
\newcommand\mytag[1]{\tag{\textsf{#1}}}
\newcommand\inparVI[6]{\begin{array}{@{}l@{~~}||@{~~}l@{~~}||@{~~}l@{~~}||@{~~}l@{~~}||@{~~}l@{~~}||@{~~}l@{}}
\inarr{#1}&\inarr{#2}&\inarr{#3}&\inarr{#4}&\inarr{#5}&\inarr{#6}\end{array}}

\newcommand\imm{\mathsf{im}}
\newcommand\SCR{\mathsf{scr}}
\newcommand\Consistent{\mathsf{Consistent}}
\newcommand\Race{\mathsf{Racy}}
\newcommand\rfrx{\mathit{rf}}

\newcommand\oll{\mathsf{LX}}
\newcommand\osc{\mathsf{SX}}
\newcommand\ollsc{\mathsf{\oll\text{-}\osc}}

\newcommand\oimm{\mathsf{Imm}}
\newcommand\getnfs{\textsc{getNFs}}
\newcommand\updnfs{\mathsf{updNFs}}

\newcommand\fdelete{\textsc{FDelete}}
\newcommand\fweaken{\textsc{FWeaken}}

\newcommand\xfelim{\textsc{x86FElim}}
\newcommand\armafelim{\textsc{ARMv8FElim}}
\newcommand\armsfelim{\textsc{ARMv7FElim}}

\newcommand\xa{\E}
\newcommand\xpo{\lpo}
\newcommand\xrf{\lrf}
\newcommand\xmo{\mo}
\newcommand{\moe}{\mathsf{\color{colorMO}moe}}

\newcommand\epo{\mathsf{epo}}

\newcommand\mosim{\mo}

\newcommand{\ii}{\mathsf{ii}}
\newcommand{\ic}{\mathsf{ic}}
\newcommand{\ci}{\mathsf{ci}}
\newcommand{\cc}{\mathsf{cc}}
\newcommand{\rdw}{\mathsf{rdw}}
\newcommand{\detour}{\mathsf{detour}}

\newcommand\fexists{\mathsf{fexists}}
\newcommand\nfs{\mathsf{nfs}}
\newcommand\postdom{\mathsf{postdom}}
\newcommand\predom{\mathsf{predom}}

\newcommand\df{\mathsf{df}}
\newcommand\ds{\mathsf{ds}}

\newcommand\code{\mathsf{code}}
\newcommand\locs{\mathsf{loc}}
\newcommand\labUpd[2]{\mathit{lab}[{#1}\,{:=}\,{#2}]}
\newcommand\Askip{\mathsf{skip}}
\newcommand\Aalloc[1]{\mathsf{A}({#1})}
\newcommand\Fxxx[1]{\mathsf{F}_{#1}}

\newcommand\Wlab {\mathsf{W}}
\newcommand\Rlab {\mathsf{R}}
\newcommand\Ulab {\mathsf{U}}
\newcommand\Mlab {\mathsf{M}}

\newcommand\Rxxx[3]{\Rlab_{#3}({#1},{#2})}
\newcommand\Wxxx[3]{\Wlab_{#3}({#1},{#2})}
\newcommand\Uxxx[4]{\Ulab_{#4}({#1},{#2},{#3})}

\newcommand\WUlab {\mathcal{W}}
\newcommand\RUlab {\mathcal{R}}
\newcommand\UUlab {\mathcal{U}}
\newcommand\RWUlab {\mathcal{RW}}
\newcommand\Flab {\mathsf{F}}
\newcommand\Elab {\mathcal{E}}

\newcommand\implid{{\color{colorSYN}\mathsf{implied}}}

\newcommand\mf {\mathcal{mf}}

\newcommand\xii {\mathsf{x}}
\newcommand\yii {\mathsf{y}}

\newcommand\adjacent{\mathsf{Adj}}
\newcommand\skips{\mathsf{skip}}
\newcommand\bS{\mathsf{bS}}

\newcommand\hbw{\mathsf{\color{colorMO}mo}_{\mathsf{strong}}}
\newcommand\co{\mathsf{\color{colorCO}co}}
\newcommand\ca{\mathsf{\color{colorCA}ca}}
\newcommand\frd{\mathsf{\color{colorFR}fr}_{\mathsf{strong}}}

\newcommand\reord{\mathsf{reord}}

\newcommand{\MOrr}{\textsc{rr}}
\newcommand{\MOrw}{\textsc{rw}}
\newcommand{\MOwr}{\textsc{wr}}
\newcommand{\MOrm}{\textsc{rm}}
\newcommand{\MOww}{\textsc{ww}}
\newcommand{\MOmw}{\textsc{mw}}
\newcommand{\MOmm}{\textsc{mm}}
\newcommand{\MOmr}{\textsc{mr}}
\newcommand{\MOwm}{\textsc{wm}}

\newcommand\conswf{\ensuremath{(\mathsf{WF})}}
\newcommand\conswfmo{\ensuremath{(\mathsf{WF})}}

\newcommand\consviswrf{\ensuremath{(\textsf{VISJ})}}

\newcommand\consncf{\ensuremath{(\textsf{CF})}}
\newcommand\consncfrf{\ensuremath{(\textsf{CFJ})}}

\newcommand\conscfhbrf{\ensuremath{(\textsf{HBCFJ})}}

\newcommand\conscfr{\ensuremath{(\textsf{ICF})}}
\newcommand\consncfew{\ensuremath{(\textsf{ICFJ})}}

\newcommand\conscoh{\ensuremath{(\textsf{COH})}}
\newcommand\conscohmo{\ensuremath{(\textsf{COH$'$})}}

\newcommand\consncfu{\ensuremath{(\textsf{NCFU})}}
\newcommand\consncfsc{\ensuremath{(\textsf{NCFSC})}}

\newcommand\store{\mathsf{store}}
\newcommand\load{\mathsf{load}}

\newcommand\rels{\mathsf{Rel}}
\newcommand\acqs{\mathsf{Acq}}

\newcommand\AcqRel{\mathsf{AcqRel}}

\newcommand\RCsc{\mathsf{RCsc}}

\newcommand\frels{\mathsf{Frel}}
\newcommand\facqs{\mathsf{Facq}}
\newcommand\fsc{\mathsf{Fsc}}
\newcommand\AT{\mathsf{AT}}

\newcommand\transitive{\mathsf{transitive}}
\newcommand\irr{\mathsf{irr}}
\newcommand\poloc{\mathsf{poloc}}
\newcommand\npoloc{\mathsf{npoloc}}

\newcommand\xpolocs{\mathsf{xpo\text{-}locs}}
\newcommand\xponlocs{\mathsf{xpo\text{-}nlocs}}

\newcommand\awr{\mathsf{WR}}
\newcommand\aww{\mathsf{WW}}
\newcommand\arw{\mathsf{RW}}
\newcommand\arr{\mathsf{RR}}
\newcommand\awra{\mathsf{WRA}}

\newcommand\isawra{\mathsf{isWR}}

\newcommand\ponlocs{\mathsf{po\text{-}nlocs}}

\newcommand\opset{\mathsf{opSet}}

\newcommand\msgto{\mathsf{to}}
\newcommand\msgfrom{\mathsf{from}}

\newcommand\mset{\mathsf{MSet}}

\newcommand\wm{\mathsf{wm}}

\newcommand{\kw}[1]{\textsf{\bfseries #1}}

\newcommand{\Cif}[1]{\kw{if}\,({#1})\ }
\newcommand{\Celse}{\kw{else}}
\newcommand\Cstore[3]{{#1}.\mathsf{store}({#2},{#3})}
\newcommand\Ccas[4]{{#1}.\mathsf{CAS}({#2},{#3},{#4})}
\newcommand\Cload[2]{{#1}.\mathsf{load}({#2})}
\newcommand\Caccess[1]{\mathsf{access}({#1})}
\newcommand\Cfence[1]{\mathsf{fence}_{#1}}

\newcommand\Rpln[2]  {\Rxxx{#1}{#2}{}}
\newcommand\Rna[2]   {\Rxxx{#1}{#2}{\MOna}}
\newcommand\Rrlx[2]  {\Rxxx{#1}{#2}{\MOrlx}}
\newcommand\Racq[2]  {\Rxxx{#1}{#2}{\MOacq}}
\newcommand\Rsc[2]   {\Rxxx{#1}{#2}{\MOsc}}
\newcommand\Racqsc[2]{\Rxxx{#1}{#2}{\sqsupseteq\MOacq}}
\newcommand\Wpln[2]  {\Wxxx{#1}{#2}{}}
\newcommand\Wna[2]   {\Wxxx{#1}{#2}{\MOna}}
\newcommand\Wrlx[2]  {\Wxxx{#1}{#2}{\MOrlx}}
\newcommand\Wrel[2]  {\Wxxx{#1}{#2}{\MOrel}}
\newcommand\Wsc[2]   {\Wxxx{#1}{#2}{\MOsc}}
\newcommand\Wrelsc[2]{\Wxxx{#1}{#2}{\sqsupseteq\MOrel}}
\newcommand\Facq{\mathsf{F}_{\MOacq}}
\newcommand\Frel{\mathsf{F}_{\MOrel}}
\newcommand\oCAS{\mathsf{CAS}}
\newcommand\usc{\mathsf{updateSC}}

\newcommand\oRna   {\mathsf{R}_{\MOna}}
\newcommand\oRrlx  {\mathsf{R}_{\MOrlx}}
\newcommand\oRacq  {\mathsf{R}_{\MOacq}}
\newcommand\oRsc   {\mathsf{R}_{\MOsc}}
\newcommand\oRacqsc {\mathsf{R}_{\sqsupseteq\MOacq}}
\newcommand\oWna  {\mathsf{W}_{\MOna}}
\newcommand\oWrlx  {\mathsf{W}_{\MOrlx}}
\newcommand\oWrel  {\mathsf{W}_{\MOrel}}
\newcommand\oWsc   {\mathsf{W}_{\MOsc}}
\newcommand\oWrelsc {\mathsf{W}_{\sqsupseteq\MOrel}}

\newcommand\ldacq  {\Rlab_{\MOacq}}
\newcommand\strel  {\Wlab_{\MOrel}}
\newcommand\ura  {\Ulab_{\MOacqrel}}
\newcommand\fra{\mathsf{F}_{\MOacqrel}}

\newcommand\oUra  {\mathsf{U}_{\racons}}
\newcommand\mocons  {\mathsf{moCons}}
\newcommand\addmo  {\mathsf{AddMO}}

\newcommand\cset  {\mathsf{cset}}
\newcommand\abort  {\mathsf{abort}}

\newcommand\Wppc {\mathcal{W}}
\newcommand\Rppc {\mathcal{R}}
\newcommand\Uppc {\mathcal{U}}
\newcommand\RWppc {\mathcal{RW}}
\newcommand\Fppc {\mathcal{F}}
\newcommand\Xppc {\mathcal{X}}
\newcommand\X {\mathsf{X}}

\newcommand\mapp {\mathcal{M}}
\newcommand\deps {\mathsf{deps}}
\newcommand\dep {\mathsf{dep}}
\newcommand\pdep {\mathsf{pdep}}
\newcommand\es {\mathsf{es}}
\newcommand\oao {\mathsf{oao}}
\newcommand\st {\mathsf{st}}
\newcommand\ld {\mathsf{ld}}

\newcommand\fldld {\mathsf{Frr}}
\newcommand\fstld {\mathsf{Fwr}}
\newcommand\fldst {\mathsf{Frw}}
\newcommand\fmst {\mathsf{Fmw}}
\newcommand\fstst {\mathsf{Fww}}
\newcommand\fmm {\mathsf{Fmm}}
\newcommand\fstm {\mathsf{Fwm}}

\newcommand\fldm {\mathsf{Frm}}
\newcommand\fmld {\mathsf{Fmr}}

\newcommand\blr {\texttt{BLR}}
\newcommand\ret {\texttt{RET}}

\newcommand\llsc {\mathsf{LL/SC}}
\newcommand\lwarx {\mathsf{lwarx}}
\newcommand\stwcx {\mathsf{stwcx}}

\newcommand\fork {\mathsf{Fork}}
\newcommand\join {\mathsf{join}}

\newcommand\mapst {\mathbb{M}}

\newcommand\xll {\mathsf{ll}}
\newcommand\xsc {\mathsf{sc}}

\newcommand\dmb {\F}
\newcommand\dsb {\mathsf{dsb}}

\newcommand\wo {\mathsf{wo}}

\newcommand\rmov{\texttt{RMOV}}
\newcommand\wmov{\texttt{WMOV}}

\newcommand\ldapr{\ldr_{\Qlab}}

\newcommand{\xarch}{\mathsf{x86}}
\newcommand{\arma}{\mathsf{ARMv8}}
\newcommand{\arms}{\mathsf{ARMv7}}
\newcommand{\armsmca}{\mathsf{ARMv7\text{-}mca}}

\newcommand{\actions}{\textsf{Actions}}
\newcommand{\EVENTS}{\mathit{AName}}
\newcommand{\Events}{\mathit{Act}}
\newcommand{\Event}{\mathit{Event}}

\newcommand{\XX}{\mathit{X}}

\newcommand{\lwsync}{\mathsf{lwsync}}
\newcommand{\hwsync}{\mathsf{hwsync}}
\newcommand{\sync}{\mathsf{sync}}
\newcommand{\cmp}{\mathsf{cmp}}
\newcommand{\bc}{\mathsf{bc}}
\newcommand{\isync}{\mathsf{isync}}
\newcommand{\fence}{\mathsf{fence}}

\newcommand{\rfex}{\mathsf{rf}_E}
\newcommand{\poex}{\mathsf{po}_E}

\newcommand{\El}{\mathsf{E_L}}
\newcommand{\Ec}{\mathsf{E_C}}

\newcommand{\com}{\mathsf{com}}
\newcommand{\mo}{\mathsf{\color{colorMO}mo}}
\newcommand{\jr}{\mathsf{jr}}
\newcommand{\rb}{\mathsf{\color{colorFR}rb}}
\newcommand{\fr}{\mathsf{\color{colorFR}fr}}
\newcommand{\frp}{\mathsf{\color{colorRF}frp}}
\newcommand{\fre}{\mathsf{\color{colorFR}fre}}
\newcommand{\fri}{\mathsf{\color{colorFR}fri}}
\newcommand{\SW}{\mathsf{\color{colorSW}sw}}
\newcommand{\swe}{\mathsf{\color{colorSW}swe}}
\newcommand{\SC}{\mathsf{SC}}

\newcommand{\scsim}{\mathsf{\color{colorSC}sc}}

\newcommand{\hbmax}{\mathsf{hbmax}}

\newcommand{\Wevent}{\mathsf{E}\!\wtomsg}

\newcommand{\lidd}{\mathsf{id}}
\newcommand{\ltid}{\mathsf{tid}}
\newcommand{\lcode}{\mathsf{code}}
\newcommand{\llab}{\mathsf{lab}}
\newcommand{\lloc}{\mathsf{loc}}
\newcommand{\lord}{\mathsf{ord}}

\newcommand{\lpo}{\mathsf{po}}
\newcommand{\lrf}{\mathsf{\color{colorRF}rf}}
\newcommand{\ljf}{\mathsf{\color{colorJF}jf}}
\newcommand{\lsc}{\mathsf{\color{colorSC}sc}}
\newcommand{\lhb}{\mathsf{\color{colorHB}hb}}
\newcommand{\hb}{\lhb}
\newcommand{\xhb}{\mathsf{\color{colorHB}hb}}

\newcommand{\rf}{\mathsf{\color{colorRF}rf}}
\newcommand{\rfe}{\mathsf{\color{colorRF}rfe}}
\newcommand{\jfe}{\mathsf{\color{colorJF}jfe}}

\newcommand{\EvType}[1]{{#1}.\llab}
\newcommand{\rVal}{\mathsf{rval}}
\newcommand{\wVal}{\mathsf{wval}}

\newcommand{\spo}{\mathsf{spo}}
\newcommand{\srf}{\mathsf{\color{colorRF}srf}}
\newcommand{\shb}{\mathsf{\color{colorHB}shb}}
\newcommand{\sfr}{\mathsf{\color{colorFR}sfr}}
\newcommand{\ssw}{\mathsf{\color{colorSW}ssw}}
\newcommand{\smo}{\mathsf{\color{colorMO}smo}}
\newcommand{\sco}{\mathsf{\color{colorECO}seco}}
\newcommand{\ssc}{\mathsf{\color{colorMO}SSC}}

\newcommand{\upred}{\mathsf{Upred}}
\newcommand{\usucc}{\mathsf{Usucc}}

\newcommand{\reorder}{\mathsf{Reorder}}

\newcommand{\reorderable}{\mathsf{Reorderable}}

\newcommand{\po}{\mathsf{po}}
\newcommand{\ppo}{\mathsf{{\color{colorPPO}ppo}}}
\newcommand{\xppo}{\mathsf{{\color{colorPPO}ppo}}}

\newcommand{\op}{\mathsf{op}}
\newcommand{\OP}{\mathsf{OP}}
\newcommand{\tid}{\mathsf{tid}}
\newcommand{\event}{e}
\newcommand{\id}{\mathsf{id}}
\newcommand{\cf}{\mathsf{cf}}
\newcommand{\ecf}{\mathsf{ecf}}
\newcommand{\rfr}{\mathsf{rf}}
\newcommand{\access}{\mathsf{Access}}
\newcommand{\events}{\mathsf{events}}
\newcommand{\ev}{\mathsf{Ev}}
\newcommand{\evstr}{\mathsf{EvStr}}
\newcommand{\pc}{\mathsf{pc}}
\newcommand{\Nw}{\mathsf{Nw}}
\newcommand{\LF}{\mathsf{LF}}
\newcommand{\lwf}{\mathcal{F}}
\newcommand{\Wrr}{\mathsf{Wr}}
\newcommand{\LE}{\RWUlab}

\newcommand{\poimm}{\imm(\po)}
\newcommand{\cfimm}{\overrightarrow{\cf}}

\newcommand{\CYC}{\mathsf{CYC}}
\newcommand{\IRW}{\mathsf{IRW}}
\newcommand{\LB}{\mathsf{LB}}
\newcommand{\RRW}{\mathsf{RRW}}
\newcommand{\OOTA}{\mathsf{OOTA}}
\newcommand{\lr}{\mathsf{LR}}
\newcommand{\lwb}{\mathsf{LWB}}
\newcommand{\lw}{\mathsf{lw}}
\newcommand{\wb}{\mathsf{wb}}
\newcommand{\concw}{\mathsf{concW}}

\newcommand{\yess}{{\color{green!70!black}\checkmark}}
\newcommand{\noo}{{\color{red!70!black}\times}}

\newcommand{\poS}{\mathsf{poSet}}
\newcommand{\rfS}{\mathsf{rfSet}}
\newcommand{\cfS}{\mathsf{cfSet}}
\newcommand{\viswr}{\mathsf{visW}}
\newcommand{\viswrsc}{\mathsf{visWsc}}
\newcommand{\visrd}{\mathsf{Rd}}
\newcommand{\eid}{\mathsf{id}}
\newcommand{\setrval}{\mathsf{setRval}}
\newcommand{\type}{\mathsf{type}}
\newcommand{\itype}{\mathsf{typ}}
\newcommand{\XC}{X_{\mathsf{C}}}
\newcommand{\XL}{X_{\mathsf{L}}}
\newcommand{\subst}{\mathsf{subst}}
\newcommand{\observe}{\mathsf{observe}}
\newcommand{\nop}{\mathsf{nop}}
\newcommand{\Ru}{\mathsf{Ru}}
\newcommand{\ru}{\mathsf{ru}}

\newcommand{\oRMW}{\mathsf{RMW}}

\newcommand{\Wd}{\mathsf{Wd}}
\newcommand{\oplusperm}{\oplus_{\rm perm}}
\newcommand{\oplusacq}{\oplus_{\rm acq}}
\newcommand{\oplusrel}{\oplus_{\rm rel}}
\newcommand{\oplusrmw}{\oplus_{\rm rmw}}
\newcommand{\oplusphi}{\oplus_{\emptyset}}

\newcommand{\rmw}{\mathsf{\color{colorRMW}rmw}}

\newcommand{\lxsx}{\mathsf{\color{colorRMW}lxsx}}
\newcommand{\amo}{\mathsf{\color{colorRMW}amo}}

\newcommand{\frange}{\mathsf{range}}

\newcommand{\uid}{\mathsf{uid}}
\newcommand{\fresh}{\mathsf{fresh}}
\newcommand{\nw}{\mathsf{new}}
\newcommand{\canadd}{\mathsf{can\_add}}
\newcommand{\canrace}{\mathsf{can\_race}}

\newcommand{\World}{\mathsf{World}}
\newcommand{\AsnName}{\mathsf{AsnName}}

\newcommand{\Assn}{\mathsf{Assn}}
\newcommand{\Contexts}{\mathsf{Contexts}}
\newcommand{\Exp}{\mathsf{Exp}}
\newcommand{\Locs}{\mathsf{Locs}}
\newcommand{\regs}{\mathsf{Reg}}
\newcommand{\Heap}{\mathsf{Heap}}
\newcommand{\HeapSpec}{\mathsf{Heap}_{\rm spec}}
\newcommand{\Stack}{\mathsf{Stack}}
\newcommand{\CExp}{\mathsf{CExp}}
\newcommand{\Val}{\mathsf{val}} 
\newcommand{\Var}{\mathsf{Var}}
\newcommand{\Perm}{\mathsf{Perm}}
\newcommand{\Bit}{\mathsf{Bit}}
\newcommand{\br}{\mathbf{br}}

\newcommand{\prv}{\mathsf{prev}}
\newcommand{\nxt}{\mathsf{next}}
\newcommand{\eco}{\mathsf{\color{colorECO}eco}}
\newcommand{\scb}{\mathsf{scb}}
\newcommand{\psc}{\mathsf{\color{colorPSC}psc}}
\newcommand{\pscb}{\psc_{\mathsf{base}}}
\newcommand{\pscf}{\psc_{\mathsf{F}}}

\newcommand{\gscb}{\mathsf{scb}}
\newcommand{\gpscb}{\mathsf{\color{colorPSC}pscb}}
\newcommand{\gpscf}{\mathsf{\color{colorPSC}pscf}}
\newcommand{\gpsc}{\mathsf{\color{colorPSC}psc}}

\newcommand\poi{\mathsf{\color{colorPO}poi}}
\newcommand\coi{\mathsf{\color{colorCO}coi}}
\newcommand\rfi{\mathsf{\color{colorRF}rfi}}
\newcommand\coe{\mathsf{\color{colorCO}coe}}
\newcommand\data{\mathsf{\color{colorPPO}data}}
\newcommand\addr{\mathsf{\color{colorPPO}addr}}
\newcommand\ctrl{\mathsf{\color{colorPPO}ctrl}}

\newcommand\polocnow{\mathsf{\color{colorDEPS}poloc\text{-}no\text{-}w}}

\newcommand\rsw{\mathsf{\color{blue}rsw}}

\newcommand\sbprg{\mathsf{SB}}

\newcommand\AQ{\mathsf{AQ}}
\newcommand\RL{\mathsf{RL}}

\newcommand\xlob{\mathsf{\color{blue}xlob}}
\newcommand\alob{\mathsf{\color{blue}alob}}
\newcommand\rlob{\mathsf{\color{blue}rlob}}
\newcommand\lob{\mathsf{\color{blue}lob}}

\newcommand\lrs{\mathsf{\color{blue}lrs}}
\newcommand\lws{\mathsf{\color{blue}lws}}
\newcommand\aob{\mathsf{aob}}
\newcommand\bob{\mathsf{bob}}
\newcommand\dob{\mathsf{dob}}
\newcommand\obs{\mathsf{obs}}
\newcommand\ob{\mathsf{\color{blue}ob}}
\newcommand\obx{\mathsf{obx}}

\newcommand\xob{\mathsf{xob}}

\newcommand\obi{\mathsf{obi}}
\newcommand\obe{\mathsf{obe}}

\newcommand\Alab{\mathsf{A}}
\newcommand\Qlab{\mathsf{Q}}
\newcommand\Llab{\mathsf{L}}
\newcommand\Nlab{\mathsf{N}}

\newcommand\ALlab{\mathsf{AL}}

\newcommand\dmbfull{\F}
\newcommand\dmbld{\F_{\textsc{ld}}}
\newcommand\dmbst{\F_{\textsc{st}}}

\newcommand{\Ctx}{\mathit{Ctx}}
\newcommand{\Cmd}{\mathit{Cmd}}
\newcommand{\Tuple}[1]{\langle{#1}\rangle}
\newcommand{\fst}{\mathit{fst}}
\newcommand{\lst}{\mathit{lst}}
\newcommand{\res}{\mathit{res}}

\newcommand{\mode}{\mathsf{mode}}

\newcommand{\parCtx}[1]{
\textcolor{blue!70!green}{\inarr{~~\textbf{\small Context:}\\\left[\inarr{
\inparII{#1}{}
}\right]}}}

\newcommand{\An}{\mathsf{An}}
\newcommand{\NameEnv}{\mathsf{NEnv}}
\newcommand{\defined}[1]{\mathsf{defined}({#1})}
\newcommand{\distXX}[1]{\mathbf{D}_{\XX}({#1})}
\newcommand{\restr}[2]{{#1}{\downharpoonright_{#2}}}

\newcommand{\dom}{\mathsf{dom}}
\newcommand{\codom}{\mathsf{codom}}
\newcommand{\onerel}{\mathsf{one}}
\newcommand{\sequence}[2]{\mathsf{sequence}_{#2}({#1})}
\newcommand{\labels}{\mathsf{labels}}

\newcommand{\set}[1]{\{#1\}}

\newcommand{\hbppc}{\mathsf{\color{colorHB}ahb}}
\newcommand{\prop}{\mathsf{{\color{colorPB}prop}}}
\newcommand{\ahb}{\mathsf{\color{colorHB}ahb}}

\newcommand{\isconsistent}{\mathsf{isCons}} 
\newcommand{\LLVM}{\mathsf{LLVM}} 
\newcommand{\cmm}{\mathsf{C11}} 
\newcommand{\maps}{\mathsf{maps}} 

\newcommand{\src}{\mathsf{src}} 
\newcommand{\tgt}{\mathsf{tgt}}
\newcommand{\final}{\mathsf{final}} 

\newcommand{\useq}{\mathsf{Useq}} 

\newcommand{\expand}{\mathsf{expand}} 
\newcommand{\collapse}{\mathsf{collapse}} 

\newcommand{\codecomment}[1]{\color{teal}\text{//}{#1}}
\newcommand{\ncodecomment}[1]{\color{dark-red}\text{ // }{#1}}

\newcommand{\collapsed}{\mathsf{collapsed}} 

\newcommand\vrel{\mathsf{rel}}
\newcommand\vacq{\mathsf{acq}}
\newcommand\vcur{\mathsf{cur}}

\newcommand{\pre}{\mathsf{pre}}

\newcommand{\elim}{\mathsf{elim}}

\newcommand{\felim}{\mathsf{felim}}

\newcommand{\diffThread}{\mathsf{diffThread}}
\newcommand{\acyclic}{\mathsf{acyclic}}
\newcommand{\any}{\_}
\newcommand{\card}[1]{|#1|}

\newcommand{\acy}{\mathsf{acy}}
\newcommand{\total}{\mathsf{total}}

\newcommand{\yes}{{\color{green!70!black}\ding{51}}}
\newcommand{\no}{{\color{red!70!black}\ding{55}}}
\newcommand{\partialmark}{{\color{orange!85!black}\ding{119}}}
\newcommand{\conditionalmark}{{\color{blue!70!black}\ding{51}\textsuperscript{*}}}
\newcommand{\noII}[2]{{\color{red!70!black}\ding{55}\ref{#1}}}
\newcommand{\yxn}[2]{{\color{blue!70!black}\yes{#1}/\textbf{?}/\no{#2}}}
\newcommand{\ynn}[3]{{\color{blue!70!black}\yes{#1}/\no{#2}/\no{#3}}}

\newcommand{\ite}[3]{\mathsf{if} \ #1 \ \mathsf{then} \ #2 \ \mathsf{else} \ #3}

\newcommand{\undeff}{\mathbf{undef}}
\newcommand{\undf}{\mathbf{u}}
\newcommand{\undfnd}{\widehat{\mathbf{u}}}
\newcommand{\indet}{\mathbf{nd}}
\newcommand{\const}{\mathbf{const}}
\newcommand{\expr}{\mathsf{expr}}
\newcommand{\EX}{\mathsf{ex}}
\newcommand{\EV}{\mathsf{ev}}

\newcommand{\symmetric}{\mathsf{Symmetric}}

\newcommand{\acc}{\mathsf{ac}}

\newcommand{\rcmm}{\mathsf{RC11}}
\newcommand{\wrcmm}{\mathsf{WRC11}}

\newcommand{\rld}{\mathsf{Ld}}
\newcommand{\wst}{\mathsf{St}}

\newcommand{\isst}{\mathsf{isSt}}
\newcommand{\isld}{\mathsf{isLd}}
\newcommand{\isrel}{\mathsf{isRel}}
\newcommand{\isacq}{\mathsf{isAcq}}
\newcommand{\isll}{\mathsf{isLL}}
\newcommand{\issc}{\mathsf{isSC}}

\newcommand{\isw}{\mathsf{isW}}
\newcommand{\isr}{\mathsf{isR}}

\newcommand{\srels}{\mathsf{Lcoi}}
\newcommand{\relacq}{\mathsf{RA}}

\newcommand{\true}{\mathsf{true}}
\newcommand{\false}{\mathsf{false}}

\newcommand{\ucfpred}{\mathsf{ucfpred}}
\newcommand{\ucfsucc}{\mathsf{ucfsucc}}

\newcommand{\apred}{\mathsf{apred}}
\newcommand{\asucc}{\mathsf{asucc}}

\newcommand{\succr}{\stackrel{\text{R}}{\succ}}
\newcommand{\precr}{\stackrel{\text{R}}{\prec}}

\newcommand{\sucr}{\mathsf{succ}}
\newcommand{\prdr}{\mathsf{pred}}

\newcommand{\isRacyRead}{\mathsf{isRacyRead}}
\newcommand{\isRacyWrite}{\mathsf{isRacyWrite}}
\newcommand{\newR}{\mathsf{newR}}
\newcommand{\isRacyWR}{\mathsf{isRacyWR}}
\newcommand{\isconcurrent}{\mathsf{isConc}}
\newcommand{\conc}{\mathsf{conc}}
\newcommand{\isconcacc}{\mathsf{isConcAcc}}%
\newcommand{\isconflict}{\mathsf{isConurrent}}
\newcommand{\ihbW}{\mathsf{ihbW}}
\newcommand{\visW}{\mathsf{visW}}

\newcommand{\invis}{\mathsf{invis}}
\newcommand{\vis}{\mathsf{vis}}

\newcommand{\bottom}{\perp}

\newcommand{\race}{\mathsf{Race}}
\newcommand{\xrace}{\mathsf{xRace}}

\newcommand{\racy}{\mathsf{Racy}}

\newcommand{\accpairs}{\mathsf{accPairs}}
\newcommand{\mfences}{\mathsf{mfences}}
\newcommand{\getfences}{\mathsf{getFs}}
\newcommand{\getmemaccs}{\mathsf{getMs}}
\newcommand{\maccs}{\mathsf{M}}
\newcommand{\unordered}{\mathsf{UnOrdered}}
\newcommand{\getmaccs}{\mathsf{getMs}}

\newcommand{\mfb}{\mathsf{mfence}}

\newcommand{\getpairs}{\mathsf{getPairs}}
\newcommand{\nelimfs}{\mathsf{nelimFs}}
\newcommand{\isnoelim}{\mathsf{isNoElim}}
\newcommand{\uord}{O}
\newcommand{\reaches}{\mathsf{Reaches}}
\newcommand{\cfgto}{\mathsf{CFGTo}}

\newcommand{\reach}{\mathsf{Reach}}
\newcommand{\reachwo}{\mathsf{ReachWO}}

\newcommand{\lbreak}{\mathsf{break}}

\newcommand{\nelim}{{\mathsf{nfs}}}
\newcommand{\intermediate}{\mathsf{Intermediate}}
\newcommand{\onelim}{{\mathsf{NFS}}}

\newcommand{\iteri}{\mathsf{I}}
\newcommand{\iterj}{\mathsf{J}}

\newcommand{\fnc}{\mathsf{f}}

\newcommand{\isLd}{\mathsf{isLd}}
\newcommand{\isSt}{\mathsf{isSt}}
\newcommand{\isF}{\mathsf{isF}}

\newcommand{\opath}{\mathsf{Path}}
\newcommand{\oelim}{\mathsf{elim}}

\newcommand{\qr}{\texttt{qatomic\_read}}
\newcommand{\qw}{\texttt{qatomic\_write}}
\newcommand{\qra}{\texttt{qatomic\_read\_acquire}}
\newcommand{\qwl}{\texttt{qatomic\_write\_release}}

\newcommand{\facq}{\texttt{smp\_mb\_acquire}}
\newcommand{\fww}{\texttt{smp\_wmb}}
\newcommand{\ff}{\texttt{smp\_mb}}

\colorlet{dark-green}{green!70!black}
\colorlet{dark-red}{red!80!black}
\colorlet{dark-blue}{blue!80!black}

\newenvironment{observation}[1]{\par\noindent{\bf Observation.}\space#1}{}
\newenvironment{explanation}[1]{\par\noindent{\bf Explanation.}\space#1}{\hfill $\square$}

\tikzset{
   every path/.style={>=latex},
   lightarrow/.style={->,thick,>=triangle 60},
   longarrow/.style={->,thick,shorten >=-1mm,shorten <=-1mm},
   box1/.style={minimum height=3mm, minimum width=4mm, rounded corners,rectangle,draw=gray, 
                thick,inner xsep=1mm,inner ysep=1mm},
   src/.style={rounded corners,rectangle,draw=gray,fill=gray!10,inner sep=1.5mm}
}

\newcommand{\Subcase}[1]{\bigskip\noindent\textbf{Subcase {#1}:}}
\newcommand{\Subsubcase}[1]{\bigskip\noindent\textbf{Subsubcase {#1}:}}

\acrodef{ISA}{Instruction Set Architecture}
\acrodef{MTE}{Memory Tagging Extension}
\acrodef{MPK}{Memory Protection Keys}
\acrodef{CFI}{Control-Flow Integrity}
\acrodef{PAC}{Pointer Authentication}
\acrodef{DDC}{Default Data Capability}
\acrodef{PTE}{Page Table Entry}
\acrodef{MPX}{Memory Protection Extensions}
\acrodef{DBMS}{database management systems}
\acrodef{CVE}{Common Vulnerabilities and Exposures}
\acrodefplural{CVE}[CVEs]{Common Vulnerabilities and Exposures}
\acrodef{LLM}{Large Language Model}
\acrodef{LoC}{Lines of Code}
\acrodef{CHERI}{Capability Hardware Enhanced RISC Instructions}
\acrodef{ART}{Adaptive Radix Tree}
\acrodef{ABI}{Application Binary Interface}
\acrodef{RSS}{resident set size}

\newcommand{\arm}{Arm\xspace}
\newcommand{\riscv}{RISC-V\xspace}
\newcommand{\qemu}{\textsc{QEMU}\xspace}
\newcommand{\llvm}{LLVM\xspace}

\newcommand{\projname}{\textsc{CHERI-DB}\xspace}
\newcommand{\system}{\projname}
\newcommand{\cherifmt}{CHERI\xspace}
\newcommand{\cheri}{\cherifmt}
\newcommand{\mtefmt}{\ac{MTE}\xspace}
\newcommand{\mte}{\mtefmt}

\newcommand{\etal}{\textit{et al.}}

\newcommand{\Circled}[1]{\raisebox{.5pt}{\textcircled{\raisebox{-.9pt} {#1}}}}

\definecolor{codegreen}{rgb}{0,0.6,0}
\definecolor{codegray}{rgb}{0.5,0.5,0.5}
\definecolor{codepurple}{rgb}{0.58,0,0.82}
\definecolor{backcolour}{rgb}{0.95,0.95,0.92}

\lstdefinestyle{custom}{
    backgroundcolor=\color{backcolour},
    commentstyle=\color{codegreen},
    keywordstyle=\color{magenta},
    numberstyle=\tiny\color{codegray},
    stringstyle=\color{codepurple},
    basicstyle=\ttfamily,
    showstringspaces=false,
    breaklines,
    tabsize=2,
    numbers=left,
    columns=fullflexible,
    keepspaces=true,
    frame=lines,
    numbersep=2pt,
    captionpos=b,
    autogobble=true,
}

\newcommand{\thinmidrule}{\specialrule{\cmidrulewidth}{1pt}{1pt}}

\mdfdefinestyle{takeawaystyle}{
  hidealllines=true,
  leftline=true,
  linewidth=2.5pt,
  leftmargin=0pt,
  rightmargin=0pt,
  innerleftmargin=2pt,
  innerrightmargin=2pt,
  innertopmargin=2pt,
  innerbottommargin=2pt,
  skipabove=1pt,
  skipbelow=1pt,
  nobreak=false,
}

\newenvironment{takeaway}[1][black!5]{%
  \begin{mdframed}[
    style=takeawaystyle,
    linecolor=#1,
    backgroundcolor=#1!15!white,
  ]%
  \sffamily
  \noindent\textbf{Takeaway:}\;
}{%
  \end{mdframed}%
}

\newcommand{\equalcontribs}{\thanks{Equal contribution.}}

\definecolor{arch_color}{rgb}{0.35, 0.70, 0.49}       %
\definecolor{comprunt_color}{rgb}{0.90, 0.62, 0.00}   %
\definecolor{alloc_color}{rgb}{0.15, 0.20, 0.70}      %
\definecolor{dbms_color}{rgb}{0.40, 0.15, 0.70}       %
\newcommand{\coloropacity}{15}

\definecolor{bcgood}{RGB}{190,230,190}
\definecolor{bcpartial}{RGB}{225,242,200}
\definecolor{bcprob}{RGB}{255,236,160}
\definecolor{bcother}{RGB}{255,212,170}
\definecolor{bcmiss}{RGB}{245,180,175}
\definecolor{bcos}{RGB}{220,220,220}
\newcommand{\BcTrap}{\cellcolor{bcgood}\ding{51}}                         %
\newcommand{\BcLate}{\cellcolor{bcpartial}(\ding{51})}                    %
\newcommand{\BcProb}[1]{\cellcolor{bcprob}#1\%}                           %
\newcommand{\BcPartial}[1]{\cellcolor{bcprob}\ding{51}\,{\scriptsize(#1)}}       %
\newcommand{\BcOptIn}{\cellcolor{bcpartial}\ding{51}\textsuperscript{\textdagger}} %
\newcommand{\BcLibc}{\cellcolor{bcother}libc}                             %
\newcommand{\BcCrash}{\cellcolor{bcother}crash}                           %
\newcommand{\BcHang}{\cellcolor{bcother}hang}
\newcommand{\BcSilent}{\cellcolor{bcmiss}\ding{55}}                       %
\newcommand{\BcNoEffect}{\cellcolor{bcmiss}\ding{55}}                     %
\newcommand{\BcOSFault}{\cellcolor{bcother}OS}                              %
\newcommand{\BcNoOS}{\cellcolor{bcos}rev.}                                %
\newcommand{\BcWorks}{\cellcolor{bcgood}\ding{51}}                        %
\newcommand{\BcBreaks}{\cellcolor{bcmiss}\ding{55}}                       %
\newcommand{\BcBreaksProb}[1]{\cellcolor{bcprob}\ding{55}\,#1\%}          %
\newcommand{\BcNA}{n/a}
\newcommand{\BcLegend}[2]{\colorbox{#1}{\makebox[1.8em]{\strut #2}}}                     %

\crefname{figure}{Figure}{Figures}
\crefname{table}{Table}{Tables}

\renewcommand{\sectionautorefname}{\S\!}%
\renewcommand{\subsectionautorefname}{\S\!}%
\renewcommand{\subsubsectionautorefname}{\S\!}%
\crefname{section}{\S\!}{\S\!}
\crefname{subsection}{\S\!}{\S\!}
\crefname{subsubsection}{\S\!}{\S\!}
\Crefname{section}{\S\!}{\S\!}
\Crefname{subsection}{\S\!}{\S\!}
\Crefname{subsubsection}{\S\!}{\S\!}
\crefname{appendix}{Appendix}{Appendices}
\crefname{subappendix}{Appendix}{Appendices}
\crefname{subsubappendix}{Appendix}{Appendices}

\title{Should Your Database Systems Use Hardware-Assisted Memory Safety Extensions in Production?}%

\author{Ilya Meignan-{}-Masson}
\authornote{Both authors contributed equally.}
\affiliation{%
  \institution{Technical University of Munich}
  \country{Germany}
}
\author{Martin Fink}
\authornotemark[2]
\affiliation{%
  \institution{Technical University of Munich}
  \country{Germany}
}
\author{Masanori Misono}
\affiliation{%
  \institution{Technical University of Munich}
  \country{Germany}
}
\author{Dimitrios Stavrakakis}
\affiliation{%
  \institution{Technical University of Munich}
  \country{Germany}
}
\author{Pramod Bhatotia}
\affiliation{%
  \institution{Technical University of Munich}
  \country{Germany}
}

\begin{abstract}
Database systems are predominantly developed in unsafe languages (e.g., C/C++) to meet the performance requirements through low-level memory management, yet this reliance renders them prone to systemic memory-safety issues that compromise reliability, consistency, security, and durability. 
Through an extensive bug analysis of prominent database systems, we show that these memory-safety issues persist in production environments despite advancements in database testing tools. While emerging hardware-assisted extensions, such as Arm’s Memory Tagging Extension (MTE) and CHERI, offer a promising mitigation path for memory safety, their practical applicability and performance overhead within the specialized constraints of database systems
remain largely unexplored.

In this paper, we evaluate these modern hardware extensions through the lens of what we define as the {\em database trilemma}: the fundamental trade-off between {\em safety}, {\em performance}, and {\em portability}, to determine their viability for production-grade database systems. \rev[rev1]{R1.O6}{We provide the \emph{first} side-by-side comparison of MTE and CHERI across a database workload suite. Our bottom-up study spans microarchitecture, the compiler/runtime/OS stack, core data structures (ART, B+Tree, hash table, skip list, linked list, and queue), and full database systems (Redis, LevelDB, SQLite, MySQL, DuckDB, and LadyBugDB) to characterize their performance, safety guarantees, and ease of adoption.}

We find that while hardware extensions can offer near-determinis\-tic protection, they introduce non-uniform performance taxes: MTE provides high portability with modest overheads ($\approx$10\%), while CHERI delivers superior safety guarantees with higher performance penalties (20–60\%) and significant porting effort.
Our study equips database systems architects with an actionable guide towards building the next generation of reliable, secure databases using hardware-assisted safety mechanisms.
\footnote{This manuscript is the extended version (with complete Appendix) of the paper accepted at SIGMOD'27~\cite{memsafedb}.}
\end{abstract}

\setcopyright{cc}
\setcctype{by}
\acmJournal{PACMMOD}
\acmYear{2026} \acmVolume{4} \acmNumber{6 (SIGMOD)} \acmArticle{378}
\acmMonth{12} \acmDOI{10.1145/3856373}
\makeatletter
\gdef\acmConference@shortname{Proc. ACM Manag. Data 4, 6 (SIGMOD)}
\gdef\acmConference@date{Article 378}
\gdef\acmConference@venue{December 2026}
\makeatother
\renewcommand{\shortauthors}{Meignan-{}-Masson et al.}
\acmPrice{}

\received{17 April 2026}
\received[revised]{19 August 2026}
\received[accepted]{12 September 2026}


\maketitle

\begin{figure}[!t]
    \centering
    \revbox[rev1]{R1.O12}{\includegraphics[width=\linewidth]{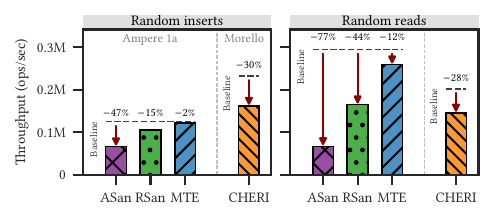}}
    \vspace{-3em}
    \caption{Overheads of ASan, \rev[rev1]{R1.O12}{RSan,} MTE, and CHERI on LevelDB.}
    \label{fig:dbms_leveldb_motiv}
    \vspace{-0.5em}
\end{figure}

\begingroup\small\noindent\raggedright\textbf{Artifact Availability:}
The artifacts are available at \url{https://github.com/TUM-DSE/memsafedb}.
\endgroup

\section{Introduction}

Database systems are predominantly implemented in memory-unsafe languages (e.g., C/C++), to meet the performance requirements of data-intensive applications.
Such languages provide developers with fine-grained control over memory management, enabling practitioners to design high-performance buffer managers, indexing structures, and query execution engines.
However, this power comes at a significant cost: {\em the lack of inherent memory safety}.

{\em Memory safety} denotes the guarantee that memory accesses are well-defined and restricted to their intended regions; violations of this property~\cite{szekeres2013sok, memory-errors} are categorized as {\em spatial} or {\em temporal}.
{\em Spatial errors} (e.g., overflows, out-of-bounds access) occur when a program accesses memory outside its intended boundaries, while
{\em temporal errors} (e.g., use-after-free, double-free) refer to when a program accesses memory after its deallocation.
In database systems, such violations constitute fundamental reliability, consistency, durability, and security issues.

The software industry has long grappled with memory safety~\cite{heartbleed, stagefright, ghost_libc}, with Microsoft and Google reporting that $\approx$70\% of their reliability and security issues stem from memory safety bugs~\cite{thechromiumprojectMemorySafety, android_vuln, msr_vuln}.
Database systems are no exception; \rev[rev1]{R1.O10}{our study across major open-source database systems (\autoref{subsec:bug_study}) reveals that memory-safety bugs remain a substantial share of all reported issues, and that even recently developed systems~\cite{duckdb, clickhouse} are not immune, confirming that memory safety remains a serious concern.}

One option to address these issues is to rewrite existing systems in memory-safe languages (e.g., Rust), which offer performance while providing built-in safety guarantees.
\rev[rev1]{R1.O15}{While such rewrites have proven successful, both for new projects~\cite{turso} and for rewriting core components of production systems~\cite{influxdb}, they remain a substantial undertaking for systems comprising millions of highly optimized lines of code, where the engineering cost is often prohibitive.}

As an alternative, database practitioners commonly resort to \emph{exhaustive testing} and \emph{software-based runtime checks}.
While formal verification can prove safety, it fails to scale to the complexity of modern database systems.
Meanwhile, \emph{software-based runtime approaches}~\cite{serebryanyGWPASanSamplingBasedDetection2024,tailcheck,asan,kroesDeltaPointersBuffer2018,camp} are prohibitively expensive (up to 79\% for lookups in LevelDB), as shown in \autoref{fig:dbms_leveldb_motiv}, and are, in general, not practical for production database systems.

To this end, hardware vendors have proposed {\em new hardware extensions} to provide memory safety with lower performance overheads.
In particular, several propositions such as ARM’s \acf{MTE}~\cite{mte}, \acf{CHERI}~\cite{cheri}, 
x86's upcoming ChkTag~\cite{chktag}, Intel's deprecated \acf{MPX}~\cite{mpx}, and Oracle’s SPARC Silicon Secured Memory~\cite{sparc} offer new primitives for memory safety.
As shown in \autoref{fig:dbms_leveldb_motiv}, such extensions reduce overhead considerably compared to software-based approaches---yet they still impose non-trivial costs, and adopting them requires non-trivial code modifications, raising portability concerns.

Together, these constraints expose a {\em ``Database Systems Trilemma''}: a fundamental conflict between \emph{safety}, \emph{performance}, and \emph{portability}.
In essence, developers must sacrifice performance for safety, or portability for a modern safe language, or, as is most common, sacrifice safety for the performance and control of unsafe languages.

\begin{figure}%
    \newcommand{\ra}[0]{\ref{sec:motivation}}
    \newcommand{\rd}[0]{\ref{sec:database-systems}}
    \newcommand{\rds}[0]{\ref{sec:datastructures}}
    \newcommand{\rc}[0]{\ref{sec:compiler-runtime}}
    \newcommand{\rh}[0]{\ref{sec:hw_analysis}}
    \newcommand{\rt}[0]{\ref{tab:takeaways}}
    \centering
    \includesvg{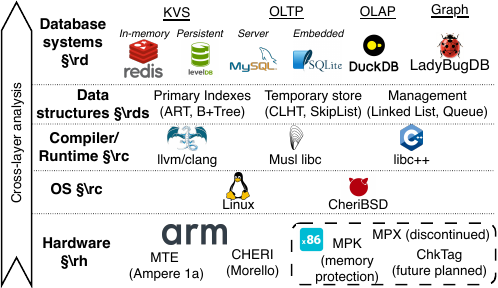}
    \vspace{-15pt}
    \caption{Analysis overview from hardware to full systems.}
    \label{fig:intro-overview}
    \vspace{-15pt}
\end{figure}

While these technologies are increasingly adopted in mobile ecosystems~\cite{android_mte, mte_android_aosp}, they have been slow to reach database systems, mainly due to the lack of a systematic analysis and experimentation with these hardware extensions in this context.
To this end, we aim to answer the central question: {\em Should your database system use hardware-assisted memory-safety extensions in production?}

To provide an answer, we conduct \rev[rev1]{R1.O6}{the first side-by-side, \emph{cross-layer} comparison} (\autoref{fig:intro-overview}) of memory-safety hardware extensions tailored for database systems, \rev[rev1]{R1.O6}{spanning from microarchitecture through data-structures to full systems}.
Our study begins with a {\em microarchitectural analysis} of the underlying hardware primitives (\autoref{sec:hw_analysis}).
Next, we evaluate how these primitives affect the {\em "building blocks" of database systems} by porting and benchmarking a set of critical data structures (\autoref{sec:datastructures}).
We scale our analysis to {\em production-grade} database systems, detailing the engineering effort and performance trade-offs of retrofitting them with hardware safety (\autoref{sec:database-systems}).
Finally, we analyze the role of the operating system and compiler toolchains for hardware-assisted memory safety (\autoref{sec:compiler-runtime}).

To ground our analysis in practice, we distill our findings into actionable takeaways for database practitioners.
At the microarchitectural level, we show that while hardware extensions effectively mask the cost of runtime checks, the overhead of managing metadata---such as CHERI's 128-bit capability loads or MTE’s tag-on-store requirements---can become a bottleneck for write-heavy workloads.
At the data structure layer, the ``performance tax'' is not uniform; cache-friendly structures (e.g., B+ trees) absorb the overhead of memory safety checks better than pointer-chasing structures (e.g., linked lists).
For full systems, there is an ``engineering-performance'' trade-off: MTE offers high portability and modest overheads ($<$10\%) but provides only probabilistic safety, whereas CHERI offers deterministic, fine-grained protection at the cost of higher porting complexity and performance penalties (20\%--60\%).

To summarize, our study serves as a bridge between architectural research and industrial database engineering. 
By quantifying the performance-safety trade-offs of modern hardware, we provide a concrete roadmap for building the next generation of ``reliable/secure-by-default'' data infrastructure. 
Our work demonstrates that hardware-assisted safety is finally approaching the maturity required for production database systems, offering a viable path to resolving the long-standing database trilemma.

Overall, this paper makes the following contributions:
\begin{itemize}[leftmargin=*]

\item {\bf Empirical bug study:} \rev[rev1]{R1.O10}{We quantify the prevalence of memory-safety-related \acp{CVE} and bugs in open-source database systems, showing they remain a substantial share of reported issues.}

\item {\bf  Hardware-assisted memory safety extensions:} We port six data structures and six database systems.
We detail the best practices and pitfalls of using the \acs{MTE} and \acs{CHERI} extensions.

\item {\bf  Cross-layer experimental analysis:} We evaluate the performance across four layers: {\em (i)~microarchitecture}, {\em (ii)~data structures}, {\em (iii)~database systems}, and {\em (iv)~compilers, runtime, and OS}.

\item {\bf The database trilemma synthesis:} We formalize and evaluate the trade-offs between safety, performance, and portability. 

\item {\bf Open-source evaluation framework:} To facilitate further research into secure and reliable data infrastructure, we make our entire evaluation pipeline, including the ``\acs{MTE}-tagged'' and ``\acs{CHERI}-fied''  database systems source code, publicly available.
\end{itemize}

\begin{table}[t]
    \footnotesize
    \centering
    \setlength{\tabcolsep}{2.8pt}
    \revbox[rev1][0pt]{R1.O11}{\begin{tabular}{@{}lcc|cccccccc@{}}
        \toprule
        & \textbf{Total} & \textbf{Memory} & \multicolumn{8}{c}{\textbf{Implication}} \\
        \textbf{} & \textbf{issues} & \textbf{safety (\%)} 
            & \textbf{Cr} & \textbf{DC} & \textbf{IQ} 
            & \textbf{RE} & \textbf{DoS} & \textbf{Co} & \textbf{Rp} & \textbf{DL} \\
        \midrule
        DuckDB     &  5,981 &   549 (9.2\%)  & 493 &  0 & 13 &  4 &  2 &  1 &  0 &  1 \\
        Redis      &  3,433 &   393 (11.4\%) & 370 &  2 &  0 &  9 &  1 &  0 &  0 &  0 \\
        RocksDB    &  2,432 &   227 (9.3\%)  & 205 &  3 &  0 &  6 &  0 &  0 &  0 &  0 \\
        ClickHouse & 22,295 & 1,681 (7.5\%)  & 1,434 &  6 & 57 & 24 &  3 & 23 &  0 &  0 \\
        MySQL      & 21,503 &   614 (2.9\%)  & 476 &  6 & 10 & 42 &  1 &  1 &  1 &  0 \\
        LevelDB    &    306 &    13 (4.2\%)  &  12 &  0 &  0 &  1 &  0 &  0 &  0 &  0 \\
        MariaDB    & 30,536 & 3,290 (10.8\%) & 2,905 & 41 & 60 & 70 &  1 &  1 &  1 &  0 \\
        \hdashline
        CVEs       &  1,448 &    92 (6.4\%)  &  60 &  0 &  1 &  0 &  3 &  0 &  0 &  0 \\
        \midrule
        \textbf{Total} & \textbf{87,934} & \textbf{6,859 (7.8\%)}
            & \textbf{5,955} & \textbf{58} & \textbf{141}
            & \textbf{156} & \textbf{11} & \textbf{26} & \textbf{2} & \textbf{1} \\
        \bottomrule
        \multicolumn{11}{@{}l}{\textbf{Cr}:Crash; \textbf{DC}:Data corruption; \textbf{IQ}:Incorrect query results; \textbf{RE}:Resource exhaustion} \\
        \multicolumn{11}{@{}l}{\textbf{DoS}:Denial of service; \textbf{Co}:Concurrency hazard; \textbf{Rp}:Replication issue; \textbf{DL}:Data loss} \\
    \end{tabular}}
    \caption{Memory safety bug and CVE analysis (2015--2025).}
    \label{tab:bug-cve-totals}
    \vspace{-25pt}
\end{table}

\section{Background and Motivation}
\label{sec:background}

We first establish why memory safety violations are a pressing concern for database systems, and why existing approaches fail to resolve the tension between safety, performance, and portability.

\subsection{Memory Safety in Database Systems}
\label{subsec:bug_study}
Despite the performance benefits of manual memory management, database systems written in unsafe languages (e.g., C/C++) are inherently exposed to spatial memory safety violations (e.g., buffer overflows, out-of-bounds accesses) and temporal violations (e.g., use-after-free, double-free)~\cite{depths-c,szekeres2013sok,van2012memory}, whose consequences go well beyond simple crashes.
A single erroneous memory access can cause \emph{silent data corruption} by overwriting adjacent records before they are flushed to storage, produce \emph{incorrect query results} by corrupting in-memory index structures, \textit{exhaust system resources}, introduce \emph{concurrency hazards} by corrupting shared metadata, \textit{expose sensitive user records} to unauthorized parties, or allow unauthorized transactions---ultimately threatening the correctness and durability guarantees that database systems are designed to provide.

\begin{figure}
    \centering
    \includegraphics[width=\linewidth]{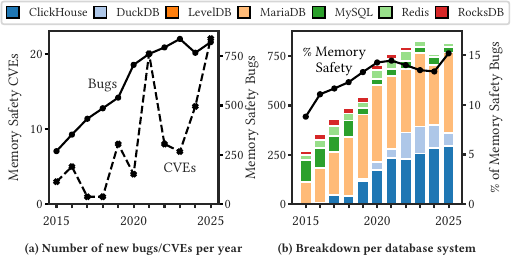}
    \vspace{-0.7cm}
    \caption{Time-based memory safety bug and CVE analysis.}
    \label{fig:bug_analysis}
    \vspace{-2em}
\end{figure}

To quantify the prevalence of memory safety bugs in real-world database systems, we conduct a systematic study of publicly reported \acp{CVE} and bug reports in the last decade (2015--2025) across \rev[rev3]{R3.O4}{seven open-source database systems---ClickHouse, DuckDB,
LevelDB, MariaDB, MySQL, Redis, and RocksDB}---spanning different deployment models and storage engines.
The volume of bugs and \acp{CVE} (\autoref{tab:bug-cve-totals}), and the non-expert nature of bug report writers
render manual classification impractical and simple keyword matching unreliable.

Instead, we propose using a \ac{LLM} to classify bug reports and \acp{CVE}.
We employ a locally hosted GPT-OSS 120B model on an NVIDIA H100 GPU to classify each report into \emph{bug}, \emph{programmer error}, or \emph{other}, as some bug trackers include additional reports (e.g., user errors, feature requests, and more).
Given the higher selectivity, we consider all \acp{CVE} to be bugs.
Further, the LLM determines whether a bug is memory-safety-related and, if so, identifies its primary operational implication, given a set of options. 
Each classification is repeated three times independently.
We treat a bug as a confirmed memory safety issue only when \textit{all} three repetitions agree.
Further details on the methodology are provided in \cref{sec:appendix-bug-study}.

\noindent
\rev[rev1]{R1.O8}{\myparagraph{Limitations} We do not aim to establish a new state-of-the-art vulnerability classifier, but to provide a system-specific overview complementing related work~\cite{liWhatTheyFix2026,torkamaniStreamliningSecurityVulnerability2025,ahmedSecVulEvalBenchmarkingLLMs2025,talibzadeUsingLLMsVulnerability2025} using more robust pipelines and frontier models.
Our pipeline has several limitations, and we report its results as \emph{motivating estimates}, not exact counts.
First, we run a quantized model without ground-truth validation. %
Second, requiring agreement across all three runs excludes inconsistently flagged borderline bugs.}
\rev[rev1]{R1.O9}{However, as the model tends to repeat errors~\cite{byerlySelfConsistencyFallsShort2026}, agreement provides less confidence than three independent judgments.}
\rev[rev1]{R1.O17}{Third, we classify only issue-tracker reports and \acp{CVE}, excluding fixes landed without corresponding issues;
our claims therefore concern memory-safety \emph{reports}, not all memory-safety bugs.}
\rev[rev1]{R1.O16}{Finally, MySQL redacts security-sensitive tracker issues, which may affect its memory-safety share.}
\rev[rev1]{R1.O10}{The per-year timeline is also affected by project age, the model's familiarity with recent terminology, and gaps in \ac{CVE} enrichment; we therefore interpret it as evidence of sustained presence, not a precise trend.}

Our analysis yields three key findings:
\begin{inparaenum}[(i)]
\item Memory safety bugs affect all database systems.
Even recent systems (e.g., DuckDB, ClickHouse) are not immune, highlighting that memory-safety violations remain prevalent despite good practices and modern software engineering techniques.
\item \rev[rev1]{R1.O10}{Memory-safety issues remain a substantial source of bugs in these systems today: through the period, they consistently account for a non-trivial share of all reported bugs and \acp{CVE} every year (\autoref{fig:bug_analysis}), despite significant advances in testing, fuzzing, and bug-finding tools~\cite{sqlsmith,afl,squirrel,fucci,dynsql,sqlancer,sqlancer_plus_plus, rags, apollo, testing_db_pivoted_query_synthesis,finding_bugs_query_partitioning}.
By contrast, large vendors such as Microsoft and Android identified memory safety as the source of $\approx70\%$ of security issues~\cite{thechromiumprojectMemorySafety,msr_vuln,android_vuln} and report a \emph{declining} share after deploying mitigations, e.g., Android from $76\%$ to $24\%$ between 2019--2024~\cite{GoogleOnlineSecurity2024}.
We observe no comparable decline in our dataset.}
\item Memory safety violations can cause diverse harm, as shown in \autoref{tab:bug-cve-totals}.
The most common implication is a system crash ($\approx$87\%), followed by resource exhaustion, and incorrect query results.
\end{inparaenum}

\subsection{Scope: The Database System Trilemma}
\label{sec:trilemma}

Memory safety solutions are governed by three competing dimensions: \emph{performance overhead},
\emph{protection coverage}, and \emph{portability}.
High-performance systems typically reject solutions that introduce non-trivial latency. 
Consequently, developers often employ best-effort safety models that prioritize execution speed over exhaustive protection.
However, the primary determinant for adoption is often the engineering effort required for integration.
For large-scale systems, the cost of manual refactoring often outweighs the benefits of improved memory safety.
Several software-based approaches have been proposed to prevent memory safety violations~\cite{RangeSanitizer,abadiControlflowIntegrityPrinciples2009,akritidisBaggyBoundsChecking2009,asan,bozdoganSafePMSanitizerPersistent2022,camp,catamaran,cecsan,cmasan,cucatch,dhurjatiSAFECodeEnforcingAlias2006,duckHeapBoundsProtection2016,duckStackBoundsProtection2017,giantsan,hasabnisLightweightBoundsChecking2012,jimCycloneSafeDialect2002,kroesDeltaPointersBuffer2018,kuvaiskiiSGXBOUNDSMemorySafety2017,kuznetsovCodepointerIntegrity2014,nagarakatteSoftBoundHighlyCompatible2009,neculaCCuredTypesafeRetrofitting2005,nethercoteValgrindFrameworkHeavyweight2007,serebryanyGWPASanSamplingBasedDetection2024,serebryanyMemoryTaggingHow2018,spp,tailcheck,wpbound}, with all facing the challenge of reconciling the tensions within this trilemma.
Their runtime overheads remain the most significant hindrance to production deployment.
\rev[rev1]{R1.O15}{The alternative of memory-safe languages shows promising results both in new projects~\cite{turso} and rewrites of core production components~\cite{influxdb}.
However, rewriting mature systems with millions of lines of highly optimized code remains a substantial and often prohibitively costly undertaking.}
In contrast, hardware-assisted memory safety promises significantly reduced overhead and, by extension, broader industry adoption.
In this paper, we conduct a detailed empirical analysis of two such extensions---ARM \ac{MTE} and \cheri---designed to provide hardware-backed memory-safety.

\begin{figure}
    \includegraphics[width=\columnwidth]{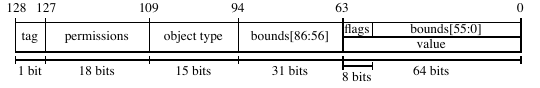}
    \vspace{-2.5em}
    \caption{CHERI Morello capability format.}
    \label{fig:arch-cheri-morello-ptr-layout}
    \vspace{-10pt}
\end{figure}
\begin{figure}
        \includegraphics[width=\columnwidth]{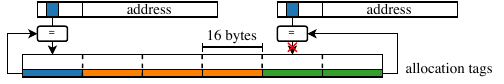}
        \vspace{-2.5em}
        \caption{MTE tag checking mechanism.}
        \label{fig:arch-mte-tag-check}
        \vspace{-15pt}
\end{figure}

\section{Architectural Analysis}
\label{sec:hw_analysis}

After establishing hardware-assisted memory safety as a promising direction, we describe and evaluate the microarchitectural foundations of two modern memory safety extensions implemented in \rev[rev1]{R1.O2}{research and production} hardware (\acs{CHERI} (\cref{subsec:hw-cheri}) and \acs{MTE} (\cref{subsec:hw-mte})) to realize the costs and constraints that any high-level system will inherit.
\rev[rev1]{R1.O2}{These two extensions differ in maturity: \ac{MTE} ships in production silicon today, whereas \cheri is available only on research hardware.}
We also analyze Intel \acs{MPK} in \cref{subsec:hw-mpk}.

\myparagraph{Experimental setup}
We conduct our experiments on the Morello Evaluation Platform~\cite{grisenthwaiteArmMorelloEvaluation2023a}, \rev[rev1]{R1.O2}{a prototype implementing CHERI}, and an Ampere 1a server, \rev[rev1]{R1.O2}{a commercially available server}.
\rev[rev1]{R1.O13}{We run all experiments with synchronous MTE (\cref{subsec:hw-mte}), as this is the only mode the Ampere 1a supports~\cite{kaushikOptimizedMemoryTagging2025}.}
\rev[rev1]{R1.O2}{For \cheri, we use Morello LLVM~\cite{MorelloLlvmproject2025}; when compiling in \emph{purecap} mode (\cref{subsec:hw-cheri}), we link against a capability-aware port of the Musl C standard library~\cite{MorelloMuslLibc2025}. For \ac{MTE}, we use upstream LLVM and glibc with \ac{MTE} enabled.}
The detailed testbed is described in our supplementary material (\autoref{appendix:arch_analysis}).
We evaluate instructions by executing them \qty{1.6e9}{times} in an unrolled loop of \qty{16}{instructions}.
We measure \emph{(i)}~\textit{Latency} with data dependencies between instructions and \emph{(i)}~\textit{Throughput} without dependencies.

\begin{table}[t]
    \footnotesize
    \centering
    \begin{tabularx}{\columnwidth}{l|Xcc}
        \toprule
        \textbf{Category} & \textbf{Instruction} & \textbf{Lat} & \textbf{Tp} \\
        \midrule
        \multirow{11}{*}{\textbf{Memory}}
        & ldr & 4.00 & 2.00 \\
        & str, str (in place) & --- & 2.00 \\
        & ldp, stp,str (cap), ldp (cap) & --- & 1.00 \\
        & ldr (cap) & 4.46 & 2.00 \\
        & stp (cap) & --- & 0.50 \\
        & ldxr, ldxr (cap) & --- & 0.18 \\
        & str (cap, in place), stp (in place) & --- & 0.33 \\
        & stp (cap, in place) & --- & 0.17 \\
        & stxr & --- & 0.13 \\
        & stxr (cap) & --- & 0.11 \\
        & cas (cap), swp (cap) & --- & 0.07 \\
        \thinmidrule
        \multirow{2}{*}{\textbf{Conversion}}
        & cvtd\_tocap, cvtp\_tocap, cvt\_tocap & 2.00 & 1.00 \\
        & cvtd\_toptr, cvtp\_toptr, cvt\_toptr & 1.00 & 1.00 \\
        \thinmidrule
        \multirow{8}{*}{\textbf{Cap Ops}}
        & add (cap), sub (cap), orrflgs, bicflgs, gcperm, gctype, gcvalue, gcflgs, gcseal, gctag, \mbox{clrperm}, cpy & --- & 2.82 \\
        & cfhi, cthi & 1.00 & 2.82 \\
        & cvtdz\_tocap, cvtpz\_tocap, cvtz\_tocap, cpytype, cpyvalue, rrlen, rrmask, alignd, alignu & 1.00 & 1.00 \\
        & gcbase, gclen, gclim, gcoff, scoff, scflgs, scvalue, scbnds, scbndse, seal, cseal, unseal & --- & 1.00 \\
        \thinmidrule
        \multirow{2}{*}{\textbf{Cap Checks}}
        & chkss, chkssu & --- & 0.50 \\
        & chktgd, chkeq & --- & 1.00 \\
        \bottomrule
        \multicolumn{4}{l}{\textbf{Lat}: Latency (cycles), higher is better; \textbf{Tp}: Throughput (inst/cycle), lower is better}
    \end{tabularx}
    \caption{Morello CHERI instructions}
    \label{tab:morello-insts}
    \vspace{-22pt}
\end{table}

\subsection{CHERI}
\label{subsec:hw-cheri}

\myparagraph{Mechanism}
CHERI~\cite{woodruffCHERICapabilityModel2014} replaces 64-bit pointers with 128+1-bit hardware-enforced capabilities that encode a memory address, the accessible boundaries, its permissions and object type, and a validity bit (\autoref{fig:arch-cheri-morello-ptr-layout}).
Capabilities are \emph{unforgeable} and \emph{monotonic}: overwriting a capability with arbitrary data clears its validity bit.
New capabilities can only be derived by narrowing existing ones.
Morello implements CHERI for Armv8.2-A, storing the validity bit alongside data in memory.
CHERI protects against spatial violations and enables control-flow integrity, while temporal safety can be achieved through capability revocation~\cite{xiaCHERIvokeCharacterisingPointer2019,wesleyfilardoCornucopiaTemporalSafety2020,filardoCornucopiaReloadedLoad2024}.
\rev[rev1]{R1.O1}{Code can be compiled in two modes: \emph{purecap}, where every pointer is a capability, or \emph{hybrid}, where only explicitly annotated pointers are.}

\myparagraph{Microarchitectural Analysis}
We evaluate a representative set of Morello instructions in \cref{tab:morello-insts}.
For memory operations, we distinguish three groups: plain loads/stores (ldr, str) moving 64-bit integers via capability-based addresses, \emph{cap} variants loading or storing 128+1-bit capabilities, and \emph{in place} variants that repeatedly access the same address rather than advancing the pointer.

Notably, loading one capability achieves the same throughput as a 64-bit integer load (\qty{2}{inst/cycle}), even though the external data bus remains 64-bit wide.
Loading two capabilities and two integers both reach \qty{\approx1}{op/cycle}.
Stores, however, scale with the amount of data written; \textit{stp~(cap)} reaches only \qty{0.5}{inst/cycle}, half that of \textit{stp}.

\begin{figure*}[t]
    \begin{subfigure}[t]{0.33\textwidth}
        \revbox[rev1]{R1.O1}{\includegraphics{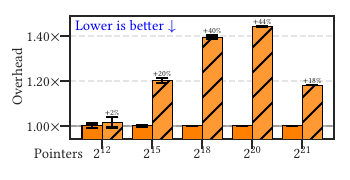}}
        \vspace{-2.7em}
        \caption{Pointer chasing.}
        \label{fig:arch-cheri-ptr-chase}
    \end{subfigure}
    \begin{subfigure}[t]{0.33\textwidth}
        \revbox[rev1]{R1.O1}{\includegraphics{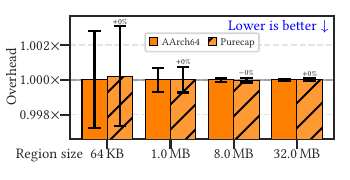}}
        \vspace{-2.7em}
        \caption{Memory copy.}
        \label{fig:arch-cheri-memcpy}
    \end{subfigure}
    \begin{subfigure}[t]{0.33\textwidth}
        \revbox[rev1]{R1.O1}{\includegraphics{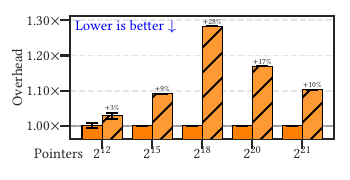}}
        \vspace{-2.7em}
        \caption{Atomic operations.}
        \label{fig:arch-atomic-table}
    \end{subfigure}
    \vspace{-2.2em}
    \caption{Linear and non-linear memory access patterns with and without CHERI.}
    \label{fig:arch-cheri-access-patterns}
    \vspace{-1.4em}
\end{figure*}

\begin{figure*}[t]
    \begin{minipage}[t]{0.66\textwidth}
        \vspace{0pt}
        \centering
        \begin{subfigure}[t]{0.49\textwidth}
            \includegraphics{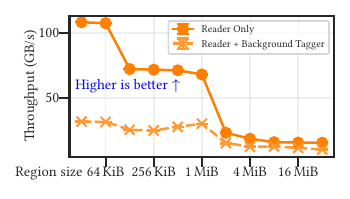}
            \vspace{-2.5em}
            \caption{CHERI validity tag.}
            \label{fig:arch-cheri-bg-tagger}
        \end{subfigure}
        \begin{subfigure}[t]{0.49\textwidth}
            \includegraphics{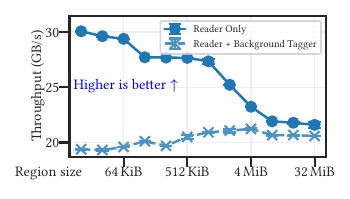}
            \vspace{-2.5em}
            \caption{MTE allocation tag.}
            \label{fig:arch-mte-bg-tagger}
        \end{subfigure}
        \vspace{-1.5em}
        \caption{Load throughput with/without a background thread writing memory tags.}
        \label{fig:arch-bg-tagger}
        \vspace{-1.5em}
    \end{minipage}
    \hfill
    \begin{minipage}[t]{0.32\textwidth}
        \vspace{0pt}
        \centering
        \revbox[rev3]{R3.O6}{\includegraphics[width=\linewidth]{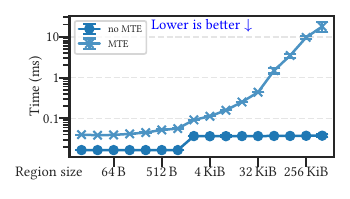}}
        \vspace{-3.05em}
        \caption{Glibc malloc time with and without \ac{MTE} tagging.}
        \label{fig:runtime-mte-alloc}
        \vspace{-1.5em}
    \end{minipage}
\end{figure*}

\begin{figure*}[t]
    \begin{subfigure}[t]{0.33\textwidth}
        \revbox[rev1]{R1.O1}{\includegraphics{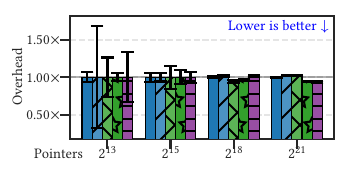}}
        \vspace{-2.7em}
        \caption{Pointer chasing.}
        \label{fig:arch-mte-ptr-chase}
    \end{subfigure}
    \begin{subfigure}[t]{0.33\textwidth}
        \revbox[rev1]{R1.O1}{\includegraphics{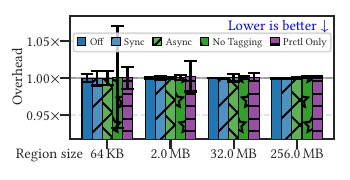}}
        \vspace{-2.7em}
        \caption{Memory read.}
        \label{fig:arch-mte-memread}
    \end{subfigure}
    \begin{subfigure}[t]{0.33\textwidth}
        \revbox[rev1]{R1.O1}{\includegraphics{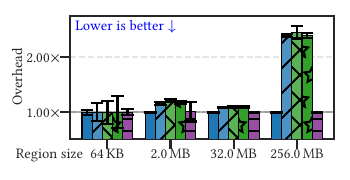}}
        \vspace{-2.7em}
        \caption{Memory write.}
        \label{fig:arch-mte-memwrite}
    \end{subfigure}
    \vspace{-2.2em}
    \caption{Linear and non-linear memory access patterns with and without \ac{MTE}. \rev[rev1]{R1.O1}{Variance is high for the smallest working sets because they fit entirely in cache, resulting in short execution times dominated by fixed overheads and noise.}}
    \label{fig:arch-mte-access-patterns}
    \vspace{-1.5em}
\end{figure*}

\myparagraph{Memory}
We benchmark Morello in two configurations:
\emph{(i)}~\textit{AArch64}, running unmodified code and 
\emph{(ii)}~\textit{Purecap}, where all accesses use fine-grained capabilities.
We perform a set of benchmarks for different access patterns:
\cref{fig:arch-cheri-ptr-chase} presents pointer chasing in a randomized list, \cref{fig:arch-cheri-memcpy} reflects the memory-region copy, and \cref{fig:arch-atomic-table} shows the atomic accesses, where three readers and one writer operate on pointers/capabilities concurrently.
In \cref{fig:arch-cheri-ptr-chase}, the Purecap list contains half the entries of the AArch64 variant, as capabilities occupy 128 bits.
In \cref{fig:arch-cheri-access-patterns}, we observe a significant overhead, caused by the change from pointers to capabilities, which decreases at larger working sets: once accesses miss in cache, a full 64-byte cache line accommodates both a pointer and a capability, equalizing the cache-miss penalty.
For copies, we see no notable overhead, as both variants copy the same amount of data, and the baseline performs checks using the \ac{DDC}~\cite[\S{}1.3.5]{armltd.ArmArchitectureReference2022}.
In \cref{fig:arch-cheri-bg-tagger}, concurrent tag writes cause a significant throughput drop; writing the validity bit invalidates the entire cache line, since tags are stored inline with data, forcing readers to re-fetch both. \rev[rev1]{R1.O1}{As a result, all configurations converge to the RAM access throughput.}

\begin{takeaway}[arch_color]
\cheri doubles the pointer size, thus halving the cache capacity for pointer-heavy workloads.
While the hardware manages single-capability loads efficiently, concurrent metadata updates incur overhead from cache line invalidations.
\end{takeaway}

\begin{table}
    \footnotesize
    \centering
    \begin{tabularx}{\columnwidth}{lXcc}
        \toprule
        \textbf{Inst} & \textbf{Description} & \textbf{Lat} & \textbf{Tp} \\
        \midrule
        irg & Generate random tag. & 3.41 & 0.29 \\
        addg & Add with tag, skipping excluded tags. & 0.98 & 2.04 \\
        subg & Subtract with tag, skipping excluded tags. & 0.98 & 2.04 \\
        subp & Subtract pointer, preserving the tag. & 0.98 & 3.88 \\
        subps & Subtract pointer, preserving the tag, setting flags. & 0.98 & 1.94 \\
        \midrule
        stg & Store allocation tag. & --- & 1.02 \\
        st2g & Store two allocation tags. & --- & 0.51 \\
        stzg & Store allocation tag, zeroing. & --- & 1.02 \\
        stz2g & Store two allocation tags, zeroing. & --- & 0.51 \\
        stgp & Store allocation tag and pair of registers. & --- & 1.02 \\
        ldg & Load allocation tag. & --- & 1.99 \\
        \bottomrule
        \multicolumn{4}{l}{\textbf{Lat}: Latency (cycles), higher is better; \textbf{Tp}: Throughput (inst/cycle), lower is better}\\
    \end{tabularx}
    \caption{MTE instructions}
    \label{tab:mte-insts}
    \vspace{-24pt}
\end{table}

\begin{table}[t]
    \footnotesize
    \centering
    \begin{tabular}{lccccc}
        \toprule
        Mode & Off & Sync & Async & No Tagging & Prctl Only \\
        \midrule
        read & 9.5\,M & 10.2\,M & 9.5\,M & 10.5\,M & 14.3\,M \\
        write & 179.4\,M & \textbf{468.1\,M} & \textbf{467.4\,M} & \textbf{468.2\,M} & 184.0\,M \\
        \bottomrule
    \end{tabular}
    \caption{Data Cache Misses for \ac{MTE}.}
    \label{tab:mte-dcache-misses}
    \vspace{-24pt}
\end{table}

\subsection{MTE}
\label{subsec:hw-mte}

\myparagraph{Mechanism}
Arm's \acf{MTE}~\cite{armltd.Armv85AMemoryTagging2019}\cite[D.10]{armltd.ArmArchitectureReference2025} \rev[rev1]{R1.O5}{assigns tags to memory regions and pointers and checks if they match on every access.}
Memory is divided into aligned \qty{16}{byte} \emph{tag granules}, each with a 4-bit \emph{allocation tag} stored separately or alongside the data~\cite{kaushikOptimizedMemoryTagging2025}.
Pointers embed a \emph{logical tag} in bits 56--59.
On memory access (\cref{fig:arch-mte-tag-check}), the hardware checks that tags match or triggers a fault: synchronously at the offending instruction, or asynchronously via a status register checked on kernel entry.
\rev[rev1]{R1.O5}{This design has two consequences:}
By assigning different tags to adjacent or freed regions, \ac{MTE} protects against spatial and temporal violations, but not intra-object overflows.
Additionally, \ac{MTE} limits the number of tags and only provides probabilistic safety.

\begin{figure}[t]
    \centering
    \revbox[rev3]{R3.O6}{\includegraphics[width=\columnwidth]{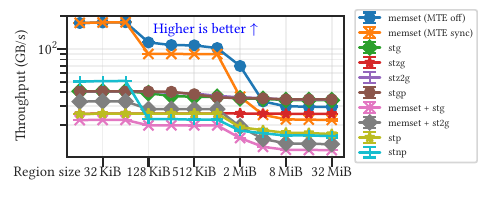}}
    \vspace{-3.1em}
    \caption{Memory and tag initialization.}
    \label{fig:arch-mte-init-memory}
    \vspace{-1.2em}
\end{figure}

\myparagraph{Microarchitectural Analysis}
\cref{tab:mte-insts} lists the performance of \ac{MTE} instructions.
In \cref{fig:arch-mte-access-patterns}, we measure \ac{MTE} overhead across five configurations: \textit{Off}, \textit{Sync} \rev[rev1]{R1.O5}{(faults on offending instruction)}, \textit{Async} \rev[rev1]{R1.O5}{(faults on kernel entry)}, \textit{No Tagging} (sync, zero tag), and \textit{Prctl Only} (sync, MTE disabled for accessed pages).
On the Ampere~1a, allocation tags are stored alongside data; reads therefore fetch both tag and data in a single operation, incurring no overhead (\cref{fig:arch-mte-ptr-chase,fig:arch-mte-memread}).
Stores (\cref{fig:arch-mte-memwrite}) must load the existing allocation tag before writing; this nearly triples the number of D-cache misses under MTE (179.4\,M $\to$ 468.1\,M, \cref{tab:mte-dcache-misses}).
As our server only implements synchronous \ac{MTE}, \textit{sync}, \textit{async}, and \textit{no tagging} perform similarly.
\cref{fig:arch-mte-bg-tagger} shows memory throughput with a second thread changing allocation tags, showing a similar trend as \cheri. %

\begin{takeaway}[arch_color]
MTE provides probabilistic safety with an asymmetrical performance profile: reads remain efficient, but writes incur a higher performance penalty.
\end{takeaway}

\noindent
\rev[rev3]{R3.O6}{
\myparagraph{Memory Allocations}
We evaluate glibc's ptmalloc implementation, with and without \ac{MTE}, by allocating/freeing chunks of various sizes in \autoref{fig:runtime-mte-alloc}.
Without \ac{MTE}, allocation complexity is $\mathcal{O}(1)$, with a slow-down above \qty{1}{KiB} where glibc switches from extending its contiguous heap via \textit{sbrk} to obtaining separately mapped memory via \textit{mmap}.
With \ac{MTE}, small allocations lose the fast \textit{sbrk} allocation path, and the allocation time grows linearly ($\mathcal{O}(n)$) as the entire region must be tagged, evicting cache lines in the process.
Thus, large, short-lived allocations may benefit from different protection mechanisms (e.g., guard pages).
}

\rev{}{
When initializing memory, choosing the right primitive also matters for performance.
\cref{fig:arch-mte-init-memory} compares several initialization strategies.
For buffers that fit in data caches, \textit{memset} performs the best.
For larger allocations, tag-setting instructions (\textit{stzg}, \textit{stgp}) outperform plain stores (\textit{stp}, \textit{stnp}, \textit{memset}) because the CPU implicitly fetches the allocation tag before stores, leading to \qty{95.87}{\percent} D-cache miss rate (\cref{tab:perf-mte}).
Tag-setting instructions write the tag without the implicit fetch, reducing the miss rate to \qty{0.08}{\percent}.
}

\begin{table}[t]
    \footnotesize
    \centering
    \revbox[rev3][0pt]{R3.O6}{
    \hspace{1em}
    \begin{tabular}{lcccc}
        \toprule
        Metric & memset & stzg/stgp & stnp & stp \\
        \midrule
        Cycles & \num{2.48e9} & \num{1.6e9} & \num{3.47e9} & \num{3.22e9} \\
        Cache Miss Rate & 95.87\% & \textbf{0.08\%} & 24.79\% & 24.80\% \\
        \bottomrule
    \end{tabular}
    \hspace{1em}
    }
    \caption{Memory initialization perf stats (32MB).}
    \label{tab:perf-mte}
    \vspace{-25pt}
\end{table}

\section{Analysis of Data Structures}
\label{sec:datastructures}
Building upon our %
analysis, we now investigate how \ac{MTE} and \cheri overheads propagate to the core data structures of modern database systems, chosen within a representative taxonomy.

\subsection{Data Structures Taxonomy}
To this end, we propose a use-case-driven taxonomy to ensure coverage of the distinct memory access patterns found in practice.

\myparagraph{Primary access to data}
Primary index data structures express relationships across large parts of the data and are used to improve the performance of typical operations.
For Key-Value Store workloads, systems typically use \textit{Log-Structured Merge (LSM)} trees or \textit{hash tables}. 
\textit{$B^{*}$-trees} and \textit{tries} are common for transaction-oriented systems, whereas analytics-oriented systems usually opt for \textit{block range} indexes. 
Finally, graphs are typically represented in \textit{adjacency lists} using the Compressed Sparse Row (CSR) format.

\myparagraph{Temporary buffering}
Database systems use additional structures to hold temporary data. 
Temporary data can be buffered before writing to storage, especially in write-heavy workloads, or when executing long-running, complex queries. 
Key-Value Store systems commonly use \textit{Skip lists} to buffer writes in memory, while transactional systems usually perform writes directly in the primary data structure.
Analytical workloads require complex selections and joins, which are typically performed using \textit{hash tables}.

\myparagraph{Internal organization} 
Database systems employ auxiliary data structures internally to organize transient data and manage state and operations (e.g., \textit{queues}). 
For our analysis, we choose established representatives of the above categories, listed in \autoref{table:ds_operations}, along with their operations and memory access types.
For primary indexes, we select the B+ tree from the Leanstore~\cite{leisVirtualMemoryAssistedBuffer2023} implementation that uses optimistic lock coupling, and the Adaptive Radix Tree (ART)~\cite{leisAdaptiveRadixTree2013}. 
For temporary data buffering, we consider the cache-line-aligned hash table (CLHT)~\cite{davidAsynchronizedConcurrencySecret2015} and LevelDB's Skip list~\cite{googleinc.LevelDB2026}. 
Finally, we examine Boost's implementations of a queue and a linked list.

\subsection{Porting to Memory Safe Hardware}

\begin{figure*}
\centering
\revbox[rev3][0pt]{R3.O2}{\includegraphics[width=\textwidth]{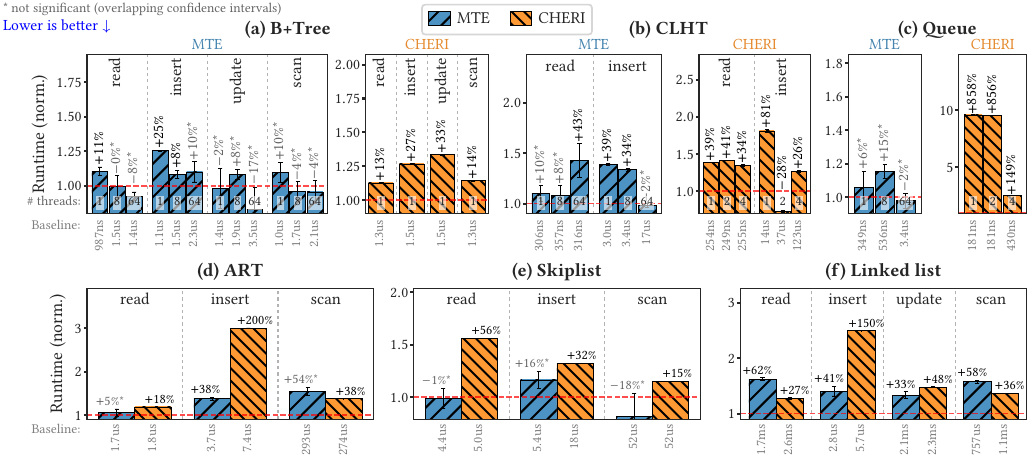}}
\vspace{-20pt}
\caption{Overhead of \ac{MTE} and \cheri on commonly-used data structures in database systems.}
\label{fig:datastructures_all}
\vspace{-15pt}
\end{figure*}

\myparagraph{\cheri}
For the \cheri \texttt{purecap} target, the compiler handles most of the portable C/C++ code.
The compiler natively inserts capability-aware atomic operations, used in lock-free data structures.

Three classes of issues require manual intervention.
First, \cheri enforces 16-byte capability alignment, disrupting cache-aware designs designed to prevent false sharing or pack pointers densely.
Second, traditional bit-wise pointer manipulation for custom alignment discards capability metadata, triggering runtime violations. 
Third, pre-allocating a large region and subdividing it at runtime holds only a capability covering the entire range by default, providing no sub-region safety.
We resolve these issues by replacing alignment-dependent code (e.g., in Boost's queue) with \cheri-native alignment functions, and narrowing the returned capability to each allocated block at the point of subdivision.

\myparagraph{\mte}
Porting to \ac{MTE} requires no changes beyond the memory allocator.
For data structures that rely on the system allocator, we use an \ac{MTE}-enabled allocator (e.g., glibc's Ptmalloc); the data structure code itself remains intact.
Turning on tagging mandates glibc to support memory tagging (>2.33) and to set the \textit{GLIBC\_TUNABLES} environment variable to \textit{glibc.mem.tagging=3}.

However, high-performance data structures often define their own allocators.
We port these by modifying alloc/free functions to align pointers to 16-byte granule boundaries, assign a random tag to the pointer (using \textit{irg}), and, for larger allocations, iterate over the allocated memory range and set the allocation tag (with \textit{stg}).

Another non-trivial case arises in data structures that encode links between elements as custom indices rather than pointers---our B-tree, for instance, stores parent-to-child links as Page IDs (PIDs) and reconstructs the pointer on every traversal.
To minimize changes to the data structure design, we first reconstruct the pointer and then fetch the allocation tag using the \textit{ldg} instruction.

\begin{takeaway}[comprunt_color]
Retrofitting data structures for hardware safety extensions requires careful porting due to the pointer length and alignment assumptions.
Porting to \ac{MTE} is simpler, as most modifications are in the allocators, whereas \cheri requires invasive refactoring to maintain capability provenance and alignment.
\end{takeaway}

\newcolumntype{y}{>{\hsize=0.03\hsize\linewidth=\hsize}X}
\newcolumntype{s}{>{\hsize=0.13\hsize\linewidth=\hsize}X}
\newcolumntype{o}{>{\hsize=0.1\hsize\linewidth=\hsize}X}
\newcolumntype{d}{>{\hsize=0.66\hsize\linewidth=\hsize}X}
\begin{table}
\label{tab:ds_all}
\footnotesize
\begin{tabularx}{\linewidth}{@{}ysod@{}}
    \toprule
    \textbf{} & \textbf{Structure} & \textbf{Op} & \textbf{Description} \\ \midrule
    \multirow{9}{*}{\rotatebox[origin=c]{90}{Primary}} &
        \multirow{5}{*}{B+Tree} &  read  & Follow fences down the tree \textit{(Rnd)} \\
                         & &  scan  & Read \textit{(Rnd)} and iterate in last level node(s) \textit{(Seq)} \\
                         & & \multirow{2}{*}{insert} & Find location \textit{(Rnd)} then emplace \textit{(CAS W)} \\ 
                         & &        & and potential rebalancing \textit{(Rnd)} \\
                         & & update & Find location \textit{(Rnd)} then replace \textit{(CAS W)} \\ \cmidrule{2-4}
        & \multirow{4}{*}{ART} &  read  & Follow partial keys down the tree \textit{(Rnd)}    \\
                         & &  scan  & Read \textit{(Rnd)} and iterate in last level node(s) \textit{(Seq)} \\
                         & & \multirow{2}{*}{insert} & Find location \textit{(Rnd)} then emplace \textit{(W)}  \\
                         & &        & and potential node growth/rebalancing \textit{(Rnd)} \\ \thinmidrule
    \multirow{5}{*}{\rotatebox[origin=c]{90}{Temporary}} &
        \multirow{2}{*}{CLHT} 
                         & insert & Compute hash then emplace \textit{(Rnd + CAS W)}  \\
                         & &  read  & Compute hash then retrieve \textit{(Rnd)} \\ \cmidrule{2-4}
        & \multirow{3}{*}{Skiplist}
                         &  read  & Use index and access element \textit{(Rnd)} \\
                         & &  scan  & Use index \textit{(Rnd)} then iterate \textit{(Seq)} \\
                         & & insert & Use index \textit{(Rnd)} then emplace \textit{(W)}\\ \thinmidrule
    \multirow{6}{*}{\rotatebox[origin=c]{90}{Management}} &
        \multirow{2}{*}{Queue}
                         & produce & Put element and advance head (Rnd CAS) \\
                         & & consume & Get element and advance tail (Rnd CAS) \\ \cmidrule{2-4}
        & \multirow{4}{*}{\parbox[l]{1.5cm}{Linked\\list}}
                            &  read  & Iterate through the list \textit{(Rnd)} \\
                         & &  scan  & Iterate through for each element \textit{(Rnd)} \\
                         & & insert & Emplace at the tail \textit{(W)} \\ 
                         & & update & Find location \textit{(Rnd)} then replace \textit{(W)} \\ \bottomrule
    \end{tabularx}
\caption{Data structures operations by memory access types \textit{Random (Rnd), Sequential (Seq), Write (W), and Atomic (CAS)}}
\label{table:ds_operations}
\vspace{-22pt}
\end{table}

\subsection{Experimental Analysis}

\myparagraph{Methodology}
We evaluate the performance overhead of \ac{MTE} and \cheri on each data structure operation using microbenchmarks.
\rev[rev1]{R1.O13}{Unless noted otherwise, all experiments use the same experimental setup as described in \cref{sec:hw_analysis}.}
We plot the latency relative to the respective native baseline in \autoref{fig:datastructures_all}.
\rev[rev3]{R3.02}{We report configurations whose baseline and safety-enabled 95\% confidence intervals overlap as not statistically significant and mark them with an asterisk.}
\autoref{table:ds_operations} lists each structure's supported operations (insert, read, update, scan) and their memory access patterns.
For the queue, we pair a producer and consumer thread and measure the latency of a single enqueue/dequeue round-trip.
\rev[rev3]{R3.O2}{We evaluate concurrent data structures (B+Tree, CLHT, and queue) across thread counts, \textit{i.e.,} 1, 8, and 64 on the \ac{MTE} platform and 1, 2, and 4 on the 4-core Morello board.}

\myparagraph{B+Tree}
The B+Tree shows a moderate overhead (\autoref{fig:datastructures_all}a), \rev[rev3]{R3.O2}{ranging up to 25\% on \ac{MTE} and from 13\% to 33\% on \cheri when single-threaded}.
Inserts and updates are the most affected operations, as pointer chasing combined with optimistic lock coupling requires expensive atomic synchronization.
Scans, however, are efficient (\rev[rev3]{R3.O2}{10\% \ac{MTE}}, 14\% \cheri), since the large 4\,KiB nodes frequently contain the entire 100-key scan range, minimizing inter-node traversals.
\rev[rev3]{R3.O2}{With rising thread counts, the \ac{MTE} overhead shrinks and, at 64 threads, is no longer statistically significant for any operation: cache-coherence traffic and lock-coupling contention occupy a larger portion of the runtime and hide the tagging cost.
We do no report multi-threaded \cheri configuration: the B+Tree uses C++ exceptions to implement optimistic lock coupling, and we could not get exception unwinding working under purecap (\cref{sec:compiler-runtime}).}

\myparagraph{ART}
For ART, \ac{MTE} overhead is 5--38\%, while \cheri reaches 18--200\%, with the gap widening for write-heavy operations (\autoref{fig:datastructures_all}\rev[rev3]{R3.O2}{d}).
Insertions are the most expensive (38\% \ac{MTE}, 200\% \cheri) as node growth and rebalancing trigger extra random accesses and amplify write penalties.
Scans incur moderate overhead on \cheri (38\%) when traversal crosses node boundaries to locate siblings, whereas intra-node sequential access remains largely unaffected.

\smallskip
\myparagraph{CLHT}
In CLHT operations (\autoref{fig:datastructures_all}\rev[rev3]{R3.O2}{b}), \rev[rev3]{R3.O2}{\ac{MTE} overhead is modest at up to 43\%, while \cheri is more pronounced at 26--81\%}, reflecting the sensitivity of lock-free designs to atomic operation costs.
Read-only workloads show minimal \ac{MTE} impact \rev[rev3]{R3.O2}{at 1--8 threads and 43\% at 64 threads,} but incur \rev[rev3]{R3.O2}{34--41\%} overhead under \cheri, where even pointer loads carry a validation cost.
Inserts are the most expensive (\rev[rev3]{R3.O2}{up to 39\% \ac{MTE}, 81\% \cheri}), as they rely on multiple atomic compare-and-swap operations, more costly under \cheri.

\smallskip
\myparagraph{Skip list}
The skip list \rev[rev3]{R3.O2}{shows no statistically significant \ac{MTE} overhead, while \cheri adds 15--56\%} (\autoref{fig:datastructures_all}\rev[rev3]{R3.O2}{e}).
Although the multi-level index reduces search depth, random pointer chasing across levels still incurs a 56\% penalty on \cheri.
\rev[rev3]{R3.O2}{Insertions cost 32\% on \cheri}, as maintaining the skip list multiplies pointer updates.

\smallskip
\myparagraph{Queue}
The queue (\autoref{fig:datastructures_all}\rev[rev3]{R3.O2}{c}) starkly contrasts the two extensions.
\ac{MTE} almost matches the baseline, as the queue stores integer values that bypass tag checking.
Conversely, \cheri incurs
\rev[rev1]{R1.O14}{bigger overhead (up to 858\% in the worst-case scenario), since pure-capability mode protects every internal pointer and the lock-free produce/consume logic requires frequent atomic writes, an operation heavily penalized by \cheri.}
\rev[rev1]{R1.O14}{However, in a more realistic scenario with more threads, contention on the head and tail represents a bigger portion of the latency, and the \cheri overhead drops to 149\%.}

\smallskip
\myparagraph{Linked list} 
For the linked list (\autoref{fig:datastructures_all}f), \ac{MTE} overhead (33--62\%) is primarily due to random pointer chasing during full-list traversals, where each node access may incur a cache miss.
Under \cheri, insertions peak at 150\% due to heavy write penalties, while read and scan overheads (\rev[rev3]{R3.O2}{27--36\%}) are comparable to \ac{MTE}, as the bottleneck shifts from capability validation to memory latency.
\rev[rev3]{R3.O6}{The non-linear scaling we observe for allocation-heavy operations stems from the $\mathcal{O}(n)$ allocation cost of \ac{MTE} tagging (\cref{sec:hw_analysis}).}

\begin{takeaway}[dbms_color]
The overhead of \ac{MTE} and \cheri is dictated by the data structure memory access patterns.
Cache-aligned, contiguous data structures, \textit{e.g.}, B+ Trees, incur less overhead.
Conversely, lock-free structures and those that rely heavily on pointer chasing %
experience more performance degradation.
\end{takeaway}

\section{Database Systems}
\label{sec:database-systems}

After seeing the overhead at the data structure level, we analyze the impact of \cheri and \ac{MTE} across a representative set of database systems to examine whether these costs remain acceptable when \ac{MTE} and \cheri are applied to full-fledged database systems under realistic workloads and how much engineering effort porting each system requires. 
\rev[rev1]{R1.O4}{Our ports are best-effort: we are not the maintainers of these systems, so our changes may still contain bugs and could be sub-optimal.}
\autoref{tab:dbms} lists the chosen database systems, the workloads used, and the modified Lines of Code (LoC) required to port them to \ac{MTE} and \cheri.
\rev[rev1]{R1.O13}{All experiments in this section use the experimental setup as described in \cref{sec:hw_analysis}.}
Note that all systems work natively out of the box on Arm architectures (\textit{AArch64}).

\myparagraph{Experimental parameters}
\rev[rev1]{R1.O4}{We sweep both the working-set size and the thread count.
Working sets span three points an order of magnitude apart: \qty{100}{K}--\qty{10}{M} records for LevelDB, \qty{10}{K}--\qty{1}{M} keys for Redis, \qty{10}{K}--\qty{10}{M} rows for MySQL, 1--50 warehouses for SQLite, and scale factors 1--100 for DuckDB and LadyBugDB.
Thread counts are bounded by the available cores: 1, 8, and 64 on the 192-core Ampere~1a, and 1, 2, and 4 on the 4-core Morello board.}

\subsection{In-Memory Key-Value Store (Redis)}
Redis is a popular in-memory key-value store that caches clients' data in an ordered fashion.
Redis uses a hash table with separate chaining for collision handling and relies on \textit{jemalloc} for superior performance and memory utilization in database workloads~\cite{durner2019impact}.

\myparagraph{Porting}
To port Redis to \cheri, we use the purecap port of Musl libc, with static linking to avoid mixed-ABI symbols.
The main code changes involve replacing pointer casts to 64-bit integer types (e.g., \textit{unsigned long}) with \textit{(u)intptr\_t} to preserve capability metadata, and ensuring 16-byte capability alignment.
One non-trivial issue arises when Redis exploits the contiguity of \textit{argv[0]} through \textit{argv[argc]} to write to a shared buffer via \textit{argv[0]}.
This trick is forbidden by the capabilities model, as each pointer is bound to its own allocation.
We fix this by restricting the access to \textit{argv[0]} alone.

\begin{table}
    \footnotesize
    \centering
    \setlength{\tabcolsep}{2pt}
    \begin{tabular}{lccccc} 
        \toprule
        \multirow{2}{*}{Type} & \multirow{2}{*}{System} & \multirow{2}{*}{Workload} & C/C++ & \multicolumn{2}{c}{LoC changed} \\
        & & & LoC & \ac{MTE} & \cheri \\
        \midrule
        In-memory KVS & Redis~\cite{redisRedisRealtimeData} & YCSB~\cite{cooperBenchmarkingCloudServing2010} & \qty{281}{K} & +61/-22 & +140/-33 \\
        Persistent KVS & LevelDB~\cite{googleinc.LevelDB2026} & YCSB~\cite{cooperBenchmarkingCloudServing2010} & \qty{93}{K} & +13/-12 & +12/-9 \\
        OLTP (server) & MySQL~\cite{oracleinc.MySQL} & TPC-C~\cite{TPCC} & \qty{4.07}{M} & +56/-5 & --- \\
        OLTP (embedded) & SQLite~\cite{sqlite} & TPC-C~\cite{TPCC} & \qty{404}{K} & +15/-3 & +35/-3 \\
        OLAP & DuckDB~\cite{raasveldtDuckDBEmbeddableAnalytical2019} & TPC-H~\cite{TPCH} & \qty{986}{K} & +225/-10 & ---\\
        Graphs & LadyBugDB~\cite{fengKUZUGraphDatabase2023} & LDBC SNB~\cite{ladybugdbLadybugDB} & \qty{370}{K} & +63/-2 & +55/-10 \\
        \bottomrule
    \end{tabular}
    \caption{Database systems ported to \ac{MTE} and \cheri.}
    \label{tab:dbms}
    \vspace{-20pt}
\end{table}

\begin{table}[t]
    \footnotesize
    \centering
    \revbox[rev1][0pt]{R1.O4}{\setlength{\tabcolsep}{0.5pt}\setlength{\fboxsep}{0pt}\renewcommand{\arraystretch}{1.05}
  \begin{tabular}{@{}lrrrrrrrrrrrrrr@{}}
    \toprule
    \multicolumn{3}{c}{\textcolor{blue}{Higher is better $\uparrow$}} & \multicolumn{2}{c}{YCSB-A} & \multicolumn{2}{c}{YCSB-B} & \multicolumn{2}{c}{YCSB-C} & \multicolumn{2}{c}{YCSB-D} & \multicolumn{2}{c}{YCSB-E} & \multicolumn{2}{c}{YCSB-F} \\
    \cmidrule(lr){1-3} \cmidrule(lr){4-5} \cmidrule(lr){6-7} \cmidrule(lr){8-9} \cmidrule(lr){10-11} \cmidrule(lr){12-13} \cmidrule(lr){14-15}
    Store & Size & thr & base & prot. & base & prot. & base & prot. & base & prot. & base & prot. & base & prot. \\
    \midrule
    \multicolumn{15}{c}{\textbf{(a) Redis}}\\
    \multirow{1}{*}{\textbf{\ac{MTE}}\,\colorbox[HTML]{FEFEBE}{\textbf{+1\%}}} & all & 1 & 16K & \cellcolor[HTML]{FEFEBE}16K & 15K & \cellcolor[HTML]{FEFEBE}15K & 15K & \cellcolor[HTML]{FEFEBE}15K & 14K & \cellcolor[HTML]{FEFEBE}14K & 357 & \cellcolor[HTML]{FEFEBE}351 & 11K & \cellcolor[HTML]{FEFEBE}11K \\
    \cmidrule(l){1-15}
    \multirow{1}{*}{\textbf{\cheri}\,\rlap{\colorbox[HTML]{FEFAB8}{\textbf{+5\%}}}} & all & 1 & 24K & \cellcolor[HTML]{FEF9B6}22K & 25K & \cellcolor[HTML]{FEFDBC}25K & 26K & \cellcolor[HTML]{FEF9B6}24K & 25K & \cellcolor[HTML]{FEF8B4}23K & 744 & \cellcolor[HTML]{FEF7B3}680 & 18K & \cellcolor[HTML]{FEFEBE}17K \\
    \midrule
    \multicolumn{15}{c}{\textbf{(b) LevelDB}}\\
    \rlap{\textbf{\ac{MTE}}}\multirow{5}{*}{mem\,\colorbox[HTML]{FEF9B6}{\textbf{+7\%}}} & 100K & 64 & 152K & \cellcolor[HTML]{FEFAB8}144K & 624K & \cellcolor[HTML]{FEFEBE}623K & 723K & \cellcolor[HTML]{FEFEC0}724K & 685K & \cellcolor[HTML]{FDFEC2}688K & 371K & \cellcolor[HTML]{FEE699}284K & 156K & \cellcolor[HTML]{FEF9B6}147K \\
    & 1M & 1 & 26K & \cellcolor[HTML]{FEF7B3}23K & 54K & \cellcolor[HTML]{FEF5B1}49K & 88K & \cellcolor[HTML]{FEF3AD}77K & 60K & \cellcolor[HTML]{FEF7B3}55K & 9K & \cellcolor[HTML]{FDD283}6K & 31K & \cellcolor[HTML]{FEF3AD}27K \\
    & 1M & 8 & 149K & \cellcolor[HTML]{FEFDBC}146K & 426K & \cellcolor[HTML]{FEF7B3}391K & 812K & \cellcolor[HTML]{FEFAB8}770K & 474K & \cellcolor[HTML]{FEF9B6}442K & 55K & \cellcolor[HTML]{FEE191}40K & 145K & \cellcolor[HTML]{FEFDBC}141K \\
    & 1M & 64 & 136K & \cellcolor[HTML]{FEFBBA}132K & 562K & \cellcolor[HTML]{FEFEBE}558K & 853K & \cellcolor[HTML]{E1F3F5}907K & 609K & \cellcolor[HTML]{FEFEC0}611K & 198K & \cellcolor[HTML]{FEF9B6}184K & 135K & \cellcolor[HTML]{FEFBBA}130K \\
    & 10M & 64 & 67K & \cellcolor[HTML]{CDE9F2}73K & 314K & \cellcolor[HTML]{FEFBBA}300K & 411K & \cellcolor[HTML]{FEF7B3}378K & 465K & \cellcolor[HTML]{9ACCE1}542K & 163K & \cellcolor[HTML]{FEF1A9}139K & 40K & \cellcolor[HTML]{6AA2CB}52K \\
    \multirow{1}{*}{disk\,\colorbox[HTML]{FEFEBE}{\textbf{+1\%}}} & 1M & 64 & 115K & \cellcolor[HTML]{F8FCCB}116K & 549K & \cellcolor[HTML]{FEFBBA}525K & 861K & \cellcolor[HTML]{ECF7E1}895K & 580K & \cellcolor[HTML]{FAFDC6}585K & 183K & \cellcolor[HTML]{FEFBBA}176K & 123K & \cellcolor[HTML]{FEFAB8}117K \\
    \cmidrule(l){1-15}
    \rlap{\textbf{\cheri}}\multirow{4}{*}{mem\rlap{\,\colorbox[HTML]{FDCA7B}{\textbf{+53\%}}}} & 100K & 4 & 43K & \cellcolor[HTML]{FDD081}29K & 44K & \cellcolor[HTML]{FDDE8E}32K & 39K & \cellcolor[HTML]{FDD888}27K & 48K & \cellcolor[HTML]{FEE191}35K & 1K & \cellcolor[HTML]{FDBA6C}820 & 31K & \cellcolor[HTML]{FDCC7D}20K \\
    & 1M & 1 & 32K & \cellcolor[HTML]{FEE191}23K & 82K & \cellcolor[HTML]{FDCC7D}54K & 102K & \cellcolor[HTML]{FDDA8A}72K & 83K & \cellcolor[HTML]{FDDE8E}60K & 4K & \cellcolor[HTML]{FDDC8C}3K & 30K & \cellcolor[HTML]{FDD484}21K \\
    & 1M & 2 & 38K & \cellcolor[HTML]{FEE79B}29K & 87K & \cellcolor[HTML]{FBA55C}48K & 97K & \cellcolor[HTML]{EE613D}41K & 83K & \cellcolor[HTML]{FDC678}54K & 3K & \cellcolor[HTML]{EE613D}1K & 32K & \cellcolor[HTML]{FDDC8C}23K \\
    & 1M & 4 & 37K & \cellcolor[HTML]{FDCC7D}24K & 42K & \cellcolor[HTML]{FEE191}31K & 40K & \cellcolor[HTML]{FDD686}27K & 44K & \cellcolor[HTML]{FEE395}33K & 1K & \cellcolor[HTML]{FDB86B}815 & 27K & \cellcolor[HTML]{FDC879}17K \\
    \multirow{1}{*}{disk\,\colorbox[HTML]{FDD081}{\textbf{+48\%}}} & 1M & 4 & 36K & \cellcolor[HTML]{FDD283}25K & 41K & \cellcolor[HTML]{FDDE8E}30K & 41K & \cellcolor[HTML]{FDD888}28K & 43K & \cellcolor[HTML]{FDDC8C}31K & 1K & \cellcolor[HTML]{FDB265}803 & 27K & \cellcolor[HTML]{FDD081}18K \\
    \bottomrule
    \multicolumn{15}{c}{\includegraphics[width=2in]{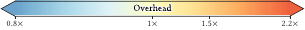}}
  \end{tabular}
}
    \caption{Overhead of \ac{MTE} and \cheri on Redis and LevelDB.}
    \label{tab:dbms_redis_leveldb}
    \vspace{-20pt}
\end{table}

\begin{figure*}
\centering
\revbox[rev1][0.5em]{R1.O4}{%
\begin{tabular}{@{}c@{}}
  \includegraphics[width=\textwidth]{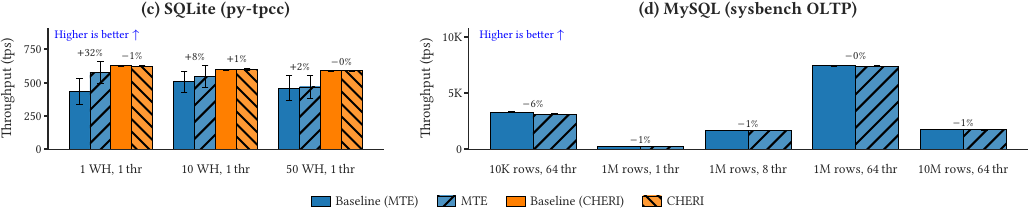} \\[1pt]
  \includegraphics[width=\textwidth]{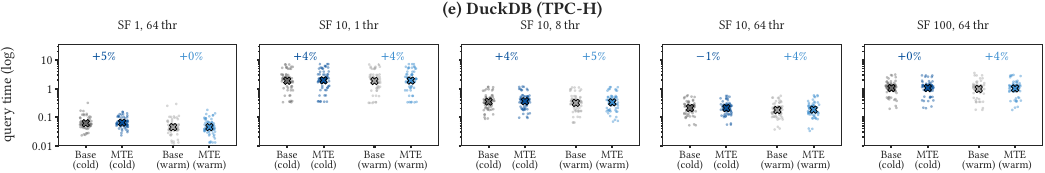} \\[1pt]
  \includegraphics[width=\textwidth]{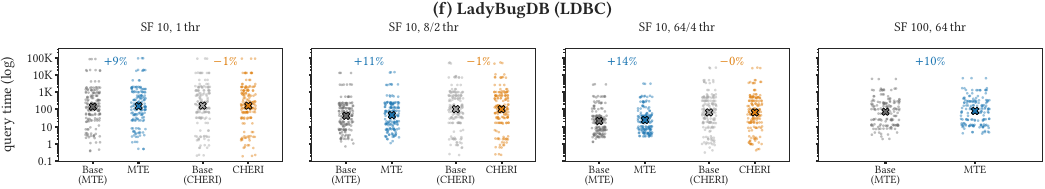} \\
\end{tabular}%
}
\vspace{-1.8em}
\caption{Overhead of \ac{MTE} and \cheri on database systems (MySQL and DuckDB could only be ported to \ac{MTE}).}
\label{fig:dbms_all}
\vspace{-1.2em}
\end{figure*}

For \ac{MTE}, we port the \texttt{jemalloc} allocator used by Redis. 
When allocating, \texttt{jemalloc} generates a new tag using \textit{irg} and assigns it to the region, ensuring adjacent regions receive different tags.
It reserves tag 0 for freed regions, and tags 1--15 for used memory.

\myparagraph{Results}
We use the YCSB workloads~\cite{cooperBenchmarkingCloudServing2010} \rev[rev1]{R1.O4}{with \qty{100}{K} operations, sweeping the working set from \qty{10}{K} to \qty{1}{M} keys; since the single-threaded \textit{redis-server} saturates fast, we run it with one client.
The overheads are insensitive to the working-set size, so \autoref{tab:dbms_redis_leveldb}a reports the average across the sweep: both extensions introduce modest average overheads (1\% for \ac{MTE}, 5\% for \cheri).}
This can be attributed to Redis's relatively low operational throughput, which limits the impact of per-operation safety checks on end-to-end latency.

\subsection{Persistent Key-Value Store (LevelDB)}
LevelDB is a persistent key-value store library that uses a Log-Structured Merge Tree (LSM).
Writes are buffered in skiplist-based MemTables, and eventually flushed to storage in SSTables.

\myparagraph{Porting}
For \cheri, we use the same Musl-based purecap toolchain as Redis, with LLVM's \textit{libc++} library provided by Morello.
We disable automatic linking to the \textit{stdlibc++} and manually add the required libraries.
For \ac{MTE}, LevelDB only needs a \ac{MTE}-aware \textit{glibc} for the memory allocator and setting \textit{glibc.mem.tagging=3}.
Overall, LevelDB requires no code changes beyond toolchain configuration.

\myparagraph{Results}
\rev[rev1]{R1.O4}{We execute the YCSB workloads with \qty{1}{M} operations across the full working-set (\qty{100}{K}--\qty{10}{M} records) and thread-count sweeps.
Across workloads (\autoref{tab:dbms_redis_leveldb}b), \ac{MTE} incurs a mean overhead of 7\%, largely stable across working-set sizes and thread counts, while \cheri averages 53\%, with the largest penalties on the insert-heavy workloads (D, E), confirming our skiplist microbenchmarks (\cref{sec:datastructures}).}
\ac{MTE} results at 10M keys, 64 threads that report a speedup are within the confidence interval of the baseline.
\rev[rev1]{R1.O3}{We compare the \emph{in-memory} results to a \emph{disk} configuration, where the SSTables reside on an SSD and the block cache is capped at 1\% of the dataset size, forcing cache replacement (disk rows in \autoref{tab:dbms_redis_leveldb}b).
The out-of-core overhead is a bit inferior to the in-memory results: \ac{MTE}'s overhead decreases from $\approx$7\% to 1\% , while \cheri incurs 48\% overhead on disk versus 53\% in memory, indicating that I/O operations have modest impact on the overhead of \ac{MTE}/\cheri.}

\subsection{Transactional Database Server (MySQL)}
MySQL is a transactional (OLTP) relational database server that enables applications to store structured data and execute complex SQL queries.
It uses a B+ Tree for indexing%
and a hash-table-based buffer manager to map Page IDs to memory locations.

\myparagraph{Porting}
MySQL relies on the OpenSSL library to provide cryptographic functionalities to connections.
At the time of writing, no \cheri purecap port of OpenSSL exists, preventing us from porting MySQL to \cheri. 
For \ac{MTE}, we patch MySQL's memory allocator to tag allocated regions before returning tagged pointers to the callee and re-tag them on free, invalidating dangling pointers and hardening against use-after-free.
To detect double-free errors, we probe the region before re-tagging, which triggers a fault when a free is repeated.
As with any \ac{MTE}-based allocator, tag collisions remain possible due to the limited tag space; quarantining freed memory can mitigate this at the cost of higher memory usage.

\myparagraph{Results}
We execute a TPC-C-like workload~\cite{TPCC} \rev[rev1]{R1.O4}{sweeping the table size from \qty{10}{K} to \qty{10}{M} rows and the client count from 1 to 64}, using \rev[rev3]{R3.O1}{\texttt{sysbench\_oltp}~\cite{kopytovSysbench2026} load generator}%
, and measure throughput after 60 seconds.
As shown in \autoref{fig:dbms_all}\rev[rev1]{R1.O4}{d, \ac{MTE} reduces MySQL's throughput by at most 6\%---on the smallest, fully cached table under 64 clients---and by ${\leq}1\%$ in every other configuration.}

\subsection{Embedded Database System (SQLite)}
SQLite is an embedded relational (OLTP) database system.
It provides a SQL engine that interacts directly with disk files.%
and uses a B+Tree for table storage, with a default block size of 4\,KiB.

\myparagraph{Porting}
For \cheri, we follow the same process as in Redis.
A notable challenge lies in SQLite's allocator, which adds 64 bits of metadata%
to each allocation and returns a pointer to the region.%
Under \cheri, this pointer %
retains capabilities over the entire region %
, with the metadata, allowing buggy code to corrupt the allocator's state.
Restricting the capability to the user region prevents this, but requires the allocator to maintain a separate capability for metadata access.
To minimize code changes, we retain the unrestricted pointer returned by the allocator, which allows metadata access.

For \ac{MTE}, since SQLite relies on the system allocator, we use \ac{MTE}-enabled glibc and avoid code modifications.
However, \ac{MTE} initialization via \textit{prctl} must occur exactly once per process.
Standalone database systems can perform this initialization at startup; for embedded libraries such as SQLite, initialization must be handled either by the host application or by the library itself, in which case it should first check the active \ac{MTE} state to avoid conflicts.

\myparagraph{Results}
We execute a \rev[rev3]{R3.01}{synthetic transaction workload} on SQLite using tpcc-python~\cite{pavloPytpcc2025}, \rev[rev1]{R1.O4}{with 1--50 warehouses and measure} transaction throughput after 60 seconds.
\rev[rev1]{R1.O4}{The throughput differences of both the \ac{MTE} and \cheri ports remain within the measurements' confidence intervals at all sizes, as depicted in \autoref{fig:dbms_all}c.}

\begin{figure*}[t]
    \begin{minipage}[t]{0.32\textwidth}
        \definecolor{good}{rgb}{0.35, 0.70, 0.49}
        \definecolor{bad}{rgb}{0.80, 0.20, 0.15}
        \definecolor{warn}{rgb}{0.90, 0.62, 0.00}
        \vspace{0pt}
        \centering
        \scriptsize
        \begin{tabularx}{\linewidth}{lXX}
            \toprule
            \textbf{Feature} & \textbf{CHERI} & \textbf{MTE} \\
            \midrule
            \multicolumn{3}{l}{\textbf{Compiler Optimizations}} \\
            \thinmidrule
            Ptr Aliasing & \cellcolor{bad!15!white}Disabled & \cellcolor{good!15!white}Unaffected \\
            Addr. comparisons & \cellcolor{bad!15!white}Disabled & \cellcolor{good!15!white}Unaffected \\
            \texttt{memcpy} & \cellcolor{warn!15!white}Needs 16\,b align & \cellcolor{good!15!white}Unaffected \\
            \thinmidrule
            \multicolumn{3}{l}{\textbf{OS \& Runtime Support}} \\
            \thinmidrule
            Stack Allocations & \cellcolor{good!15!white}Prot. by compiler & \cellcolor{bad!15!white}Android only \\
            Global Variables & \cellcolor{good!15!white}Prot. by linker & \cellcolor{bad!15!white}Android only \\
            Anon. \texttt{mmap} & \cellcolor{good!15!white}Supported. & \cellcolor{good!15!white}Supported \\
            File-backed \texttt{mmap} & \cellcolor{good!15!white}Supported. & \cellcolor{bad!15!white}Not supported \\
            \bottomrule
        \end{tabularx}
        \captionsetup{margin=0.5em}
        \vspace{8pt} %
        \captionof{table}{Comparison of CHERI and MTE features}
        \label{tab:opt-memory-support}

    \end{minipage}
    \begin{minipage}[t]{0.32\textwidth}
        \vspace{0pt}
        \centering
        \includegraphics{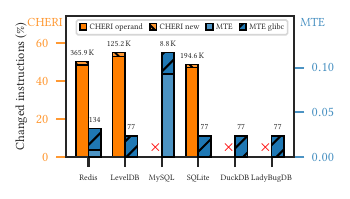}
        \captionsetup{margin=0.5em}
        \vspace{-22pt}
        \caption{Changed instructions when compiling for MTE or CHERI.}
        \label{fig:compiler-insts}
    \end{minipage}
    \begin{minipage}[t]{0.33\textwidth}
        \vspace{0pt}
        \centering
        \scriptsize
        \definecolor{mte}{rgb}{0.12, 0.47, 0.71}
        \definecolor{cheri}{rgb}{1.00, 0.50, 0.00}
        \begin{tabularx}{\linewidth}{p{0pt}Xr}
            \toprule
            & \textbf{Instruction} & \textbf{Count} \\
            \midrule
            \multirow{4}{*}{\rotatebox[origin=c]{90}{\tiny CHERI op}} & \text{ldr} & \cellcolor{cheri!31!white} 176,152 (26.7\%) \\
            & \text{mov} & \cellcolor{cheri!23!white} 100,409 (15.2\%) \\
            & \text{str} & \cellcolor{cheri!19!white} 68,118 (10.3\%) \\
            & \text{add}, \text{ldp}, \text{adrp}, \text{stp}, \text{ldrb}, \dots & \cellcolor{cheri!45!white} 315,358 (47.8\%) \\
            \hline
            \multirow{4}{*}{\rotatebox[origin=c]{90}{\tiny CHERI new}} & \text{scbnds} & \cellcolor{cheri!58!white} 16,827 (65.6\%) \\
            & \text{clrperm} & \cellcolor{cheri!24!white} 4,455 (17.4\%) \\
            & \text{scvalue} & \cellcolor{cheri!17!white} 1,942 (7.6\%) \\
            & \text{seal}, \text{gcvalue}, \text{gctag}, \text{gclen}, \dots & \cellcolor{cheri!19!white} 2,425 (9.5\%) \\
            \hline
            \multirow{3}{*}{\rotatebox[origin=c]{90}{\tiny MTE}} & \text{irg} & \cellcolor{mte!46!white} 4,391 (47.9\%) \\
            & \text{stg} & \cellcolor{mte!46!white} 4,386 (47.9\%) \\
            & \text{ldg}, \text{st2g}, \text{stz2g}, \text{gmi}, \text{stzg} & \cellcolor{mte!15!white} 385 (4.2\%) \\
            \bottomrule
        \end{tabularx}
        \captionsetup{margin=0.5em}
        \vspace{0pt} %
        \captionof{table}{Instruction breakdown aggregated across evaluated systems.}
        \label{tab:compiler-instruction-breakdown}
    \end{minipage}
    \vspace{-20pt}
\end{figure*}

\subsection{Analytical Database System (DuckDB)}
DuckDB is an embedded analytical (OLAP) system optimized for complex analytical queries.
It uses a columnar format, with fixed-size row groups and min-max indexes to skip unnecessary blocks.%

\myparagraph{Porting}
DuckDB uses external dependencies bundled via CMake.
We define a toolchain file that realizes most of the build system modifications.
DuckDB's reliance on functions absent from Musl libc (e.g., \textit{strtoll\_l()}) requires a compatibility shim layer that translates missing functions to their Musl counterparts.
Another three classes of issues require further manual fixes. 
First, five locations in the codebase assume pointers are exactly 8 bytes; we update them to accommodate \cheri's 16-byte pointers, while accounting for silent assumptions in the buffer manager where \textit{size\_t} and pointer sizes are used interchangeably.
Second, integer pointer arithmetic using \textit{uint64\_t} or \textit{uint128\_t} discards capability metadata, leading to provenance issues. 
We replace them with \textit{uintptr\_t} and \textit{intptr\_t}, which preserve the capability metadata and the validity bit. 
Third, the \textit{yyjson} library strips \textit{const} qualifiers via a \textit{size\_t} cast, which drops capability bits; we replace it with the 128-bit \textit{uintptr\_t} in the \cheri version.
Unfortunately, even after this effort, we only succeed in porting DuckDB to \cheri for simple queries.

For \ac{MTE}, we port DuckDB's \textit{jemalloc}, as performed for Redis. 
One additional issue arises: DuckDB stores cryptographic salts in the upper bits of pointers to protect hash tables against denial-of-service attacks.
These bits overlap with the \ac{MTE} tag field, corrupting either the tag or the salt. 
We patch DuckDB to place the salt in non-overlapping bits and mask out tag bits before hash calculations.

\myparagraph{Results}
We run the TPC-H workload~\cite{TPCH} \rev[rev1]{R1.O4}{sweeping the scale factor from 1--100 and the thread count from 1--64}.
To measure the impact of IO operations on the performance overhead of \ac{MTE}, we evaluate two configurations: \emph{(i)~cold}, where the OS caches are flushed before each query, and \emph{(ii)~warm}, where data resides already in the cache.
\rev[rev1]{R1.O4}{As shown in \autoref{fig:dbms_all}e, the geometric-mean overhead of \ac{MTE} stays within 5\% across all scale factors, thread counts, and both cache states, although individual queries deviate more strongly.}

\rev[rev3]{R3.O3}{A single query, Q1, significantly deviates from the average overheads with 55\% of overhead against a 2\% median over the remaining queries.
This overhead is due to the combination of the low aggregation cardinality and the number of rows processed, confirmed by profiling data: 97\% of the additional cycles fall in DuckDB's aggregation methods.
A low cardinality alone is not sufficient: Q8 and Q12 also group into two states, yet only add 3--4\% overhead, as a join or a selective filter cuts their input first.}

\subsection{Graph Database System (LadyBugDB)}
LadyBugDB is a graph database system, based on KuzuDB.
It uses Compressed Sparse Row (CSR) indexing to  store the graph topology, and uses analytical columnar format for node and edge properties.
To avoid memory explosions and spills during graph traversals and joins, LadyBugDB leverages a factorized representation of intermediate results, keeping data compressed in memory.

\myparagraph{Porting}
For \cheri, we use the same CMake toolchain file that streamlines build system modifications and manually disable Scalable Vector Extension (SVE) support, since the Morello board does not implement it, but the build system fails to detect it automatically.
Beyond these configuration changes, LadyBugDB runs without code modifications under the \cheri purecap architecture.
For \ac{MTE}, we instrument LadyBugDB's buffer manager's existing \textit{getFrame()} function---which maps a frame or a page ID to a memory location---to tag the returned pointer before passing it to the rest of the system, making \ac{MTE} fully transparent to the codebase.

\myparagraph{Results}
We evaluate LadyBugDB using queries from the official LDBC Social Network Benchmark (SNB)\rev[rev1]{R1.O4}{, sweeping scale factors 10--100 and the thread count,} which includes standard SNB queries and graph-specific operations (e.g., shortest-path search).
\rev[rev1]{R1.O4}{As shown in \autoref{fig:dbms_all}f, \ac{MTE} introduces a geo-mean overhead of 9--14\%, growing mildly with the thread count.
In comparison, the mean overhead of \cheri is negligible across all configurations.}

\rev[rev3]{R3.O3}{The outliers can be explained by the value types.
Queries over variable-length string attributes, which are dictionary-encoded and must be materialized in heap-allocated memory, cost 10\% overhead, against 4\% for queries over fixed-width numeric attributes of the same table.
55\% of the additional cycles are spent in the dictionary scan and a further 29\% in the allocator, whose cost doubles because every allocation must have its region tagged and re-tagged on free.
}

\myparagraph{\rev[rev4]{R4.02}{Comparison with data structure analysis}}
\rev[rev4]{R4.02}{Across all six systems, overheads are lower than for their underlying data structures (\cref{sec:datastructures}).
The reason is that end-to-end execution also encompasses other tasks, including query parsing and planning, logging, \textit{etc.}, which are typically less memory-intensive and amortize safety-check costs.
System workloads exercise the whole database system, whereas our microbenchmarks isolate a single data structure operation.
}

\begin{takeaway}[dbms_color]
Porting full database systems exposes the difference between \ac{MTE} and \cheri in terms of portability and performance.
\ac{MTE} offers a highly portable integration path with minimal modifications to the allocator and maintains a low performance overhead of 5-10\%.
Conversely, \cheri demands deeper refactoring to correct memory alignments and capability provenance, which can hinder adoption if external dependencies are incompatible.
Performance-wise, \cheri imposes a higher penalty, 20-60\%, particularly on write-heavy key-value stores.

\end{takeaway}

\section{Compiler and Runtime Analysis}
\label{sec:compiler-runtime}

The performance and portability results (\autoref{sec:datastructures}, \autoref{sec:database-systems}) are not determined solely by the hardware mechanisms themselves.
Several observed effects (e.g., disabled compiler optimizations, ABI incompatibilities) originate from the toolchain and OS layers above the hardware.
Therefore, we analyze these layers to identify which constraints are fundamental to the hardware design and which are artifacts of current toolchain maturity.
\cref{tab:opt-memory-support} summarizes the key findings.
\rev[rev1]{R1.O1}{Overall, it shows a trade-off: \cheri requires disabling or constraining several pointer-related optimizations (aliasing, address comparisons, and \texttt{memcpy}), whereas \ac{MTE} leaves these unaffected because it preserves the standard pointer representation.
Conversely, \cheri benefits from support for protecting stack allocations, globals, and both anonymous and file-backed \texttt{mmap} regions, while \ac{MTE} protection for stacks and globals is currently limited to Android and does not extend to file-backed \texttt{mmap}.}

\subsection{Compiler Analysis}
\label{subsec:compiler_analysis}
For C and C++, \textit{clang}~\cite{MorelloLlvmproject2025} and \textit{gcc}~\cite{MorelloGcc2023} forks exist for \cheri and Morello; \ac{MTE} is also supported by both upstream compilers.
Ongoing work extends \cheri support to other languages (i.e., Rust~\cite{harrisRustMorelloAlwaysOn2023}).

\myparagraph{Programming model}
\cheri compilers support both \emph{purecap} mode, where all pointers are capabilities, and \emph{hybrid} mode, where only explicitly annotated pointers are.
To manipulate capabilities, the compiler  exposes intrinsics for operations such as setting or clearing the validity bit, adjusting bounds or permissions, and sealing/unsealing.
In contrast, \ac{MTE} requires no pointer annotation, preserving the type system and ABI.
The compiler exposes intrinsics to manipulate logical tags and load or store allocation tags.

\myparagraph{IR representation and optimizations}
\cheri's LLVM fork represents capabilities using a separate address space, allowing them to coexist with normal pointers in hybrid code while preserving most optimization passes.
Certain passes that assume 64-bit pointers must be disabled when compiling for capabilities.
Further, at the time of writing, LLVM may miscompile capability comparisons by checking only addresses. %
Correct behavior requires treating capabilities with potentially differing bounds, permissions, or validity as distinct.
Otherwise, optimizations may be missed or produce miscompilations (\autoref{fig:compiler-cheri-miscompile}).
\rev[rev1]{R1.O1}{In this example, \texttt{foo} is called with two capabilities with the same address but differing tags.
Since \texttt{q == p} only checks addresses, the compiler concludes that \texttt{q} and \texttt{p} are interchangeable and may dereference the untagged \texttt{q} in place of \texttt{p}.}
Another constraint appears when storing capabilities in untyped byte arrays.
For small copies, the compiler may replace \textit{memcpy} with a load-store sequence if 16-byte alignment can be proven statically.
Code that stores capabilities in untyped byte buffers loses this guarantee, forcing a fallback to \textit{memcpy} to preserve correctness.

\begin{figure}[t]
    \footnotesize
    \centering
    \begin{lstlisting}[
        style=custom,
        language=c,
    ]
        void foo(char *q, char *p) {
            if (q == p) { // `q == p` succeeds when called from bar
                *p = 0; // compiler assumes it may dereference `q`
            }
        }
        void bar(char *p) {
            foo(cheri_tag_clear(p), p); // same address, differing tag
        }
    \end{lstlisting}
    \vspace{-13pt}
    \caption{Code being miscompiled by LLVM. By default, comparisons of capabilities ignore metadata.}
    \label{fig:compiler-cheri-miscompile}
    \vspace{-16pt}
\end{figure}

\myparagraph{Compiler-driven safety}
Typically, compilers insert instructions to automatically provide safety when compiling to memory-safe hardware.
For \cheri, the compiler transparently creates capabilities with appropriate bounds for stack allocations; the dynamic linker handles references to globals, whose addresses are unavailable at compile time.
Subobject bounds narrowing (i.e., restricting a reference to a struct field or array element to its own bounds) is disabled by default for compatibility but is supported and can be enabled.
This affects common database code patterns in which nodes store keys, payloads, and child pointers as subobjects.
Our microbenchmarks, which create references to objects within an array, show no measurable performance difference between the two configurations; thus, enabling it globally is safe for most database code.
Users can still opt out of subobject bounds narrowing for structs or fields where this leads to compatibility problems.

For \ac{MTE}, LLVM provides a stack and globals sanitizer, while GCC support is under development.
Rather than tagging each stack variable individually, which would require padding every object to the 16-byte tag granularity, the entire stack frame receives a single tag, preventing stack overflows.
Correct stack and global tagging require support in the entire toolchain. 
The compiler must insert tagging and untagging code on function entries, returns, and when unwinding stack frames after non-local control flow (e.g., exceptions, \textit{setjmp/longjmp} instructions), while the dynamic linker must be \ac{MTE}-aware to enable \ac{MTE} for stack and global regions.
Currently, only the Android toolchain supports this end-to-end.

\cref{fig:compiler-insts} and \Cref{tab:compiler-instruction-breakdown} quantify the resulting code-generation changes across the evaluated systems.
For \cheri, we see changes to instructions (e.g., \textit{ldr}) which now operate on capabilities, as well as new instructions (e.g., \textit{scbdns}) that manipulate capabilities.
The large share of modified instructions means that the entire tooling stack, including debuggers and disassemblers, must be \cheri-aware to correctly interpret binaries.
For \ac{MTE}, in contrast, the binaries contain few explicit instructions; in several systems, most \ac{MTE}-specific instructions reside in glibc rather than in the application.
This highlights a key difference between the two mechanisms: \cheri changes pointer-carrying code throughout the binary, whereas \ac{MTE} is limited to allocator and runtime support.

\begin{takeaway}[comprunt_color]
Toolchain maturity is a major factor in the porting complexity.
\ac{MTE} is highly portable by retaining existing ABIs and relies mostly on libraries.
\cheri%
alters the entire binary, requiring specialized toolchains and increasing the porting effort.
\end{takeaway}

\subsection{Runtime Libraries}
\label{subsec:runtime_libraries}

\rev[rev1]{R1.O5}{Hardware memory safety is only effective if the software that manages memory cooperates with it.
Two components are central: the \ac{ABI}, the binary contract that fixes pointer size and layout across compiled modules, and the \emph{memory allocator}, which hands out pointers (or capabilities).
We discuss the constraints each mechanism places on these components.}

\myparagraph{ABI and binary compatibility}
\cheri purecap binaries are ABI-incompatible with standard binaries. 
The generated ELF marks the ABI in a flag, and all linked libraries must be recompiled using the purecap ABI.
As a partial solution, a hybrid ABI eases migration by allowing developers to introduce capabilities into unchanged code via pointer annotations.
In contrast, \ac{MTE} leaves the ABI unchanged, enabling adoption without recompiling dependencies.

\myparagraph{\rev[rev3]{R3.O2}{Exception handling}}
\rev[rev3]{R3.O2}{Stack unwinding recognizes signal frames by reading instructions at the return address and matching them against the kernel's \textit{rt\_sigreturn} trampoline.
Purecap forbids this because the program counter carries \textit{execute} but not \textit{load} permission, so inspecting code as data faults.
Unwinders must instead obtain the bytes indirectly, \textit{e.g.}, through a system call, which requires significant changes to the library implementation.
As a result, we could not get exception handling working in our purecap builds.
The database systems are unaffected, as they compile with \textit{-fno-exceptions}.}

\myparagraph{Memory allocation}
For heap protection, \cheri allocators must narrow the capabilities they return.
Further, \rev[rev1]{R1.O5}{allocators keep bookkeeping metadata adjacent to each allocation.}
When freeing memory, allocators cannot widen the user-facing, non-privileged pointer to access this metadata; instead, they must retain a separate, privileged capability for this.
They must also pad allocations and avoid placing adjacent objects within loosely fitting bounds, as capability bounds are imprecise and may over-approximate, potentially leading to undetected illegal accesses.

For \ac{MTE}, allocators must align allocations to 16-byte granules, generate tags, and assign them to memory.
On Linux, \ac{MTE} is limited to \textit{mmap}-backed pages \rev[rev1]{R1.O5}{(the alternative to \textit{sbrk}, which extends the heap contiguously)}, preventing allocators from placing objects in \textit{sbrk}-based pages.
Glibc's \textit{ptmalloc} enables \ac{MTE} through \textit{GLIBC\_TUNABLES}, assigning a random tag to the reserved region while leaving allocator metadata untagged.
Other allocators may use deterministic schemes based on allocation classes.
On free, memory is retagged with zero, so access through the old pointer triggers a mismatch and enables double-free detection.

\begin{figure}[t]
    \centering
    \includegraphics[width=\columnwidth]{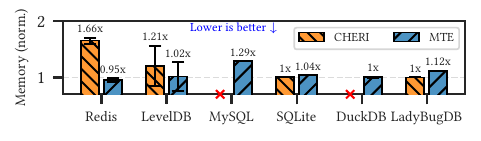}
    \vspace{-30pt}
    \caption{Memory overheads of \ac{MTE} and \cheri.}
    \label{fig:memory-overhead}
    \vspace{-15pt}
\end{figure}

\myparagraph{Memory footprint overhead}
We measure peak \ac{RSS} under the workloads described in \cref{sec:database-systems} and report the normalized memory footprint relative to their baselines for \ac{MTE} and \cheri in \cref{fig:memory-overhead}.
\ac{MTE} imposes negligible overhead on systems only using custom allocators.
However, systems that extensively use \textit{malloc} for small allocations (e.g., MySQL) or the standard library containers to index the data (e.g., LadyBugDB) experience higher, though moderate, overhead from MTE's alignment and padding requirements.
Conversely, \cheri's 128-bit capabilities heavily penalize systems with pointer-heavy data structures:
Redis' and LevelDB's extensive use of hash tables and skip lists results in up to 66\% memory overhead.
In contrast, systems that maintain indexes using mostly integers, such as B+ trees in SQLite or CSR in LadyBugDB, experience negligible memory overhead with \cheri.

\myparagraph{Temporal memory safety}
\cheri can enforce temporal safety via capability revocation~\cite{xiaCHERIvokeCharacterisingPointer2019,wesleyfilardoCornucopiaTemporalSafety2020,filardoCornucopiaReloadedLoad2024}. 
Freed regions are quarantined until a privileged capability revocation pass invalidates their capabilities, preventing \emph{use-after-reallocation}. %
In contrast, \ac{MTE} can retag freed memory, probabilistically invalidating existing pointers. 
As tags are finite, a reallocated region may receive the same tag as a dangling pointer, leaving stale accesses undetected.
Allocators can reduce, but not eliminate, by delaying reuse through quarantine.

\begin{takeaway}[alloc_color]
Hardware-assisted memory safety extensions place a significant performance and complexity burden on the memory allocator.
\cheri inflates the memory footprint of pointer-heavy database systems by up to 66\%.
Similarly, \ac{MTE} changes the complexity of memory allocations to $\mathcal{O}(n)$, which can impact allocation-heavy database systems.
\end{takeaway}

\subsection{Operating System Support}
\label{subsec:os_support}

In terms of operating system support, there are currently both FreeBSD~\cite{CheriBSD} and Linux~\cite{MorelloLinux} forks providing \cheri support, while \ac{MTE} is supported in mainline Linux since version 5.9.

\myparagraph{Process initialization}
For \cheri, the language runtime must constrain capabilities before transferring control to the program, including bounding the stack capability register and appropriately revoking permissions. %
For \ac{MTE}, the process must call \textit{prctl}\rev[rev1]{R1.O5}{, the process-control syscall,} to enable tag checking.
Individual memory regions must also enable \ac{MTE} when they are mapped with \textit{mmap} or \rev[rev1]{R1.O5}{when their protection is later changed with} \textit{mprotect}.

\myparagraph{OS interfaces}
CheriBSD~\cite{CheriBSD} and Morello Linux~\cite{MorelloLinux} both expose capability and hybrid system-call interfaces; the hybrid is the default on Morello. %
Upon detecting an invalid access, both \ac{CHERI} and \ac{MTE} raise signals to the userspace process.

On Linux, \ac{MTE} is restricted to anonymous memory mappings \rev[rev1]{R1.O5}{(memory not backed by a file, such as the heap)} and must be explicitly enabled; file-backed mappings \rev[rev1]{R1.O5}{, which expose a file's contents as memory (as used for memory-mapped I/O),} cannot be tagged.
This restriction directly impacts database systems that rely on memory-mapped files for buffer management or storage access.
If \ac{MTE} safety is desired, data must be copied into an anonymous memory region, which may result in undesirable performance and memory penalties.
Therefore, in our experiments (\cref{sec:database-systems}), \ac{MTE} is limited to heap-allocated data and does not cover file-backed memory.

\begin{takeaway}[comprunt_color]
Although there is foundational OS support for \ac{MTE} and \cheri, current implementations lack important features for database systems.
Because Linux restricts MTE protection to anonymous memory, database systems cannot natively tag file-backed memory mappings.
Meanwhile, enforcing a full pure-capability OS interface is challenging, leading CHERI-enabled OSes to default currently to a hybrid system-call interface.
\end{takeaway}

\subsection{Coverage of Memory-Safety Bug Classes}
\label{subsec:bug_classes}

Finally, we evaluate how \ac{MTE} and \cheri protect against the classes of bugs that we identified in our bug study \cref{subsec:bug_study}.
We reproduce each class that our classifier considers memory-safety-related (\cref{sec:appendix-bug-study}), \textit{i.e.} including buffer overflows, out-of-bounds accesses, use-after-free, null pointer dereferences, as small test cases.
We run each case 200 times without protection, with \ac{MTE}, and with \cheri, and report in \cref{tab:bug_classes} whether each class is detected at the faulting access.

\begin{table}[t]
    \centering
    \footnotesize
    \setlength{\tabcolsep}{3pt}
    \resizebox{\columnwidth}{!}{\begin{tabular}{@{}lccc@{}}
\toprule
\textbf{Bug class} & \textbf{Baseline} & \textbf{MTE} & \textbf{CHERI} \\
\midrule
Heap buffer overflow & \BcSilent & \BcPartial{into next object} & \BcTrap \\
Stack/global buffer overflow & \BcSilent & \BcSilent & \BcTrap \\
Out-of-bounds read/write & \BcSilent & \BcPartial{89\%} & \BcTrap \\
Integer overflow $\to$ corruption & \BcLibc & \BcTrap & \BcTrap \\
Use-after-free & \BcSilent & \BcPartial{88\%} & \BcNoOS \\
Double/invalid free & \BcLibc & \BcPartial{double free only} & \BcTrap \\
Null pointer dereference & \BcOSFault & \BcOSFault & \BcTrap \\
Uninitialized memory access & \BcSilent & \BcSilent & \BcSilent \\
\bottomrule
\end{tabular}
}
    \\[4pt]
    {\scriptsize\setlength{\fboxsep}{1.5pt}%
    \BcLegend{bcgood}{\ding{51}} detected\hspace{0.8em}
    \colorbox{bcprob}{\makebox[2.6em]{\strut\ding{51}\,(\dots)}} partially detected\hspace{0.8em}
    \BcLegend{bcother}{libc} allocator check\\[2pt]
    \BcLegend{bcother}{OS} unmapped page\hspace{0.8em}
    \BcLegend{bcmiss}{\ding{55}} undetected\hspace{0.8em}
    \BcLegend{bcos}{rev.} needs revocation\par}
    \captionsetup[table]{position=bottom,skip=4pt}
    \caption{Detection of the memory-safety bug classes of our bug study (200 runs per test case).}
    \label{tab:bug_classes}
\end{table}

\myparagraph{Spatial safety}
\cheri deterministically detects heap, stack, and global overflows and out-of-bounds accesses.
In comparison, \ac{MTE} protects only the heap, at the 16-byte granularity of its tags, and detects an access landing in another object only if both objects carry different tags.
This behavior leads to probabilistic safety, where \ac{MTE} catches 89\% of the out-of-bounds accesses in our test scenario, due to tag collisions.

\myparagraph{Temporal safety}
\ac{MTE} retags freed memory, so a use-after-free faults deterministically until the memory is reallocated and probabilistically afterwards.
\cheri requires capability revocation to detect use-after-free, which CheriBSD supports but Morello Linux does not (\cref{subsec:os_support}).
Both mechanisms catch double frees.

\myparagraph{Null pointers and uninitialized memory}
Without protection and with \ac{MTE}, which only checks tags within tagged mappings, a null pointer dereference is stopped only because the OS leaves the first page unmapped.
\cheri turns it into a capability fault at the faulting access: a null pointer carries no valid capability, so even accesses at large offsets from \texttt{NULL}, which may land in mapped memory without protection, are rejected.
Uninitialized memory accesses are not detected by any mechanism, as neither tracks whether memory was written: a new allocation still exposes the previous owner's data.
Allocators can close this gap by zeroing memory on allocation, for example by using special \ac{MTE} instructions that tag and zero memory in a single operation, \textit{i.e.} \texttt{stzg} (\cref{subsec:hw-mte}).

\begin{takeaway}[comprunt_color]
Without protection, the reproduced bugs silently corrupt or leak data, or are caught late by allocator checks, consistent with the crashes, incorrect results, and corruption of our bug study.
\ac{MTE} and \cheri instead stop most of these bug classes at the faulting access, before corrupted data propagates.
\cheri deterministically enforces spatial safety but needs OS support for temporal safety, whereas \ac{MTE} covers spatial and temporal heap bugs, but probabilistically and without protecting stack and global memory.
Neither detects uninitialized memory accesses.
\end{takeaway}

\newcolumntype{T}{>{\hsize=0.5\hsize\linewidth=\hsize}X}
\newcolumntype{I}{>{\hsize=0.5\hsize\linewidth=\hsize}X}

\begin{table*}[t]
    \small
    \centering
    \begin{tabularx}{\textwidth}{TI}
         \toprule
         \multicolumn{1}{c}{\textbf{Takeaway}} & \multicolumn{1}{c}{\textbf{Implications}} \\
         \hline
         \rowcolor{arch_color!\coloropacity!white}
         \multicolumn{2}{c}{\textbf{Architectural}} \\
         \hline
         \rowcolor{arch_color!\coloropacity!white}
         \ac{MTE} tag must be loaded before a store & \ac{MTE} loads have little overhead; stores are more expensive \\
         \rowcolor{arch_color!\coloropacity!white}
         \cheri capabilities double the memory to load on pointer access & \cheri memory accesses are more expensive than without capabilities \\
         \rowcolor{arch_color!\coloropacity!white}
         The hardware can prefetch tags/capabilities & Cache-aware data structures incur less performance overhead \\
         \rowcolor{arch_color!\coloropacity!white}
         \cheri imposes a large overhead for atomic operations on capabilities & Increases the performance tax on lock-free data structures  \\
         \rowcolor{arch_color!\coloropacity!white}
         Concurrent modifications to tags/capabilities incur significant overheads & False sharing has more impact on memory-safe hardware \\
         \rowcolor{arch_color!\coloropacity!white}
         \cheri capabilities and \ac{MTE} allocations must be aligned to \qty{16}{bytes} & \ac{MTE} and \cheri may increase memory fragmentation \\
         \hline
         \rowcolor{comprunt_color!\coloropacity!white}
         \multicolumn{2}{c}{\textbf{Compiler/Runtime}} \\
         \hline
         \rowcolor{comprunt_color!\coloropacity!white}
         \ac{MTE} requires minimal changes to the build toolchain and runtime & \ac{MTE} is mostly compatible with existing software and programming tools \\ 
         \rowcolor{comprunt_color!\coloropacity!white}
         \cheri purecap requires deep modifications of core software components & \cheri development lacks support for many common programming tools \\
         \rowcolor{comprunt_color!\coloropacity!white}
         \cheri temporal safety requires additional revocation pass & Background revocation may create latency spikes in intensive workloads\\
         \rowcolor{comprunt_color!\coloropacity!white}
         \ac{MTE} temporal safety faces the tag exhaustion challenge & Rapid memory reuse can lead to safety failures because of stale pointers\\
         \rowcolor{comprunt_color!\coloropacity!white}
         OSes provide exclusively hybrid or purecap APIs & Hybrid-API OSes offer flexibility at the cost of library dependence \\
         \rowcolor{comprunt_color!\coloropacity!white}
         \ac{MTE} must be initialized exactly once per process & Host application must handle initialization for embedded database systems \\
         \rowcolor{comprunt_color!\coloropacity!white}
         \ac{MTE} requires anonymous memory regions & Database systems cannot protect file-backed memory regions with \ac{MTE} \\ 
         \hline
         \rowcolor{alloc_color!\coloropacity!white}
         \multicolumn{2}{c}{\textbf{Memory Allocation}} \\
         \hline
         \rowcolor{alloc_color!\coloropacity!white}
         \ac{MTE}/\cheri are supported by system allocators, \textit{e.g.} glibc malloc & Database systems can benefit from safe allocations with minimal changes \\
         \rowcolor{alloc_color!\coloropacity!white}
         Custom allocators must explicitly handle tag (\ac{MTE}) or bounds (\cheri) & \ac{MTE}/\cheri increase code complexity for custom allocations\\
         \rowcolor{alloc_color!\coloropacity!white}
         \ac{MTE} padding fragments memory; \cheri pointers bloat data structures & Increased memory footprints incur additional hardware costs \\
         \rowcolor{alloc_color!\coloropacity!white}
         \ac{MTE} allocation complexity is $\mathcal{O}(n)$ & Large, short-lived allocations should be protected by another mechanism \\
         \rowcolor{alloc_color!\coloropacity!white}
         \ac{MTE}-aware allocators must initialize tags & Allocators should use dedicated instructions for initialization (\textit{stzg}/\textit{stgp}). \\
          \cellcolor{alloc_color!\coloropacity!white}& \cellcolor{alloc_color!\coloropacity!white}\cheri allocators must sacrifice protection or maintain an additional pointer \\
          \multirow{-2}{*}{\cellcolor{alloc_color!\coloropacity!white}Custom allocators typically maintain metadata for allocated blocks} & \cellcolor{alloc_color!\coloropacity!white} \ac{MTE} tags can be loaded, simplifying the porting of legacy custom allocators\\
         
         \hline
         \rowcolor{dbms_color!\coloropacity!white}
         \multicolumn{2}{c}{\textbf{Database Systems}} \\
         \hline
         \rowcolor{dbms_color!\coloropacity!white}
         Pointers carry address and metadata & Integer casts and metadata storage in unused pointer bits are disallowed \\
         \rowcolor{dbms_color!\coloropacity!white}
         Ease of porting correlates with standard adherence & Legacy code assuming 64-bit pointers requires significant manual updates\\
         \rowcolor{dbms_color!\coloropacity!white}
         \rev[rev3]{R3.O5}{Pointer indexes have higher overhead with CHERI than integer indexes} & \rev[rev3]{R3.O5}{Indexing designers should consider the performance \textit{vs.} safety tradeoff}  \\
         \rowcolor{dbms_color!\coloropacity!white}
         Reads/scans incur relatively small overheads (\ac{MTE}:2--6\%; \cheri: 15--20\%) & Read-heavy workloads are well-suited for early hardware safety adoption\\
         \rowcolor{dbms_color!\coloropacity!white}
         Inserts lead to larger overheads (\ac{MTE}: 20--40\%; \cheri: up to 200\%) & Reduces the advantage of in-memory systems for write-heavy workloads\\
         \rowcolor{dbms_color!\coloropacity!white}
         Overheads of OLTP workloads are lower than OLAP and KVS workloads & Memory-bound workloads face the highest performance tax \\
         \bottomrule
    \end{tabularx}
    \caption{\cheri and \ac{MTE} analysis takeaways and their implications for database systems practitioners.}
    \label{tab:takeaways}
    \vspace{-20pt}
\end{table*}

\section{Related Work}
\label{sec:related}

\myparagraph{Software-based approaches}
Significant work has been done to retrofit memory safety into existing codebases written in unsafe languages. 
Trip-wire-based approaches pad allocations with inaccessible guard regions to catch out-of-bounds accesses~\cite{serebryanyGWPASanSamplingBasedDetection2024,tailcheck,asan}. 
Object-based approaches associate metadata, stored in disjoint tables or shadow memory, with an object, enforcing spatial safety via bounds checks on pointer manipulation and temporal safety by invalidating stale metadata~\cite{akritidisBaggyBoundsChecking2009,hasabnisLightweightBoundsChecking2012,bozdoganSafePMSanitizerPersistent2022,nethercoteValgrindFrameworkHeavyweight2007,camp,giantsan,RangeSanitizer,cmasan}.
Pointer-based approaches embed metadata inline (e.g., fat pointers) or in separate memory regions, tracking all pointers to enforce temporal safety, and checking bounds on every access~\cite{kroesDeltaPointersBuffer2018,duckHeapBoundsProtection2016,duckStackBoundsProtection2017,spp,nagarakatteSoftBoundHighlyCompatible2009,kuvaiskiiSGXBOUNDSMemorySafety2017,wpbound}; some approaches also combine per-pointer and per-object metadata~\cite{cucatch,cecsan}.
Orthogonal techniques include static analysis~\cite{dhurjatiSAFECodeEnforcingAlias2006}, parallelization of checks~\cite{catamaran}, and protection against memory error exploits via safe control-flow~\cite{abadiControlflowIntegrityPrinciples2009,kuznetsovCodepointerIntegrity2014}  and software-based fault isolation~\cite{wahbeEfficientSoftwarebasedFault1993,anselLanguageindependentSandboxingJustintime2011,yedidiaLightweightFaultIsolation2024,haasBringingWebSpeed2017,yeeNativeClientSandbox2010}.
Additionally, safe languages or dialects enforce memory safety at the type system~\cite{neculaCCuredTypesafeRetrofitting2005,jimCycloneSafeDialect2002,RustProgrammingLanguage}.
Software approaches incur performance overheads that are prohibitive in production database deployments, a gap our work quantifies and motivates through hardware-assisted alternatives.

\myparagraph{Hardware-based approaches}
Hardware-based solutions reduce the overhead that makes software solutions impractical.
Capability architectures embed bounds and permissions in pointers.
\cheri has been applied across a range of systems~\cite{lowtherCHERIPerformanceEnhancement2023,wangCherifyingLinuxPractical2024,xiaCHERIvokeCharacterisingPointer2019,wesleyfilardoCornucopiaTemporalSafety2020,filardoCornucopiaReloadedLoad2024,amarCHERIoTRTOSOS2025,amarCHERIoTCompleteMemory2023}, extended to safe languages~\cite{harrisRustMorelloAlwaysOn2023}, while other alternatives have also been proposed~\ifthenelse{\boolean{publicversion}}{\cite{capstone,c3,chenCHERILiteMemorySafety}}{\cite{capstone,c3}\rev[rev1]{R1.O6}{~\cite{chenCHERILiteMemorySafety}}}.
Further works leverage \ac{MTE} for deterministic memory tagging or isolation~\cite{stickytags,momeuIUBIKIsolatingUser2025}.
Hardware accelerated bounds checking and pointer integrity have also been explored~\cite{hardbound,safeproc,watchdog,watchdoglite,kwonLowfatPointersCompact2013,pacmem,always_on_heap_memory_safety,liljestrandPACItPointer2019,crypt_mem_safety},
as has hardware-based compartmentalization~\cite{cage,narayanGoingLimitsSFI2023,xuPANICPANassistedIntraprocess2023}.
These works show the potential of hardware mechanisms but do not address database-specific challenges such as custom allocators.
\rev[rev1]{R1.O6}{We provide the first side-by-side comparison of \ac{MTE} and \cheri in this setting.}

\myparagraph{Hardware extension analysis} 
Prior surveys cover hardware improvements for memory-safe program execution~\cite{Survey_of_Hardware_Improvement_to_Secure_Program_Execution} and the limitations and opportunities of modern isolation mechanisms~\cite{chenLimitationsOpportunitiesModern2024}.
Specifically, memory safety has been addressed through extensions such as Intel MPX~\cite{oleksenkoIntelMPXExplained2018b} and Intel MPK~\cite{mpk,parkLibmpkSoftwareAbstraction2019}. 
Adjacent work analyzes confidential computing hardware (e.g., Intel TDX, AMD SEV)~\cite{misonoConfidentialVMsExplained2025} and confidential GPUs~\cite{dissecting_cgpu,ibarra2025performanceconfidentialcomputinggpus}.
\rev[rev1]{R1.O6}{Concurrent work analyzes the Arm \ac{MTE}~\cite{noh2026armmteperformancepractice,multitag,kaushikOptimizedMemoryTagging2025} and \cheri~\cite{sunSweetSourCHERI2025,kresselSoftwareCompartmentalizationTradeOffs2023} in isolation, treating database systems as black-box applications in benchmarks.}
Our work is complementary: rather than analyzing a single extension or a general-purpose workload, we conduct a \emph{cross-layer} comparison of \ac{MTE} and \cheri specifically for database systems, \rev[rev1]{R1.O6}{linking overheads to the data structures and microarchitectural access patterns.}

\myparagraph{Database testing} 
Efforts to improve database reliability have progressed from crash-oriented fuzzers~\cite{sqlsmith, afl} to logic-bug detection, where a system returns incorrect results without crashing.
Key advances include Pivoted Query Synthesis (PQS)~\cite{testing_db_pivoted_query_synthesis} and Query Partitioning~\cite{finding_bugs_query_partitioning}, popularized by SQLancer~\cite{sqlancer,sqlancer_plus_plus}.
Differential testing~\cite{rags,apollo} and reference engine construction approaches isolate optimizer and correctness bugs~\cite{optimization_bugs_in_dbs}, while modern frameworks leverage coverage-guided techniques~\cite{squirrel,fucci,dynsql,ratel} and relational constraint solvers~\cite{query_aware_test_generation,systematic_testing_of_dbms,detecting_logical_bugs} to explore complex query execution paths.
Such approaches focus on detecting bugs after they occur, while %
we focus on \emph{preventing} %
memory-safety exploits%
and our bug study (\cref{subsec:bug_study}) shows that such bugs exist even in well-tested systems.
\rev[rev1]{R1.O7}{Unlike prior work on discovering database-system bugs~\cite{gaoComprehensiveSurveyDatabase2023}, we classify existing reports and \acp{CVE} by their memory-safety implications, complementing recent \ac{LLM}-based classification of security-relevant bugs and patches~\cite{liWhatTheyFix2026,torkamaniStreamliningSecurityVulnerability2025,ahmedSecVulEvalBenchmarkingLLMs2025}.}

\begin{figure}[t] %
    \vspace{-5pt}
  \begin{center}
    \includegraphics[width=\linewidth]{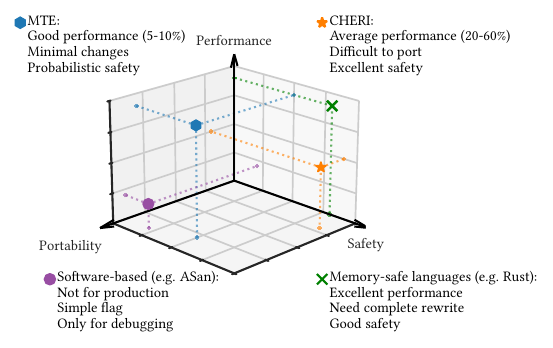}
  \end{center}
  \vspace{-7pt}
  \caption{The database systems trilemma}
  \label{fig:conclusion}
  \vspace{-8pt}
\end{figure}

\section{The Database Systems Trilemma}
\label{sec:conclusion}

\rev[rev1]{R1.O6}{This paper presents the first side-by-side, cross-layer comparison of \ac{MTE} and \cheri on a database workload suite, spanning from microarchitecture through an indexed-data-structure layer to full systems.}
We investigate the answers that \ac{MTE} and \cheri offer to the memory safety trilemma (\autoref{sec:trilemma}) across micro-architectural analysis, data structures, and full-fledged database systems.
We summarize our findings in \autoref{tab:takeaways} and compare against software-based approaches and memory-safe languages in \autoref{fig:conclusion}.

\rev[rev3]{R3.O7}{The two extensions take different approaches.
\cheri provides strong, deterministic guarantees by enforcing pointer bounds, permissions, and unforgeability.
It detects spatial violations, including out-of-bounds accesses and buffer overflows, provides temporal safety through capability revocation, and catches intra-object overflows when the compiler narrows sub-object bounds.
By contrast, \ac{MTE} detects the same bug classes probabilistically and cannot detect intra-object overflows.
Neither mechanism catches type confusion, valid-but-incorrect pointer use, or other logic bugs.
}

This divergence inversely correlates with performance: \cheri imposes significant overheads (20--60\%) across database system workloads, whereas \ac{MTE}'s average overhead is moderate (5--10\%).
Finally, the invasiveness of the two solutions dictates their practical portability.
\ac{MTE} requires minimal changes to the build and deployment pipeline because it is fully supported by current toolchains.
On the other hand, the \cheri toolchain introduces considerable complexity due to the fundamental change in pointer size.
Its purecap ABI also requires compatible builds of all dependent libraries, which demands significant engineering effort for use in large, dependency-heavy applications like database systems.
Overall, \ac{MTE} and \cheri show that the Pareto frontier has moved beyond what software-based approaches can offer, making hardware enforcement a viable path to production-grade memory safety in database systems.

\myparagraph{Acknowledgements}
We thank the anonymous reviewers for their helpful comments. We also thank Viktor Leis for his feedback on the project.
This work was supported in part by an ERC Starting Grant (ID: 101077577) and the Chips Joint Undertaking (JU), European Union (EU) HORIZON-JU-IA, under grant agreement No. 101140087 (SMARTY), the Intel Trustworthy Data Center of the Future (TDCoF), Google Research Grants, and the TUM Innovation Network Resilient, Trustworthy, Sustainable (ReTruSt).

\bibliographystyle{ACM-Reference-Format}
\bibliography{main,cheridb-zotero-export}


\begin{thebibliography}{153}


\ifx \showCODEN    \undefined \def \showCODEN     #1{\unskip}     \fi
\ifx \showISBNx    \undefined \def \showISBNx     #1{\unskip}     \fi
\ifx \showISBNxiii \undefined \def \showISBNxiii  #1{\unskip}     \fi
\ifx \showISSN     \undefined \def \showISSN      #1{\unskip}     \fi
\ifx \showLCCN     \undefined \def \showLCCN      #1{\unskip}     \fi
\ifx \shownote     \undefined \def \shownote      #1{#1}          \fi
\ifx \showarticletitle \undefined \def \showarticletitle #1{#1}   \fi
\ifx \showURL      \undefined \def \showURL       {\relax}        \fi
\providecommand\bibfield[2]{#2}
\providecommand\bibinfo[2]{#2}
\providecommand\natexlab[1]{#1}
\providecommand\showeprint[2][]{arXiv:#2}
\makeatletter
\@ifundefined{NAT@parse@date}{}{\let\NAT@parse@date@orig\NAT@parse@date}
\@ifundefined{NAT@parse@date}{}{\def\NAT@parse@date#1#2#3#4#5#6@@{\NAT@parse@date@orig#1#2#3#4#5#6@@\def\NAT@tempyear{0000}\def\NAT@tempexlab{{?}}\ifx\NAT@year\NAT@tempyear\ifx\NAT@exlab\NAT@tempexlab\def\NAT@date{[n.\,d.]}\else\edef\NAT@date{[n.\,d.]\NAT@exlab}\fi\fi}}
\makeatother

\bibitem[Che(0000)]%
        {CheriBSD}
 \bibinfo{year}{[n.\,d.]}\natexlab{}.
\newblock \bibinfo{title}{{{CheriBSD}}}.
\newblock
\urldef\tempurl%
\url{https://www.cheribsd.org/}
\showURL{%
\tempurl}


\bibitem[Mor(0000)]%
        {MorelloLinux}
 \bibinfo{year}{[n.\,d.]}\natexlab{}.
\newblock \bibinfo{title}{Morello {{Linux}}}.
\newblock
\urldef\tempurl%
\url{https://linux.morello-project.org/}
\showURL{%
\tempurl}


\bibitem[Rus(0000)]%
        {RustProgrammingLanguage}
 \bibinfo{year}{[n.\,d.]}\natexlab{}.
\newblock \bibinfo{title}{Rust {{Programming Language}}}.
\newblock
\urldef\tempurl%
\url{https://rust-lang.org/}
\showURL{%
\tempurl}


\bibitem[TPC(0000a)]%
        {TPCC}
 \bibinfo{year}{[n.\,d.]}\natexlab{a}.
\newblock \bibinfo{title}{{{TPC-C}}}.
\newblock
\urldef\tempurl%
\url{https://www.tpc.org/tpcc/}
\showURL{%
\tempurl}


\bibitem[TPC(0000b)]%
        {TPCH}
 \bibinfo{year}{[n.\,d.]}\natexlab{b}.
\newblock \bibinfo{title}{{{TPC-H}}}.
\newblock
\urldef\tempurl%
\url{https://www.tpc.org/tpch/}
\showURL{%
\tempurl}


\bibitem[spa(2015)]%
        {sparc}
 \bibinfo{year}{2015}\natexlab{}.
\newblock \bibinfo{title}{Hardware-Assisted Checking Using Silicon Secured
  Memory (SSM)}.
\newblock
  \bibinfo{howpublished}{\url{https://docs.oracle.com/cd/E37069_01/html/E37085/gphwb.html}}.
\newblock
\newblock
\shownote{Accessed: 2026-03-04}.


\bibitem[sta(2015)]%
        {stagefright}
 \bibinfo{year}{2015}\natexlab{}.
\newblock \bibinfo{title}{Poject Zero - Stagefrightened?}
\newblock
  \bibinfo{howpublished}{\url{https://googleprojectzero.blogspot.com/2015/09/stagefrightened.html}}.
\newblock
\newblock
\shownote{Accessed: 2026-03-04}.


\bibitem[msr(2019)]%
        {msr_vuln}
 \bibinfo{year}{2019}\natexlab{}.
\newblock \bibinfo{title}{A proactive approach to more secure code}.
\newblock
  \bibinfo{howpublished}{\url{https://msrc-blog.microsoft.com/2019/07/16/a-proactive-approach-to-more-secure-code/}}.
\newblock
\newblock
\shownote{Accessed: 2026-03-04}.


\bibitem[and(2019)]%
        {android_vuln}
 \bibinfo{year}{2019}\natexlab{}.
\newblock \bibinfo{title}{Queue the Hardening Enhancements}.
\newblock
  \bibinfo{howpublished}{\url{https://security.googleblog.com/2019/05/queue-hardening-enhancements.html}}.
\newblock
\newblock
\shownote{Accessed: 2026-03-04}.


\bibitem[hea(2020)]%
        {heartbleed}
 \bibinfo{year}{2020}\natexlab{}.
\newblock \bibinfo{title}{The Heartbleed Bug}.
\newblock \bibinfo{howpublished}{\url{https://heartbleed.com/}}.
\newblock
\newblock
\shownote{Accessed: 2026-03-04}.


\bibitem[Mor(2023)]%
        {MorelloGcc2023}
 \bibinfo{year}{2023}\natexlab{}.
\newblock \bibinfo{title}{Morello Gcc}.
\newblock
\urldef\tempurl%
\url{https://git.morello-project.org/morello/gnu-toolchain/gcc}
\showURL{%
\tempurl}


\bibitem[mte(2025)]%
        {mte}
 \bibinfo{year}{2025}\natexlab{}.
\newblock \bibinfo{title}{Arm memory tagging extension}.
\newblock
  \bibinfo{howpublished}{\url{https://source.android.com/docs/security/test/memory-safety/arm-mte}}.
\newblock
\newblock
\shownote{Accessed: 2026-03-04}.


\bibitem[mpx(2025)]%
        {mpx}
 \bibinfo{year}{2025}\natexlab{}.
\newblock \bibinfo{title}{Intel(R) Memory Protection Extensions (MPX)}.
\newblock
  \bibinfo{howpublished}{\url{https://www.kernel.org/doc/html/v5.2/x86/intel_mpx.html}}.
\newblock
\newblock
\shownote{Accessed: 2026-03-04}.


\bibitem[mpk(2025)]%
        {mpk}
 \bibinfo{year}{2025}\natexlab{}.
\newblock \bibinfo{title}{Memory Protection Keys}.
\newblock
  \bibinfo{howpublished}{\url{https://docs.kernel.org/core-api/protection-keys.html}}.
\newblock
\newblock
\shownote{Accessed: 2026-03-04}.


\bibitem[Mor(2025a)]%
        {MorelloLlvmproject2025}
 \bibinfo{year}{2025}\natexlab{a}.
\newblock \bibinfo{title}{Morello Llvm-Project}.
\newblock
\urldef\tempurl%
\url{https://git.morello-project.org/morello/llvm-project}
\showURL{%
\tempurl}


\bibitem[Mor(2025b)]%
        {MorelloMuslLibc2025}
 \bibinfo{year}{2025}\natexlab{b}.
\newblock \bibinfo{title}{Morello Musl Libc}.
\newblock
\urldef\tempurl%
\url{https://git.morello-project.org/morello/musl-libc}
\showURL{%
\tempurl}


\bibitem[Abadi et~al\mbox{.}(2009)]%
        {abadiControlflowIntegrityPrinciples2009}
\bibfield{author}{\bibinfo{person}{Mart{\'i}n Abadi}, \bibinfo{person}{Mihai
  Budiu}, \bibinfo{person}{{\'U}lfar Erlingsson}, {and} \bibinfo{person}{Jay
  Ligatti}.} \bibinfo{year}{2009}\natexlab{}.
\newblock \showarticletitle{Control-Flow Integrity Principles, Implementations,
  and Applications}.
\newblock \bibinfo{journal}{\emph{ACM Trans. Inf. Syst. Secur.}}
  \bibinfo{volume}{13}, \bibinfo{number}{1} (\bibinfo{date}{Nov.}
  \bibinfo{year}{2009}), \bibinfo{pages}{4:1--4:40}.
\newblock
\showISSN{1094-9224}
\href{https://doi.org/10.1145/1609956.1609960}{doi:\nolinkurl{10.1145/1609956.1609960}}


\bibitem[Abdul~Khalek et~al\mbox{.}(2008)]%
        {query_aware_test_generation}
\bibfield{author}{\bibinfo{person}{Shadi Abdul~Khalek}, \bibinfo{person}{Bassem
  Elkarablieh}, \bibinfo{person}{Yai O.~Laleye}, {and} \bibinfo{person}{Sarfraz
  Khurshid}.} \bibinfo{year}{2008}\natexlab{}.
\newblock \showarticletitle{Query-Aware Test Generation Using a Relational
  Constraint Solver}. In \bibinfo{booktitle}{\emph{2008 23rd IEEE/ACM
  International Conference on Automated Software Engineering}}.
  \bibinfo{pages}{238--247}.
\newblock
\href{https://doi.org/10.1109/ASE.2008.34}{doi:\nolinkurl{10.1109/ASE.2008.34}}


\bibitem[Abdul~Khalek and Khurshid(2010)]%
        {systematic_testing_of_dbms}
\bibfield{author}{\bibinfo{person}{Shadi Abdul~Khalek} {and}
  \bibinfo{person}{Sarfraz Khurshid}.} \bibinfo{year}{2010}\natexlab{}.
\newblock \showarticletitle{Automated SQL query generation for systematic
  testing of database engines}. In \bibinfo{booktitle}{\emph{Proceedings of the
  25th IEEE/ACM International Conference on Automated Software Engineering}}
  (Antwerp, Belgium) \emph{(\bibinfo{series}{ASE '10})}.
  \bibinfo{publisher}{Association for Computing Machinery},
  \bibinfo{address}{New York, NY, USA}, \bibinfo{pages}{329–332}.
\newblock
\showISBNx{9781450301169}
\href{https://doi.org/10.1145/1858996.1859063}{doi:\nolinkurl{10.1145/1858996.1859063}}


\bibitem[Ahmed et~al\mbox{.}(2025)]%
        {ahmedSecVulEvalBenchmarkingLLMs2025}
\bibfield{author}{\bibinfo{person}{Md~Basim~Uddin Ahmed},
  \bibinfo{person}{Nima~Shiri Harzevili}, \bibinfo{person}{Jiho Shin},
  \bibinfo{person}{Hung~Viet Pham}, {and} \bibinfo{person}{Song Wang}.}
  \bibinfo{year}{2025}\natexlab{}.
\newblock \bibinfo{title}{{{SecVulEval}}: {{Benchmarking LLMs}} for
  {{Real-World C}}/{{C}}++ {{Vulnerability Detection}}}.
\newblock
\showeprint[arxiv]{2505.19828}~[cs.SE]
\href{https://doi.org/10.48550/arXiv.2505.19828}{doi:\nolinkurl{10.48550/arXiv.2505.19828}}


\bibitem[Akritidis et~al\mbox{.}(2009)]%
        {akritidisBaggyBoundsChecking2009}
\bibfield{author}{\bibinfo{person}{Periklis Akritidis}, \bibinfo{person}{Manuel
  Costa}, \bibinfo{person}{Miguel Castro}, {and} \bibinfo{person}{Steven
  Hand}.} \bibinfo{year}{2009}\natexlab{}.
\newblock \showarticletitle{Baggy Bounds Checking: An Efficient and
  Backwards-Compatible Defense against out-of-Bounds Errors}. In
  \bibinfo{booktitle}{\emph{Proceedings of the 18th Conference on {{USENIX}}
  Security Symposium}} \emph{(\bibinfo{series}{{{SSYM}}'09})}.
  \bibinfo{publisher}{USENIX Association}, \bibinfo{address}{USA},
  \bibinfo{pages}{51--66}.
\newblock


\bibitem[Amar et~al\mbox{.}(2025)]%
        {amarCHERIoTRTOSOS2025}
\bibfield{author}{\bibinfo{person}{Saar Amar}, \bibinfo{person}{Tony Chen},
  \bibinfo{person}{David Chisnall}, \bibinfo{person}{Nathaniel~Wesley Filardo},
  \bibinfo{person}{Ben Laurie}, \bibinfo{person}{Hugo Lefeuvre},
  \bibinfo{person}{Kunyan Liu}, \bibinfo{person}{Simon~W. Moore},
  \bibinfo{person}{Robert {Norton-Wright}}, \bibinfo{person}{Margo Seltzer},
  \bibinfo{person}{Yucong Tao}, \bibinfo{person}{Robert N.~M. Watson}, {and}
  \bibinfo{person}{Hongyan Xia}.} \bibinfo{year}{2025}\natexlab{}.
\newblock \showarticletitle{{{CHERIoT RTOS}}: {{An OS}} for {{Fine-Grained
  Memory-Safe Compartments}} on {{Low-Cost Embedded Devices}}}. In
  \bibinfo{booktitle}{\emph{Proceedings of the {{ACM SIGOPS}} 31st
  {{Symposium}} on {{Operating Systems Principles}}}}
  \emph{(\bibinfo{series}{{{SOSP}} '25})}. \bibinfo{publisher}{Association for
  Computing Machinery}, \bibinfo{address}{New York, NY, USA},
  \bibinfo{pages}{67--84}.
\newblock
\showISBNx{979-8-4007-1870-0}
\href{https://doi.org/10.1145/3731569.3764844}{doi:\nolinkurl{10.1145/3731569.3764844}}


\bibitem[Amar et~al\mbox{.}(2023)]%
        {amarCHERIoTCompleteMemory2023}
\bibfield{author}{\bibinfo{person}{Saar Amar}, \bibinfo{person}{David
  Chisnall}, \bibinfo{person}{Tony Chen}, \bibinfo{person}{Nathaniel~Wesley
  Filardo}, \bibinfo{person}{Ben Laurie}, \bibinfo{person}{Kunyan Liu},
  \bibinfo{person}{Robert Norton}, \bibinfo{person}{Simon~W. Moore},
  \bibinfo{person}{Yucong Tao}, \bibinfo{person}{Robert N.~M. Watson}, {and}
  \bibinfo{person}{Hongyan Xia}.} \bibinfo{year}{2023}\natexlab{}.
\newblock \showarticletitle{{{CHERIoT}}: {{Complete Memory Safety}} for
  {{Embedded Devices}}}. In \bibinfo{booktitle}{\emph{56th {{Annual
  IEEE}}/{{ACM International Symposium}} on {{Microarchitecture}}}}.
  \bibinfo{publisher}{ACM}, \bibinfo{address}{Toronto ON Canada},
  \bibinfo{pages}{641--653}.
\newblock
\showISBNx{979-8-4007-0329-4}
\href{https://doi.org/10.1145/3613424.3614266}{doi:\nolinkurl{10.1145/3613424.3614266}}


\bibitem[{Amazon}(2026)]%
        {redshift}
\bibfield{author}{\bibinfo{person}{{Amazon}}.} \bibinfo{year}{2026}\natexlab{}.
\newblock \bibinfo{title}{{Amazon Redshift}}.
\newblock
  \bibinfo{howpublished}{\url{https://docs.aws.amazon.com/redshift/latest/mgmt/welcome.html}}.
\newblock


\bibitem[{Android Open Source Project}(2013)]%
        {mte_android_aosp}
\bibfield{author}{\bibinfo{person}{{Android Open Source Project}}.}
  \bibinfo{year}{2013}\natexlab{}.
\newblock \bibinfo{title}{{Arm} {Memory} {Tagging} {Extension}}.
\newblock
  \bibinfo{howpublished}{\url{https://source.android.com/docs/security/test/memory-safety/arm-mte}}.
\newblock
\newblock
\shownote{Accessed: 2026-03-06}.


\bibitem[Ansel et~al\mbox{.}(2011)]%
        {anselLanguageindependentSandboxingJustintime2011}
\bibfield{author}{\bibinfo{person}{Jason Ansel}, \bibinfo{person}{Petr
  Marchenko}, \bibinfo{person}{{\'U}lfar Erlingsson}, \bibinfo{person}{Elijah
  Taylor}, \bibinfo{person}{Brad Chen}, \bibinfo{person}{Derek~L. Schuff},
  \bibinfo{person}{David Sehr}, \bibinfo{person}{Cliff~L. Biffle}, {and}
  \bibinfo{person}{Bennet Yee}.} \bibinfo{year}{2011}\natexlab{}.
\newblock \showarticletitle{Language-Independent Sandboxing of Just-in-Time
  Compilation and Self-Modifying Code}. In
  \bibinfo{booktitle}{\emph{Proceedings of the 32nd {{ACM SIGPLAN Conference}}
  on {{Programming Language Design}} and {{Implementation}}}}
  \emph{(\bibinfo{series}{{{PLDI}} '11})}. \bibinfo{publisher}{Association for
  Computing Machinery}, \bibinfo{address}{New York, NY, USA},
  \bibinfo{pages}{355--366}.
\newblock
\showISBNx{978-1-4503-0663-8}
\href{https://doi.org/10.1145/1993498.1993540}{doi:\nolinkurl{10.1145/1993498.1993540}}


\bibitem[Bozdo{\u g}an et~al\mbox{.}(2022)]%
        {bozdoganSafePMSanitizerPersistent2022}
\bibfield{author}{\bibinfo{person}{Kartal~Kaan Bozdo{\u g}an},
  \bibinfo{person}{Dimitrios Stavrakakis}, \bibinfo{person}{Shady Issa}, {and}
  \bibinfo{person}{Pramod Bhatotia}.} \bibinfo{year}{2022}\natexlab{}.
\newblock \showarticletitle{{{SafePM}}: A Sanitizer for Persistent Memory}. In
  \bibinfo{booktitle}{\emph{Proceedings of the {{Seventeenth European
  Conference}} on {{Computer Systems}}}} \emph{(\bibinfo{series}{{{EuroSys}}
  '22})}. \bibinfo{publisher}{Association for Computing Machinery},
  \bibinfo{address}{New York, NY, USA}, \bibinfo{pages}{506--524}.
\newblock
\showISBNx{978-1-4503-9162-7}
\href{https://doi.org/10.1145/3492321.3519574}{doi:\nolinkurl{10.1145/3492321.3519574}}


\bibitem[Byerly and Khashabi(2026)]%
        {byerlySelfConsistencyFallsShort2026}
\bibfield{author}{\bibinfo{person}{Adam Byerly} {and} \bibinfo{person}{Daniel
  Khashabi}.} \bibinfo{year}{2026}\natexlab{}.
\newblock \showarticletitle{Self-{{Consistency Falls Short}}! {{The Adverse
  Effects}} of {{Positional Bias}} on {{Long-Context Problems}}}.
\newblock \bibinfo{journal}{\emph{Transactions of the Association for
  Computational Linguistics}}  \bibinfo{volume}{14} (\bibinfo{date}{April}
  \bibinfo{year}{2026}), \bibinfo{pages}{292--317}.
\newblock
\showISSN{2307-387X}
\href{https://doi.org/10.1162/TACL.a.625}{doi:\nolinkurl{10.1162/TACL.a.625}}


\bibitem[Chen et~al\mbox{.}(2025)]%
        {chenCHERILiteMemorySafety}
\bibfield{author}{\bibinfo{person}{Tony Chen}, \bibinfo{person}{David
  Chisnall}, {and} \bibinfo{person}{David Chisnall}.}
  \bibinfo{year}{2025}\natexlab{}.
\newblock \showarticletitle{{{CHERI-Lite}} for {{Memory Safety}}}.
\newblock  (\bibinfo{year}{2025}).
\newblock


\bibitem[Chen et~al\mbox{.}(2024)]%
        {chenLimitationsOpportunitiesModern2024}
\bibfield{author}{\bibinfo{person}{Xiangdong Chen}, \bibinfo{person}{Zhaofeng
  Li}, \bibinfo{person}{Tirth Jain}, \bibinfo{person}{Vikram Narayanan}, {and}
  \bibinfo{person}{Anton Burtsev}.} \bibinfo{year}{2024}\natexlab{}.
\newblock \showarticletitle{Limitations and {{Opportunities}} of {{Modern
  Hardware Isolation Mechanisms}}}. In \bibinfo{booktitle}{\emph{2024 {{USENIX
  Annual Technical Conference}} ({{USENIX ATC}} 24)}}.
  \bibinfo{pages}{349--368}.
\newblock
\showISBNx{978-1-939133-41-0}
\urldef\tempurl%
\url{https://www.usenix.org/conference/atc24/presentation/chen-xiangdong}
\showURL{%
\tempurl}


\bibitem[{ClickHouse, Inc.}(2026)]%
        {clickhouse}
\bibfield{author}{\bibinfo{person}{{ClickHouse, Inc.}}}
  \bibinfo{year}{2026}\natexlab{}.
\newblock \bibinfo{title}{{ClickHouse}: A Fast Open-Source OLAP Database
  Management System}.
\newblock
  \bibinfo{howpublished}{\url{https://github.com/ClickHouse/ClickHouse}}.
\newblock


\bibitem[Cooper et~al\mbox{.}(2010)]%
        {cooperBenchmarkingCloudServing2010}
\bibfield{author}{\bibinfo{person}{Brian~F. Cooper}, \bibinfo{person}{Adam
  Silberstein}, \bibinfo{person}{Erwin Tam}, \bibinfo{person}{Raghu
  Ramakrishnan}, {and} \bibinfo{person}{Russell Sears}.}
  \bibinfo{year}{2010}\natexlab{}.
\newblock \showarticletitle{Benchmarking Cloud Serving Systems with {{YCSB}}}.
  In \bibinfo{booktitle}{\emph{Proceedings of the 1st {{ACM}} Symposium on
  {{Cloud}} Computing}} \emph{(\bibinfo{series}{{{SoCC}} '10})}.
  \bibinfo{publisher}{Association for Computing Machinery},
  \bibinfo{address}{New York, NY, USA}, \bibinfo{pages}{143--154}.
\newblock
\showISBNx{978-1-4503-0036-0}
\href{https://doi.org/10.1145/1807128.1807152}{doi:\nolinkurl{10.1145/1807128.1807152}}


\bibitem[Corporation(2025)]%
        {chktag}
\bibfield{author}{\bibinfo{person}{Intel Corporation}.}
  \bibinfo{year}{2025}\natexlab{}.
\newblock \bibinfo{booktitle}{\emph{{ChkTag}: x86 {Memory} {Safety}}}.
\newblock
\urldef\tempurl%
\url{https://community.intel.com/t5/Blogs/Tech-Innovation/open-intel/ChkTag-x86-Memory-Safety/post/1721490}
\showURL{%
\tempurl}
\newblock
\shownote{Accessed: 2026-03-10}.


\bibitem[David et~al\mbox{.}(2015)]%
        {davidAsynchronizedConcurrencySecret2015}
\bibfield{author}{\bibinfo{person}{Tudor David}, \bibinfo{person}{Rachid
  Guerraoui}, {and} \bibinfo{person}{Vasileios Trigonakis}.}
  \bibinfo{year}{2015}\natexlab{}.
\newblock \showarticletitle{Asynchronized {{Concurrency}}: {{The Secret}} to
  {{Scaling Concurrent Search Data Structures}}}.
\newblock \bibinfo{journal}{\emph{SIGARCH Comput. Archit. News}}
  \bibinfo{volume}{43}, \bibinfo{number}{1} (\bibinfo{date}{March}
  \bibinfo{year}{2015}), \bibinfo{pages}{631--644}.
\newblock
\showISSN{0163-5964}
\href{https://doi.org/10.1145/2786763.2694359}{doi:\nolinkurl{10.1145/2786763.2694359}}


\bibitem[DB(nd)]%
        {turso}
\bibfield{author}{\bibinfo{person}{Turso DB}.}
  \bibinfo{year}{[n.\,d.]}\natexlab{}.
\newblock \bibinfo{title}{{Turso, an in-process SQL database, compatible with
  SQLite.}}
\newblock
\urldef\tempurl%
\url{https://github.com/tursodatabase/turso}
\showURL{%
Retrieved 2026-02-06 from \tempurl}


\bibitem[Developers(nda)]%
        {sqlancer}
\bibfield{author}{\bibinfo{person}{SQLancer Developers}.}
  \bibinfo{year}{[n.\,d.]}\natexlab{a}.
\newblock \bibinfo{title}{{SQLancer}}.
\newblock
\urldef\tempurl%
\url{https://github.com/sqlancer/sqlancer}
\showURL{%
Retrieved 2026-02-05 from \tempurl}


\bibitem[Developers(ndb)]%
        {sqlite}
\bibfield{author}{\bibinfo{person}{SQLite Developers}.}
  \bibinfo{year}{[n.\,d.]}\natexlab{b}.
\newblock \bibinfo{title}{{SQLite Home Page}}.
\newblock
\urldef\tempurl%
\url{https://www.sqlite.org/}
\showURL{%
Retrieved 2024-07-24 from \tempurl}


\bibitem[Devietti et~al\mbox{.}(2008)]%
        {hardbound}
\bibfield{author}{\bibinfo{person}{Joe Devietti}, \bibinfo{person}{Colin
  Blundell}, \bibinfo{person}{Milo M.~K. Martin}, {and} \bibinfo{person}{Steve
  Zdancewic}.} \bibinfo{year}{2008}\natexlab{}.
\newblock \showarticletitle{Hardbound: architectural support for spatial safety
  of the C programming language}. In \bibinfo{booktitle}{\emph{Proceedings of
  the 13th International Conference on Architectural Support for Programming
  Languages and Operating Systems}} (Seattle, WA, USA)
  \emph{(\bibinfo{series}{ASPLOS XIII})}. \bibinfo{publisher}{Association for
  Computing Machinery}, \bibinfo{address}{New York, NY, USA},
  \bibinfo{pages}{103–114}.
\newblock
\showISBNx{9781595939586}
\href{https://doi.org/10.1145/1346281.1346295}{doi:\nolinkurl{10.1145/1346281.1346295}}


\bibitem[Dhurjati et~al\mbox{.}(2006)]%
        {dhurjatiSAFECodeEnforcingAlias2006}
\bibfield{author}{\bibinfo{person}{Dinakar Dhurjati}, \bibinfo{person}{Sumant
  Kowshik}, {and} \bibinfo{person}{Vikram Adve}.}
  \bibinfo{year}{2006}\natexlab{}.
\newblock \showarticletitle{{{SAFECode}}: Enforcing Alias Analysis for Weakly
  Typed Languages}. In \bibinfo{booktitle}{\emph{Proceedings of the 27th {{ACM
  SIGPLAN Conference}} on {{Programming Language Design}} and
  {{Implementation}}}} \emph{(\bibinfo{series}{{{PLDI}} '06})}.
  \bibinfo{publisher}{Association for Computing Machinery},
  \bibinfo{address}{New York, NY, USA}, \bibinfo{pages}{144--157}.
\newblock
\showISBNx{978-1-59593-320-1}
\href{https://doi.org/10.1145/1133981.1133999}{doi:\nolinkurl{10.1145/1133981.1133999}}


\bibitem[Duck and Yap(2016)]%
        {duckHeapBoundsProtection2016}
\bibfield{author}{\bibinfo{person}{Gregory~J. Duck} {and}
  \bibinfo{person}{Roland H.~C. Yap}.} \bibinfo{year}{2016}\natexlab{}.
\newblock \showarticletitle{Heap Bounds Protection with Low Fat Pointers}. In
  \bibinfo{booktitle}{\emph{Proceedings of the 25th {{International
  Conference}} on {{Compiler Construction}}}} \emph{(\bibinfo{series}{{{CC}}
  '16})}. \bibinfo{publisher}{Association for Computing Machinery},
  \bibinfo{address}{New York, NY, USA}, \bibinfo{pages}{132--142}.
\newblock
\showISBNx{978-1-4503-4241-4}
\href{https://doi.org/10.1145/2892208.2892212}{doi:\nolinkurl{10.1145/2892208.2892212}}


\bibitem[Duck et~al\mbox{.}(2017)]%
        {duckStackBoundsProtection2017}
\bibfield{author}{\bibinfo{person}{Gregory~J. Duck}, \bibinfo{person}{Roland
  H.~C. Yap}, {and} \bibinfo{person}{Lorenzo Cavallaro}.}
  \bibinfo{year}{2017}\natexlab{}.
\newblock \showarticletitle{Stack {{Bounds Protection}} with {{Low Fat
  Pointers}}}. In \bibinfo{booktitle}{\emph{Proceedings 2017 {{Network}} and
  {{Distributed System Security Symposium}}}}. \bibinfo{publisher}{Internet
  Society}, \bibinfo{address}{San Diego, CA}.
\newblock
\showISBNx{978-1-891562-46-4}
\href{https://doi.org/10.14722/ndss.2017.23287}{doi:\nolinkurl{10.14722/ndss.2017.23287}}


\bibitem[{DuckDB Labs}(2026)]%
        {duckdb}
\bibfield{author}{\bibinfo{person}{{DuckDB Labs}}.}
  \bibinfo{year}{2026}\natexlab{}.
\newblock \bibinfo{title}{{DuckDB}: An Embeddable Analytical Database}.
\newblock \bibinfo{howpublished}{\url{https://github.com/duckdb/duckdb}}.
\newblock


\bibitem[Durner et~al\mbox{.}(2019)]%
        {durner2019impact}
\bibfield{author}{\bibinfo{person}{Dominik Durner}, \bibinfo{person}{Viktor
  Leis}, {and} \bibinfo{person}{Thomas Neumann}.}
  \bibinfo{year}{2019}\natexlab{}.
\newblock \showarticletitle{On the impact of memory allocation on
  high-performance query processing}. In \bibinfo{booktitle}{\emph{Proceedings
  of the 15th International Workshop on Data Management on New Hardware}}.
  \bibinfo{pages}{1--3}.
\newblock


\bibitem[Feng et~al\mbox{.}(2023)]%
        {fengKUZUGraphDatabase2023}
\bibfield{author}{\bibinfo{person}{Xiyang Feng}, \bibinfo{person}{Guodong Jin},
  \bibinfo{person}{Ziyi Chen}, \bibinfo{person}{Chang Liu}, {and}
  \bibinfo{person}{Semih Saliho{\u g}lu}.} \bibinfo{year}{2023}\natexlab{}.
\newblock \showarticletitle{{{K\`UZU Graph Database Management System}}}. In
  \bibinfo{booktitle}{\emph{Conference on {{Innovative Data Systems Research}}
  ({{CIDR}})}}, Vol.~\bibinfo{volume}{7}. \bibinfo{pages}{25--35}.
\newblock


\bibitem[Filardo et~al\mbox{.}(2024)]%
        {filardoCornucopiaReloadedLoad2024}
\bibfield{author}{\bibinfo{person}{Nathaniel~Wesley Filardo},
  \bibinfo{person}{Brett~F. Gutstein}, \bibinfo{person}{Jonathan Woodruff},
  \bibinfo{person}{Jessica Clarke}, \bibinfo{person}{Peter Rugg},
  \bibinfo{person}{Brooks Davis}, \bibinfo{person}{Mark Johnston},
  \bibinfo{person}{Robert Norton}, \bibinfo{person}{David Chisnall},
  \bibinfo{person}{Simon~W. Moore}, \bibinfo{person}{Peter~G. Neumann}, {and}
  \bibinfo{person}{Robert N.~M. Watson}.} \bibinfo{year}{2024}\natexlab{}.
\newblock \showarticletitle{Cornucopia {{Reloaded}}: {{Load Barriers}} for
  {{CHERI Heap Temporal Safety}}}. In \bibinfo{booktitle}{\emph{Proceedings of
  the 29th {{ACM International Conference}} on {{Architectural Support}} for
  {{Programming Languages}} and {{Operating Systems}}, {{Volume}} 2}}
  \emph{(\bibinfo{series}{{{ASPLOS}} '24}, Vol.~\bibinfo{volume}{2})}.
  \bibinfo{publisher}{Association for Computing Machinery},
  \bibinfo{address}{New York, NY, USA}, \bibinfo{pages}{251--268}.
\newblock
\showISBNx{979-8-4007-0385-0}
\href{https://doi.org/10.1145/3620665.3640416}{doi:\nolinkurl{10.1145/3620665.3640416}}


\bibitem[Fink et~al\mbox{.}(2025)]%
        {cage}
\bibfield{author}{\bibinfo{person}{Martin Fink}, \bibinfo{person}{Dimitrios
  Stavrakakis}, \bibinfo{person}{Dennis Sprokholt}, \bibinfo{person}{Soham
  Chakraborty}, \bibinfo{person}{Jan-Erik Ekberg}, {and}
  \bibinfo{person}{Pramod Bhatotia}.} \bibinfo{year}{2025}\natexlab{}.
\newblock \showarticletitle{Cage: Hardware-Accelerated Safe WebAssembly}. In
  \bibinfo{booktitle}{\emph{Proceedings of the 23rd ACM/IEEE International
  Symposium on Code Generation and Optimization}} (Las Vegas, NV, USA)
  \emph{(\bibinfo{series}{CGO '25})}. \bibinfo{publisher}{Association for
  Computing Machinery}, \bibinfo{address}{New York, NY, USA},
  \bibinfo{pages}{538–552}.
\newblock
\showISBNx{9798400712753}
\href{https://doi.org/10.1145/3696443.3708920}{doi:\nolinkurl{10.1145/3696443.3708920}}


\bibitem[Gao et~al\mbox{.}(2023)]%
        {gaoComprehensiveSurveyDatabase2023}
\bibfield{author}{\bibinfo{person}{Xiyue Gao}, \bibinfo{person}{Zhuang Liu},
  \bibinfo{person}{Jiangtao Cui}, \bibinfo{person}{Hui Li},
  \bibinfo{person}{Hui Zhang}, \bibinfo{person}{Kewei Wei}, {and}
  \bibinfo{person}{Kankan Zhao}.} \bibinfo{year}{2023}\natexlab{}.
\newblock \bibinfo{title}{A {{Comprehensive Survey}} on {{Database Management
  System Fuzzing}}: {{Techniques}}, {{Taxonomy}} and {{Experimental
  Comparison}}}.
\newblock
\showeprint[arxiv]{2311.06728}~[cs.DB]
\href{https://doi.org/10.48550/arXiv.2311.06728}{doi:\nolinkurl{10.48550/arXiv.2311.06728}}


\bibitem[Gao et~al\mbox{.}(2025)]%
        {fucci}
\bibfield{author}{\bibinfo{person}{Xiyue Gao}, \bibinfo{person}{Zhuang Liu},
  \bibinfo{person}{Yiran Shen}, \bibinfo{person}{Hui Li},
  \bibinfo{person}{Yingfan Liu}, \bibinfo{person}{Hongjun Xiao},
  \bibinfo{person}{Yanguo Peng}, {and} \bibinfo{person}{Jiangtao Cui}.}
  \bibinfo{year}{2025}\natexlab{}.
\newblock \showarticletitle{Fucci: Database Transaction Fuzzing via Random
  Conflict Construction and Multilevel Constraint Solving}.
\newblock \bibinfo{journal}{\emph{Proc. VLDB Endow.}} \bibinfo{volume}{18},
  \bibinfo{number}{6} (\bibinfo{date}{Feb.} \bibinfo{year}{2025}),
  \bibinfo{pages}{1879–1891}.
\newblock
\showISSN{2150-8097}
\href{https://doi.org/10.14778/3725688.3725713}{doi:\nolinkurl{10.14778/3725688.3725713}}


\bibitem[{ggml}(2026)]%
        {llama_cpp}
\bibfield{author}{\bibinfo{person}{{ggml}}.} \bibinfo{year}{2026}\natexlab{}.
\newblock \bibinfo{title}{{Llama.cpp}}.
\newblock \bibinfo{howpublished}{\url{https://github.com/ggml-org/llama.cpp}}.
\newblock
\newblock
\shownote{Last accessed on 2026-03-02}.


\bibitem[Ghose et~al\mbox{.}(2009)]%
        {safeproc}
\bibfield{author}{\bibinfo{person}{Saugata Ghose}, \bibinfo{person}{Latoya
  Gilgeous}, \bibinfo{person}{Polina Dudnik}, \bibinfo{person}{Aneesh
  Aggarwal}, {and} \bibinfo{person}{Corey Waxman}.}
  \bibinfo{year}{2009}\natexlab{}.
\newblock \showarticletitle{Architectural support for low overhead detection of
  memory violations}. In \bibinfo{booktitle}{\emph{2009 Design, Automation \&
  Test in Europe Conference \& Exhibition}}. \bibinfo{pages}{652--657}.
\newblock
\href{https://doi.org/10.1109/DATE.2009.5090747}{doi:\nolinkurl{10.1109/DATE.2009.5090747}}


\bibitem[Google(2019)]%
        {android_mte}
\bibfield{author}{\bibinfo{person}{Google}.} \bibinfo{year}{2019}\natexlab{}.
\newblock \bibinfo{title}{Adopting the {Arm} {Memory} {Tagging} {Extension} in
  {Android}}.
\newblock
  \bibinfo{howpublished}{\url{https://security.googleblog.com/2019/08/adopting-arm-memory-tagging-extension.html}}.
\newblock
\newblock
\shownote{Accessed: 2026-03-06}.


\bibitem[Google(nd)]%
        {afl}
\bibfield{author}{\bibinfo{person}{Google}.}
  \bibinfo{year}{[n.\,d.]}\natexlab{}.
\newblock \bibinfo{title}{{AFL: american fuzzy lop}}.
\newblock
\urldef\tempurl%
\url{https://github.com/google/AFL}
\showURL{%
Retrieved 2026-02-05 from \tempurl}


\bibitem[{Google LLC}(2026)]%
        {leveldb}
\bibfield{author}{\bibinfo{person}{{Google LLC}}.}
  \bibinfo{year}{2026}\natexlab{}.
\newblock \bibinfo{title}{{LevelDB}: A Fast Key-Value Storage Library}.
\newblock \bibinfo{howpublished}{\url{https://github.com/google/leveldb}}.
\newblock


\bibitem[Gopal et~al\mbox{.}(2023)]%
        {tailcheck}
\bibfield{author}{\bibinfo{person}{Amogha Udupa~Shankaranarayana Gopal},
  \bibinfo{person}{Raveendra Soori}, \bibinfo{person}{Michael Ferdman}, {and}
  \bibinfo{person}{Dongyoon Lee}.} \bibinfo{year}{2023}\natexlab{}.
\newblock \showarticletitle{{TAILCHECK}: A Lightweight Heap Overflow Detection
  Mechanism with Page Protection and Tagged Pointers}. In
  \bibinfo{booktitle}{\emph{17th USENIX Symposium on Operating Systems Design
  and Implementation (OSDI 23)}}. \bibinfo{publisher}{USENIX Association},
  \bibinfo{address}{Boston, MA}, \bibinfo{pages}{535--552}.
\newblock
\showISBNx{978-1-939133-34-2}
\urldef\tempurl%
\url{https://www.usenix.org/conference/osdi23/presentation/gopal}
\showURL{%
\tempurl}


\bibitem[Gorter and Giuffrida(2025)]%
        {RangeSanitizer}
\bibfield{author}{\bibinfo{person}{Floris Gorter} {and}
  \bibinfo{person}{Cristiano Giuffrida}.} \bibinfo{year}{2025}\natexlab{}.
\newblock \showarticletitle{RangeSanitizer: detecting memory errors with
  efficient range checks}. In \bibinfo{booktitle}{\emph{Proceedings of the 34th
  USENIX Conference on Security Symposium}} (Seattle, WA, USA)
  \emph{(\bibinfo{series}{SEC '25})}. \bibinfo{publisher}{USENIX Association},
  \bibinfo{address}{USA}, Article \bibinfo{articleno}{232},
  \bibinfo{numpages}{19}~pages.
\newblock
\showISBNx{978-1-939133-52-6}


\bibitem[Gorter et~al\mbox{.}(2024)]%
        {stickytags}
\bibfield{author}{\bibinfo{person}{Floris Gorter}, \bibinfo{person}{Taddeus
  Kroes}, \bibinfo{person}{Herbert Bos}, {and} \bibinfo{person}{Cristiano
  Giuffrida}.} \bibinfo{year}{2024}\natexlab{}.
\newblock \showarticletitle{Sticky Tags: Efficient and Deterministic Spatial
  Memory Error Mitigation using Persistent Memory Tags}. In
  \bibinfo{booktitle}{\emph{2024 IEEE Symposium on Security and Privacy (SP)}}.
  \bibinfo{pages}{4239--4257}.
\newblock
\href{https://doi.org/10.1109/SP54263.2024.00263}{doi:\nolinkurl{10.1109/SP54263.2024.00263}}


\bibitem[Grisenthwaite et~al\mbox{.}(2023)]%
        {grisenthwaiteArmMorelloEvaluation2023a}
\bibfield{author}{\bibinfo{person}{Richard Grisenthwaite},
  \bibinfo{person}{Graeme Barnes}, \bibinfo{person}{Robert N.~M. Watson},
  \bibinfo{person}{Simon~W. Moore}, \bibinfo{person}{Peter Sewell}, {and}
  \bibinfo{person}{Jonathan Woodruff}.} \bibinfo{year}{2023}\natexlab{}.
\newblock \showarticletitle{The {{Arm Morello Evaluation
  Platform}}---{{Validating CHERI-Based Security}} in a {{High-Performance
  System}}}.
\newblock \bibinfo{journal}{\emph{IEEE Micro}} \bibinfo{volume}{43},
  \bibinfo{number}{3} (\bibinfo{date}{May} \bibinfo{year}{2023}),
  \bibinfo{pages}{50--57}.
\newblock
\showISSN{1937-4143}
\href{https://doi.org/10.1109/MM.2023.3264676}{doi:\nolinkurl{10.1109/MM.2023.3264676}}


\bibitem[Haas et~al\mbox{.}(2017)]%
        {haasBringingWebSpeed2017}
\bibfield{author}{\bibinfo{person}{Andreas Haas}, \bibinfo{person}{Andreas
  Rossberg}, \bibinfo{person}{Derek~L. Schuff}, \bibinfo{person}{Ben~L.
  Titzer}, \bibinfo{person}{Michael Holman}, \bibinfo{person}{Dan Gohman},
  \bibinfo{person}{Luke Wagner}, \bibinfo{person}{Alon Zakai}, {and}
  \bibinfo{person}{{\relax JF} Bastien}.} \bibinfo{year}{2017}\natexlab{}.
\newblock \showarticletitle{Bringing the Web up to Speed with {{WebAssembly}}}.
\newblock \bibinfo{journal}{\emph{SIGPLAN Not.}} \bibinfo{volume}{52},
  \bibinfo{number}{6} (\bibinfo{date}{June} \bibinfo{year}{2017}),
  \bibinfo{pages}{185--200}.
\newblock
\showISSN{0362-1340}
\href{https://doi.org/10.1145/3140587.3062363}{doi:\nolinkurl{10.1145/3140587.3062363}}


\bibitem[Harris et~al\mbox{.}(2023)]%
        {harrisRustMorelloAlwaysOn2023}
\bibfield{author}{\bibinfo{person}{Sarah Harris}, \bibinfo{person}{Simon
  Cooksey}, \bibinfo{person}{Michael Vollmer}, {and} \bibinfo{person}{Mark
  Batty}.} \bibinfo{year}{2023}\natexlab{}.
\newblock \showarticletitle{Rust for {{Morello}}: {{Always-On Memory Safety}},
  {{Even}} in {{Unsafe Code}}}. In \bibinfo{booktitle}{\emph{37th {{European
  Conference}} on {{Object-Oriented Programming}} ({{ECOOP}} 2023)}}
  \emph{(\bibinfo{series}{Leibniz {{International Proceedings}} in
  {{Informatics}} ({{LIPIcs}})}, Vol.~\bibinfo{volume}{263})},
  \bibfield{editor}{\bibinfo{person}{Karim Ali} {and} \bibinfo{person}{Guido
  Salvaneschi}} (Eds.). \bibinfo{publisher}{Schloss Dagstuhl -- Leibniz-Zentrum
  f\"ur Informatik}, \bibinfo{address}{Dagstuhl, Germany},
  \bibinfo{pages}{39:1--39:27}.
\newblock
\showISBNx{978-3-95977-281-5}
\showISSN{1868-8969}
\href{https://doi.org/10.4230/LIPIcs.ECOOP.2023.39}{doi:\nolinkurl{10.4230/LIPIcs.ECOOP.2023.39}}


\bibitem[Hasabnis et~al\mbox{.}(2012)]%
        {hasabnisLightweightBoundsChecking2012}
\bibfield{author}{\bibinfo{person}{Niranjan Hasabnis}, \bibinfo{person}{Ashish
  Misra}, {and} \bibinfo{person}{R. Sekar}.} \bibinfo{year}{2012}\natexlab{}.
\newblock \showarticletitle{Light-Weight Bounds Checking}. In
  \bibinfo{booktitle}{\emph{Proceedings of the {{Tenth International
  Symposium}} on {{Code Generation}} and {{Optimization}}}}
  \emph{(\bibinfo{series}{{{CGO}} '12})}. \bibinfo{publisher}{Association for
  Computing Machinery}, \bibinfo{address}{New York, NY, USA},
  \bibinfo{pages}{135--144}.
\newblock
\showISBNx{978-1-4503-1206-6}
\href{https://doi.org/10.1145/2259016.2259034}{doi:\nolinkurl{10.1145/2259016.2259034}}


\bibitem[Hat(2015)]%
        {ghost_libc}
\bibfield{author}{\bibinfo{person}{Red Hat}.} \bibinfo{year}{2015}\natexlab{}.
\newblock \bibinfo{title}{GHOST: glibc vulnerability (CVE-2015-0235)}.
\newblock
  \bibinfo{howpublished}{\url{https://access.redhat.com/articles/1332213}}.
\newblock
\newblock
\shownote{Accessed: 2026-03-04}.


\bibitem[Hong et~al\mbox{.}(2025)]%
        {cmasan}
\bibfield{author}{\bibinfo{person}{Junwha Hong}, \bibinfo{person}{Wonil Jang},
  \bibinfo{person}{Mijung Kim}, \bibinfo{person}{Lei Yu},
  \bibinfo{person}{Yonghwi Kwon}, {and} \bibinfo{person}{Yuseok Jeon}.}
  \bibinfo{year}{2025}\natexlab{}.
\newblock \showarticletitle{CMASan: Custom Memory Allocator-Aware Address
  Sanitizer}. In \bibinfo{booktitle}{\emph{2025 IEEE Symposium on Security and
  Privacy (SP)}}. \bibinfo{pages}{740--757}.
\newblock
\href{https://doi.org/10.1109/SP61157.2025.00110}{doi:\nolinkurl{10.1109/SP61157.2025.00110}}


\bibitem[Ibarra et~al\mbox{.}(2025)]%
        {ibarra2025performanceconfidentialcomputinggpus}
\bibfield{author}{\bibinfo{person}{Antonio~Martínez Ibarra},
  \bibinfo{person}{Julian~James Stephen}, \bibinfo{person}{Aurora~González
  Vidal}, \bibinfo{person}{K.~R. Jayaram}, {and} \bibinfo{person}{Antonio
  Fernando~Skarmeta Gómez}.} \bibinfo{year}{2025}\natexlab{}.
\newblock \bibinfo{title}{Performance of Confidential Computing GPUs}.
\newblock
\showeprint[arxiv]{2505.16501}~[cs.PF]
\urldef\tempurl%
\url{https://arxiv.org/abs/2505.16501}
\showURL{%
\tempurl}


\bibitem[Inc.(2026)]%
        {googleinc.LevelDB2026}
\bibfield{author}{\bibinfo{person}{Google Inc.}}
  \bibinfo{year}{2026}\natexlab{}.
\newblock \bibinfo{title}{{{LevelDB}}}.
\newblock \bibinfo{howpublished}{Google}.
\newblock
\urldef\tempurl%
\url{https://github.com/google/leveldb}
\showURL{%
\tempurl}


\bibitem[{influxdata}(2026)]%
        {influxdb}
\bibfield{author}{\bibinfo{person}{{influxdata}}.}
  \bibinfo{year}{2026}\natexlab{}.
\newblock \bibinfo{title}{InfluxDB}.
\newblock \bibinfo{howpublished}{\url{https://github.com/influxdata/influxdb}}.
\newblock


\bibitem[Jiang et~al\mbox{.}(2023)]%
        {dynsql}
\bibfield{author}{\bibinfo{person}{Zu-Ming Jiang}, \bibinfo{person}{Jia-Ju
  Bai}, {and} \bibinfo{person}{Zhendong Su}.} \bibinfo{year}{2023}\natexlab{}.
\newblock \showarticletitle{{DynSQL}: Stateful Fuzzing for Database Management
  Systems with Complex and Valid {SQL} Query Generation}. In
  \bibinfo{booktitle}{\emph{32nd USENIX Security Symposium (USENIX Security
  23)}}. \bibinfo{publisher}{USENIX Association}, \bibinfo{address}{Anaheim,
  CA}, \bibinfo{pages}{4949--4965}.
\newblock
\showISBNx{978-1-939133-37-3}
\urldef\tempurl%
\url{https://www.usenix.org/conference/usenixsecurity23/presentation/jiang-zu-ming}
\showURL{%
\tempurl}


\bibitem[Jim et~al\mbox{.}(2002)]%
        {jimCycloneSafeDialect2002}
\bibfield{author}{\bibinfo{person}{Trevor Jim}, \bibinfo{person}{J.~Greg
  Morrisett}, \bibinfo{person}{Dan Grossman}, \bibinfo{person}{Michael~W.
  Hicks}, \bibinfo{person}{James Cheney}, {and} \bibinfo{person}{Yanling
  Wang}.} \bibinfo{year}{2002}\natexlab{}.
\newblock \showarticletitle{Cyclone: {{A Safe Dialect}} of {{C}}}. In
  \bibinfo{booktitle}{\emph{Proceedings of the {{General Track}} of the Annual
  Conference on {{USENIX Annual Technical Conference}}}}
  \emph{(\bibinfo{series}{{{ATEC}} '02})}. \bibinfo{publisher}{USENIX
  Association}, \bibinfo{address}{USA}, \bibinfo{pages}{275--288}.
\newblock
\showISBNx{978-1-880446-00-3}


\bibitem[Jung et~al\mbox{.}(2019)]%
        {apollo}
\bibfield{author}{\bibinfo{person}{Jinho Jung}, \bibinfo{person}{Hong Hu},
  \bibinfo{person}{Joy Arulraj}, \bibinfo{person}{Taesoo Kim}, {and}
  \bibinfo{person}{Woonhak Kang}.} \bibinfo{year}{2019}\natexlab{}.
\newblock \showarticletitle{APOLLO: automatic detection and diagnosis of
  performance regressions in database systems}.
\newblock \bibinfo{journal}{\emph{Proc. VLDB Endow.}} \bibinfo{volume}{13},
  \bibinfo{number}{1} (\bibinfo{date}{Sept.} \bibinfo{year}{2019}),
  \bibinfo{pages}{57–70}.
\newblock
\showISSN{2150-8097}
\href{https://doi.org/10.14778/3357377.3357382}{doi:\nolinkurl{10.14778/3357377.3357382}}


\bibitem[Kaushik et~al\mbox{.}(2025)]%
        {kaushikOptimizedMemoryTagging2025}
\bibfield{author}{\bibinfo{person}{Shiv Kaushik}, \bibinfo{person}{Mahesh
  Madhav}, \bibinfo{person}{Nagi Aboulenein}, \bibinfo{person}{Jason Bessette},
  \bibinfo{person}{Sandeep Brahmadathan}, \bibinfo{person}{Ben Chaffin},
  \bibinfo{person}{Matthew Erler}, \bibinfo{person}{Stephan Jourdan},
  \bibinfo{person}{Thomas Maciukenas}, \bibinfo{person}{Ramya Masti},
  \bibinfo{person}{Jon Perry}, \bibinfo{person}{Massimo Sutera},
  \bibinfo{person}{Scott Tetrick}, \bibinfo{person}{Bret Toll},
  \bibinfo{person}{David Turley}, \bibinfo{person}{Carl Worth}, {and}
  \bibinfo{person}{Atiq Bajwa}.} \bibinfo{year}{2025}\natexlab{}.
\newblock \bibinfo{title}{Optimized {{Memory Tagging}} on {{AmpereOne
  Processors}}}.
\newblock
\showeprint[arxiv]{2511.17773}~[cs]
\href{https://doi.org/10.48550/arXiv.2511.17773}{doi:\nolinkurl{10.48550/arXiv.2511.17773}}


\bibitem[Kim et~al\mbox{.}(2020)]%
        {always_on_heap_memory_safety}
\bibfield{author}{\bibinfo{person}{Yonghae Kim}, \bibinfo{person}{Jaekyu Lee},
  {and} \bibinfo{person}{Hyesoon Kim}.} \bibinfo{year}{2020}\natexlab{}.
\newblock \showarticletitle{Hardware-based Always-On Heap Memory Safety}. In
  \bibinfo{booktitle}{\emph{2020 53rd Annual IEEE/ACM International Symposium
  on Microarchitecture (MICRO)}}. \bibinfo{pages}{1153--1166}.
\newblock
\href{https://doi.org/10.1109/MICRO50266.2020.00095}{doi:\nolinkurl{10.1109/MICRO50266.2020.00095}}


\bibitem[Kopytov(2026)]%
        {kopytovSysbench2026}
\bibfield{author}{\bibinfo{person}{Alexey Kopytov}.}
  \bibinfo{year}{2026}\natexlab{}.
\newblock \bibinfo{title}{Sysbench}.
\newblock
\urldef\tempurl%
\url{https://github.com/akopytov/sysbench}
\showURL{%
\tempurl}


\bibitem[Kressel et~al\mbox{.}(2023)]%
        {kresselSoftwareCompartmentalizationTradeOffs2023}
\bibfield{author}{\bibinfo{person}{John~Alistair Kressel},
  \bibinfo{person}{Hugo Lefeuvre}, {and} \bibinfo{person}{Pierre Olivier}.}
  \bibinfo{year}{2023}\natexlab{}.
\newblock \bibinfo{title}{Software {{Compartmentalization Trade-Offs}} with
  {{Hardware Capabilities}}}.
\newblock
\showeprint[arxiv]{2309.11332}~[cs.CR]
\href{https://doi.org/10.48550/arXiv.2309.11332}{doi:\nolinkurl{10.48550/arXiv.2309.11332}}


\bibitem[Kroes et~al\mbox{.}(2018)]%
        {kroesDeltaPointersBuffer2018}
\bibfield{author}{\bibinfo{person}{Taddeus Kroes}, \bibinfo{person}{Koen
  Koning}, \bibinfo{person}{Erik {van der Kouwe}}, \bibinfo{person}{Herbert
  Bos}, {and} \bibinfo{person}{Cristiano Giuffrida}.}
  \bibinfo{year}{2018}\natexlab{}.
\newblock \showarticletitle{Delta Pointers: Buffer Overflow Checks without the
  Checks}. In \bibinfo{booktitle}{\emph{Proceedings of the {{Thirteenth EuroSys
  Conference}}}} \emph{(\bibinfo{series}{{{EuroSys}} '18})}.
  \bibinfo{publisher}{Association for Computing Machinery},
  \bibinfo{address}{New York, NY, USA}, \bibinfo{pages}{1--14}.
\newblock
\showISBNx{978-1-4503-5584-1}
\href{https://doi.org/10.1145/3190508.3190553}{doi:\nolinkurl{10.1145/3190508.3190553}}


\bibitem[Kuvaiskii et~al\mbox{.}(2017)]%
        {kuvaiskiiSGXBOUNDSMemorySafety2017}
\bibfield{author}{\bibinfo{person}{Dmitrii Kuvaiskii}, \bibinfo{person}{Oleksii
  Oleksenko}, \bibinfo{person}{Sergei Arnautov}, \bibinfo{person}{Bohdan
  Trach}, \bibinfo{person}{Pramod Bhatotia}, \bibinfo{person}{Pascal Felber},
  {and} \bibinfo{person}{Christof Fetzer}.} \bibinfo{year}{2017}\natexlab{}.
\newblock \showarticletitle{{{SGXBOUNDS}}: {{Memory Safety}} for {{Shielded
  Execution}}}. In \bibinfo{booktitle}{\emph{Proceedings of the {{Twelfth
  European Conference}} on {{Computer Systems}}}}
  \emph{(\bibinfo{series}{{{EuroSys}} '17})}. \bibinfo{publisher}{Association
  for Computing Machinery}, \bibinfo{address}{New York, NY, USA},
  \bibinfo{pages}{205--221}.
\newblock
\showISBNx{978-1-4503-4938-3}
\href{https://doi.org/10.1145/3064176.3064192}{doi:\nolinkurl{10.1145/3064176.3064192}}


\bibitem[Kuznetsov et~al\mbox{.}(2014)]%
        {kuznetsovCodepointerIntegrity2014}
\bibfield{author}{\bibinfo{person}{Volodymyr Kuznetsov},
  \bibinfo{person}{L{\'a}szl{\'o} Szekeres}, \bibinfo{person}{Mathias Payer},
  \bibinfo{person}{George Candea}, \bibinfo{person}{R. Sekar}, {and}
  \bibinfo{person}{Dawn Song}.} \bibinfo{year}{2014}\natexlab{}.
\newblock \showarticletitle{Code-Pointer Integrity}. In
  \bibinfo{booktitle}{\emph{Proceedings of the 11th {{USENIX}} Conference on
  {{Operating Systems Design}} and {{Implementation}}}}
  \emph{(\bibinfo{series}{{{OSDI}}'14})}. \bibinfo{publisher}{USENIX
  Association}, \bibinfo{address}{USA}, \bibinfo{pages}{147--163}.
\newblock
\showISBNx{978-1-931971-16-4}


\bibitem[Kwon et~al\mbox{.}(2013)]%
        {kwonLowfatPointersCompact2013}
\bibfield{author}{\bibinfo{person}{Albert Kwon}, \bibinfo{person}{Udit Dhawan},
  \bibinfo{person}{Jonathan~M. Smith}, \bibinfo{person}{Thomas~F. Knight},
  {and} \bibinfo{person}{Andre DeHon}.} \bibinfo{year}{2013}\natexlab{}.
\newblock \showarticletitle{Low-Fat Pointers: Compact Encoding and Efficient
  Gate-Level Implementation of Fat Pointers for Spatial Safety and
  Capability-Based Security}. In \bibinfo{booktitle}{\emph{Proceedings of the
  2013 {{ACM SIGSAC}} Conference on {{Computer}} \& Communications Security}}
  \emph{(\bibinfo{series}{{{CCS}} '13})}. \bibinfo{publisher}{Association for
  Computing Machinery}, \bibinfo{address}{New York, NY, USA},
  \bibinfo{pages}{721--732}.
\newblock
\showISBNx{978-1-4503-2477-9}
\href{https://doi.org/10.1145/2508859.2516713}{doi:\nolinkurl{10.1145/2508859.2516713}}


\bibitem[LadybugDB(0000)]%
        {ladybugdbLadybugDB}
\bibfield{author}{\bibinfo{person}{LadybugDB}.}
  \bibinfo{year}{[n.\,d.]}\natexlab{}.
\newblock \bibinfo{title}{{{LadybugDB}}}.
\newblock \bibinfo{howpublished}{LadybugDB}.
\newblock
\urldef\tempurl%
\url{https://github.com/LadybugDB/ladybug}
\showURL{%
\tempurl}


\bibitem[Leis et~al\mbox{.}(2023)]%
        {leisVirtualMemoryAssistedBuffer2023}
\bibfield{author}{\bibinfo{person}{Viktor Leis}, \bibinfo{person}{Adnan
  Alhomssi}, \bibinfo{person}{Tobias Ziegler}, \bibinfo{person}{Yannick Loeck},
  {and} \bibinfo{person}{Christian Dietrich}.} \bibinfo{year}{2023}\natexlab{}.
\newblock \showarticletitle{Virtual-{{Memory Assisted Buffer Management}}}.
\newblock \bibinfo{journal}{\emph{Proc. ACM Manag. Data}} \bibinfo{volume}{1},
  \bibinfo{number}{1} (\bibinfo{date}{May} \bibinfo{year}{2023}),
  \bibinfo{pages}{7:1--7:25}.
\newblock
\href{https://doi.org/10.1145/3588687}{doi:\nolinkurl{10.1145/3588687}}


\bibitem[Leis et~al\mbox{.}(2013)]%
        {leisAdaptiveRadixTree2013}
\bibfield{author}{\bibinfo{person}{Viktor Leis}, \bibinfo{person}{Alfons
  Kemper}, {and} \bibinfo{person}{Thomas Neumann}.}
  \bibinfo{year}{2013}\natexlab{}.
\newblock \showarticletitle{The Adaptive Radix Tree: {{ARTful}} Indexing for
  Main-Memory Databases}. In \bibinfo{booktitle}{\emph{2013 {{IEEE}} 29th
  {{International Conference}} on {{Data Engineering}} ({{ICDE}})}}.
  \bibinfo{pages}{38--49}.
\newblock
\showISSN{1063-6382}
\href{https://doi.org/10.1109/ICDE.2013.6544812}{doi:\nolinkurl{10.1109/ICDE.2013.6544812}}


\bibitem[LeMay et~al\mbox{.}(2021)]%
        {c3}
\bibfield{author}{\bibinfo{person}{Michael LeMay}, \bibinfo{person}{Joydeep
  Rakshit}, \bibinfo{person}{Sergej Deutsch}, \bibinfo{person}{David~M.
  Durham}, \bibinfo{person}{Santosh Ghosh}, \bibinfo{person}{Anant Nori},
  \bibinfo{person}{Jayesh Gaur}, \bibinfo{person}{Andrew Weiler},
  \bibinfo{person}{Salmin Sultana}, \bibinfo{person}{Karanvir Grewal}, {and}
  \bibinfo{person}{Sreenivas Subramoney}.} \bibinfo{year}{2021}\natexlab{}.
\newblock \showarticletitle{Cryptographic Capability Computing}. In
  \bibinfo{booktitle}{\emph{MICRO-54: 54th Annual IEEE/ACM International
  Symposium on Microarchitecture}} (Virtual Event, Greece)
  \emph{(\bibinfo{series}{MICRO '21})}. \bibinfo{publisher}{Association for
  Computing Machinery}, \bibinfo{address}{New York, NY, USA},
  \bibinfo{pages}{253–267}.
\newblock
\showISBNx{9781450385572}
\href{https://doi.org/10.1145/3466752.3480076}{doi:\nolinkurl{10.1145/3466752.3480076}}


\bibitem[Li et~al\mbox{.}(2026)]%
        {liWhatTheyFix2026}
\bibfield{author}{\bibinfo{person}{Xingyu Li}, \bibinfo{person}{Juefei Pu},
  \bibinfo{person}{Yifan Wu}, \bibinfo{person}{Xiaochen Zou},
  \bibinfo{person}{Shitong Zhu}, \bibinfo{person}{Qiushi Wu},
  \bibinfo{person}{Zheng Zhang}, \bibinfo{person}{Joshua Hsu},
  \bibinfo{person}{Yue Dong}, \bibinfo{person}{Zhiyun Qian},
  \bibinfo{person}{Kangjie Lu}, \bibinfo{person}{Trent Jaeger},
  \bibinfo{person}{Michael De~Lucia}, \bibinfo{person}{Srikanth~V.
  Krishnamurthy}, \bibinfo{person}{Shitong Zhu}, \bibinfo{person}{Qiushi Wu},
  {and} \bibinfo{person}{Joshua Hsu}.} \bibinfo{year}{2026}\natexlab{}.
\newblock \showarticletitle{What {{Do They Fix}}? {{LLM-Aided Categorization}}
  of {{Security Patches}} for {{Critical Memory Bugs}}}. In
  \bibinfo{booktitle}{\emph{Proceedings 2026 {{Network}} and {{Distributed
  System Security Symposium}}}}. \bibinfo{publisher}{Internet Society},
  \bibinfo{address}{San Diego, CA, USA}.
\newblock
\showISBNx{979-8-9919276-8-0}
\href{https://doi.org/10.14722/ndss.2026.230328}{doi:\nolinkurl{10.14722/ndss.2026.230328}}


\bibitem[Li et~al\mbox{.}(2022)]%
        {pacmem}
\bibfield{author}{\bibinfo{person}{Yuan Li}, \bibinfo{person}{Wende Tan},
  \bibinfo{person}{Zhizheng Lv}, \bibinfo{person}{Songtao Yang},
  \bibinfo{person}{Mathias Payer}, \bibinfo{person}{Ying Liu}, {and}
  \bibinfo{person}{Chao Zhang}.} \bibinfo{year}{2022}\natexlab{}.
\newblock \showarticletitle{PACMem: Enforcing Spatial and Temporal Memory
  Safety via ARM Pointer Authentication}. In
  \bibinfo{booktitle}{\emph{Proceedings of the 2022 ACM SIGSAC Conference on
  Computer and Communications Security}} (Los Angeles, CA, USA)
  \emph{(\bibinfo{series}{CCS '22})}. \bibinfo{publisher}{Association for
  Computing Machinery}, \bibinfo{address}{New York, NY, USA},
  \bibinfo{pages}{1901–1915}.
\newblock
\showISBNx{9781450394505}
\href{https://doi.org/10.1145/3548606.3560598}{doi:\nolinkurl{10.1145/3548606.3560598}}


\bibitem[Liang et~al\mbox{.}(2022)]%
        {detecting_logical_bugs}
\bibfield{author}{\bibinfo{person}{Yu Liang}, \bibinfo{person}{Song Liu}, {and}
  \bibinfo{person}{Hong Hu}.} \bibinfo{year}{2022}\natexlab{}.
\newblock \showarticletitle{Detecting Logical Bugs of {DBMS} with
  Coverage-based Guidance}. In \bibinfo{booktitle}{\emph{31st USENIX Security
  Symposium (USENIX Security 22)}}. \bibinfo{publisher}{USENIX Association},
  \bibinfo{address}{Boston, MA}, \bibinfo{pages}{4309--4326}.
\newblock
\showISBNx{978-1-939133-31-1}
\urldef\tempurl%
\url{https://www.usenix.org/conference/usenixsecurity22/presentation/liang}
\showURL{%
\tempurl}


\bibitem[Liljestrand et~al\mbox{.}(2019)]%
        {liljestrandPACItPointer2019}
\bibfield{author}{\bibinfo{person}{Hans Liljestrand}, \bibinfo{person}{Thomas
  Nyman}, \bibinfo{person}{Kui Wang}, \bibinfo{person}{Carlos~Chinea Perez},
  \bibinfo{person}{Jan-Erik Ekberg}, {and} \bibinfo{person}{N. Asokan}.}
  \bibinfo{year}{2019}\natexlab{}.
\newblock \showarticletitle{{{PAC}} It up: Towards Pointer Integrity Using
  {{ARM}} Pointer Authentication}. In \bibinfo{booktitle}{\emph{Proceedings of
  the 28th {{USENIX Conference}} on {{Security Symposium}}}}
  \emph{(\bibinfo{series}{{{SEC}}'19})}. \bibinfo{publisher}{USENIX
  Association}, \bibinfo{address}{USA}, \bibinfo{pages}{177--194}.
\newblock
\showISBNx{978-1-939133-06-9}


\bibitem[Lin et~al\mbox{.}(2024)]%
        {camp}
\bibfield{author}{\bibinfo{person}{Zhenpeng Lin}, \bibinfo{person}{Zheng Yu},
  \bibinfo{person}{Ziyi Guo}, \bibinfo{person}{Simone Campanoni},
  \bibinfo{person}{Peter Dinda}, {and} \bibinfo{person}{Xinyu Xing}.}
  \bibinfo{year}{2024}\natexlab{}.
\newblock \showarticletitle{{CAMP}: Compiler and Allocator-based Heap Memory
  Protection}. In \bibinfo{booktitle}{\emph{33rd USENIX Security Symposium
  (USENIX Security 24)}}. \bibinfo{publisher}{USENIX Association},
  \bibinfo{address}{Philadelphia, PA}.
\newblock
\urldef\tempurl%
\url{https://www.usenix.org/conference/usenixsecurity24/presentation/lin}
\showURL{%
\tempurl}


\bibitem[Ling et~al\mbox{.}(2024)]%
        {giantsan}
\bibfield{author}{\bibinfo{person}{Hao Ling}, \bibinfo{person}{Heqing Huang},
  \bibinfo{person}{Chengpeng Wang}, \bibinfo{person}{Yuandao Cai}, {and}
  \bibinfo{person}{Charles Zhang}.} \bibinfo{year}{2024}\natexlab{}.
\newblock \showarticletitle{GiantSan: Efficient Memory Sanitization with
  Segment Folding}. In \bibinfo{booktitle}{\emph{29th ACM International
  Conference on Architectural Support for Programming Languages and Operating
  Systems (ASPLOS 2024)}}.
\newblock


\bibitem[Lowther et~al\mbox{.}(2023)]%
        {lowtherCHERIPerformanceEnhancement2023}
\bibfield{author}{\bibinfo{person}{Duncan Lowther}, \bibinfo{person}{Dejice
  Jacob}, {and} \bibinfo{person}{Jeremy Singer}.}
  \bibinfo{year}{2023}\natexlab{}.
\newblock \bibinfo{title}{{{CHERI Performance Enhancement}} for a {{Bytecode
  Interpreter}}}.
\newblock
\showeprint[arxiv]{2308.05076}~[cs]
\urldef\tempurl%
\url{http://arxiv.org/abs/2308.05076}
\showURL{%
\tempurl}


\bibitem[Ltd.(2019)]%
        {armltd.Armv85AMemoryTagging2019}
\bibfield{author}{\bibinfo{person}{Arm Ltd.}} \bibinfo{year}{2019}\natexlab{}.
\newblock \bibinfo{booktitle}{\emph{Armv8.5-{{A Memory Tagging Extension}}}}.
\newblock \bibinfo{type}{{T}echnical {R}eport}.
\newblock
\urldef\tempurl%
\url{https://developer.arm.com/documentation/102925/latest}
\showURL{%
\tempurl}


\bibitem[Ltd.(2022)]%
        {armltd.ArmArchitectureReference2022}
\bibfield{author}{\bibinfo{person}{Arm Ltd.}} \bibinfo{year}{2022}\natexlab{}.
\newblock \bibinfo{booktitle}{\emph{Arm {{Architecture Reference Manual
  Supplement}} - {{Morello}} for {{A-profile Architecture}}}}.
\newblock \bibinfo{type}{{T}echnical {R}eport}.
\newblock
\urldef\tempurl%
\url{https://developer.arm.com/documentation/ddi0606/latest}
\showURL{%
\tempurl}


\bibitem[Ltd.(2025)]%
        {armltd.ArmArchitectureReference2025}
\bibfield{author}{\bibinfo{person}{Arm Ltd.}} \bibinfo{year}{2025}\natexlab{}.
\newblock \bibinfo{booktitle}{\emph{{{Arm}}\textregistered{} {{Architecture
  Reference Manual}} for {{A-profile}} Architecture}}.
\newblock \bibinfo{type}{{T}echnical {R}eport}.
\newblock
\urldef\tempurl%
\url{https://developer.arm.com/documentation/ddi0487/latest}
\showURL{%
\tempurl}


\bibitem[{MariaDB Foundation}(2026)]%
        {mariadb}
\bibfield{author}{\bibinfo{person}{{MariaDB Foundation}}.}
  \bibinfo{year}{2026}\natexlab{}.
\newblock \bibinfo{title}{{MariaDB} Server}.
\newblock \bibinfo{howpublished}{\url{https://github.com/MariaDB/server}}.
\newblock


\bibitem[Meignan-{}-Masson et~al\mbox{.}(2026)]%
        {memsafedb}
\bibfield{author}{\bibinfo{person}{Ilya Meignan-{}-Masson},
  \bibinfo{person}{Martin Fink}, \bibinfo{person}{Masanori Misono},
  \bibinfo{person}{Dimitrios Stavrakakis}, {and} \bibinfo{person}{Pramod
  Bhatotia}.} \bibinfo{year}{2026}\natexlab{}.
\newblock \showarticletitle{Should Your Database Systems Use Hardware-Assisted
  Memory Safety Extensions in Production?}
\newblock \bibinfo{journal}{\emph{Proc. ACM Manag. Data}} \bibinfo{volume}{4},
  \bibinfo{number}{6 (SIGMOD)}, Article \bibinfo{articleno}{378}
  (\bibinfo{date}{Dec.} \bibinfo{year}{2026}).
\newblock
\href{https://doi.org/10.1145/3856373}{doi:\nolinkurl{10.1145/3856373}}


\bibitem[Memarian et~al\mbox{.}(2016)]%
        {depths-c}
\bibfield{author}{\bibinfo{person}{Kayvan Memarian}, \bibinfo{person}{Justus
  Matthiesen}, \bibinfo{person}{James Lingard}, \bibinfo{person}{Kyndylan
  Nienhuis}, \bibinfo{person}{David Chisnall}, \bibinfo{person}{Robert~NM
  Watson}, {and} \bibinfo{person}{Peter Sewell}.}
  \bibinfo{year}{2016}\natexlab{}.
\newblock \showarticletitle{Into the depths of C: elaborating the de facto
  standards}.
\newblock \bibinfo{journal}{\emph{ACM SIGPLAN Notices}} \bibinfo{volume}{51},
  \bibinfo{number}{6} (\bibinfo{year}{2016}), \bibinfo{pages}{1--15}.
\newblock


\bibitem[{memcached}(2026)]%
        {memcached}
\bibfield{author}{\bibinfo{person}{{memcached}}.}
  \bibinfo{year}{2026}\natexlab{}.
\newblock \bibinfo{title}{{memcached}: a high performance multithreaded
  event-based key/value cache store}.
\newblock \bibinfo{howpublished}{\url{https://github.com/memcached/memcached}}.
\newblock


\bibitem[{Meta Platforms, Inc.}(2026)]%
        {rocksdb}
\bibfield{author}{\bibinfo{person}{{Meta Platforms, Inc.}}}
  \bibinfo{year}{2026}\natexlab{}.
\newblock \bibinfo{title}{{RocksDB}: A Persistent Key-Value Store for Flash and
  RAM Storage}.
\newblock \bibinfo{howpublished}{\url{https://github.com/facebook/rocksdb}}.
\newblock


\bibitem[Misono et~al\mbox{.}(2025)]%
        {misonoConfidentialVMsExplained2025}
\bibfield{author}{\bibinfo{person}{Masanori Misono}, \bibinfo{person}{Dimitrios
  Stavrakakis}, \bibinfo{person}{Nuno Santos}, {and} \bibinfo{person}{Pramod
  Bhatotia}.} \bibinfo{year}{2025}\natexlab{}.
\newblock \showarticletitle{Confidential {{VMs Explained}}: {{An Empirical
  Analysis}} of {{AMD SEV-SNP}} and {{Intel TDX}}}.
\newblock \bibinfo{journal}{\emph{SIGMETRICS Perform. Eval. Rev.}}
  \bibinfo{volume}{53}, \bibinfo{number}{1} (\bibinfo{date}{June}
  \bibinfo{year}{2025}), \bibinfo{pages}{34--36}.
\newblock
\showISSN{0163-5999}
\href{https://doi.org/10.1145/3744970.3727280}{doi:\nolinkurl{10.1145/3744970.3727280}}


\bibitem[{MITRE Corporation}(2025)]%
        {cvelistv5}
\bibfield{author}{\bibinfo{person}{{MITRE Corporation}}.}
  \bibinfo{year}{2025}\natexlab{}.
\newblock \bibinfo{title}{{CVEProject/cvelistV5}: Official {CVE} List in {JSON}
  5.0 Format}.
\newblock
  \bibinfo{howpublished}{\url{https://github.com/CVEProject/cvelistV5}}.
\newblock


\bibitem[Momeu et~al\mbox{.}(2025)]%
        {momeuIUBIKIsolatingUser2025}
\bibfield{author}{\bibinfo{person}{Marius Momeu}, \bibinfo{person}{Alexander~J.
  Gaidis}, \bibinfo{person}{Jasper {v.d. Heidt}}, {and}
  \bibinfo{person}{Vasileios~P. Kemerlis}.} \bibinfo{year}{2025}\natexlab{}.
\newblock \showarticletitle{{{IUBIK}}: {{Isolating User Bytes}} in {{Commodity
  Operating System Kernels}} via {{Memory Tagging Extensions}}}. In
  \bibinfo{booktitle}{\emph{2025 {{IEEE Symposium}} on {{Security}} and
  {{Privacy}} ({{SP}})}}. \bibinfo{pages}{867--885}.
\newblock
\showISSN{2375-1207}
\href{https://doi.org/10.1109/SP61157.2025.00135}{doi:\nolinkurl{10.1109/SP61157.2025.00135}}


\bibitem[{MongoDB, Inc.}(2026)]%
        {mongodb}
\bibfield{author}{\bibinfo{person}{{MongoDB, Inc.}}}
  \bibinfo{year}{2026}\natexlab{}.
\newblock \bibinfo{title}{{MongoDB}: A Document-Oriented Database}.
\newblock \bibinfo{howpublished}{\url{https://github.com/mongodb/mongo}}.
\newblock


\bibitem[Nagarakatte et~al\mbox{.}(2012)]%
        {watchdog}
\bibfield{author}{\bibinfo{person}{Santosh Nagarakatte}, \bibinfo{person}{Milo
  M.~K. Martin}, {and} \bibinfo{person}{Steve Zdancewic}.}
  \bibinfo{year}{2012}\natexlab{}.
\newblock \showarticletitle{Watchdog: Hardware for safe and secure manual
  memory management and full memory safety}. In \bibinfo{booktitle}{\emph{2012
  39th Annual International Symposium on Computer Architecture (ISCA)}}.
  \bibinfo{pages}{189--200}.
\newblock
\href{https://doi.org/10.1109/ISCA.2012.6237017}{doi:\nolinkurl{10.1109/ISCA.2012.6237017}}


\bibitem[Nagarakatte et~al\mbox{.}(2014)]%
        {watchdoglite}
\bibfield{author}{\bibinfo{person}{Santosh Nagarakatte}, \bibinfo{person}{Milo
  M.~K. Martin}, {and} \bibinfo{person}{Steve Zdancewic}.}
  \bibinfo{year}{2014}\natexlab{}.
\newblock \showarticletitle{WatchdogLite: Hardware-Accelerated Compiler-Based
  Pointer Checking}. In \bibinfo{booktitle}{\emph{Proceedings of Annual
  IEEE/ACM International Symposium on Code Generation and Optimization}}
  (Orlando, FL, USA) \emph{(\bibinfo{series}{CGO '14})}.
  \bibinfo{publisher}{Association for Computing Machinery},
  \bibinfo{address}{New York, NY, USA}, \bibinfo{pages}{175–184}.
\newblock
\showISBNx{9781450326704}
\href{https://doi.org/10.1145/2581122.2544147}{doi:\nolinkurl{10.1145/2581122.2544147}}


\bibitem[Nagarakatte et~al\mbox{.}(2009)]%
        {nagarakatteSoftBoundHighlyCompatible2009}
\bibfield{author}{\bibinfo{person}{Santosh Nagarakatte},
  \bibinfo{person}{Jianzhou Zhao}, \bibinfo{person}{Milo~M.K. Martin}, {and}
  \bibinfo{person}{Steve Zdancewic}.} \bibinfo{year}{2009}\natexlab{}.
\newblock \showarticletitle{{{SoftBound}}: Highly Compatible and Complete
  Spatial Memory Safety for c}.
\newblock \bibinfo{journal}{\emph{SIGPLAN Not.}} \bibinfo{volume}{44},
  \bibinfo{number}{6} (\bibinfo{date}{June} \bibinfo{year}{2009}),
  \bibinfo{pages}{245--258}.
\newblock
\showISSN{0362-1340}
\href{https://doi.org/10.1145/1543135.1542504}{doi:\nolinkurl{10.1145/1543135.1542504}}


\bibitem[Narayan et~al\mbox{.}(2023)]%
        {narayanGoingLimitsSFI2023}
\bibfield{author}{\bibinfo{person}{Shravan Narayan}, \bibinfo{person}{Tal
  Garfinkel}, \bibinfo{person}{Mohammadkazem Taram}, \bibinfo{person}{Joey
  Rudek}, \bibinfo{person}{Daniel Moghimi}, \bibinfo{person}{Evan Johnson},
  \bibinfo{person}{Chris Fallin}, \bibinfo{person}{Anjo {Vahldiek-Oberwagner}},
  \bibinfo{person}{Michael LeMay}, \bibinfo{person}{Ravi Sahita},
  \bibinfo{person}{Dean Tullsen}, {and} \bibinfo{person}{Deian Stefan}.}
  \bibinfo{year}{2023}\natexlab{}.
\newblock \showarticletitle{Going beyond the {{Limits}} of {{SFI}}:
  {{Flexible}} and {{Secure Hardware-Assisted In-Process Isolation}} with
  {{HFI}}}. In \bibinfo{booktitle}{\emph{Proceedings of the 28th {{ACM
  International Conference}} on {{Architectural Support}} for {{Programming
  Languages}} and {{Operating Systems}}, {{Volume}} 3}}
  \emph{(\bibinfo{series}{{{ASPLOS}} 2023})}. \bibinfo{publisher}{Association
  for Computing Machinery}, \bibinfo{address}{New York, NY, USA},
  \bibinfo{pages}{266--281}.
\newblock
\showISBNx{978-1-4503-9918-0}
\href{https://doi.org/10.1145/3582016.3582023}{doi:\nolinkurl{10.1145/3582016.3582023}}


\bibitem[Necula et~al\mbox{.}(2005)]%
        {neculaCCuredTypesafeRetrofitting2005}
\bibfield{author}{\bibinfo{person}{George~C. Necula}, \bibinfo{person}{Jeremy
  Condit}, \bibinfo{person}{Matthew Harren}, \bibinfo{person}{Scott McPeak},
  {and} \bibinfo{person}{Westley Weimer}.} \bibinfo{year}{2005}\natexlab{}.
\newblock \showarticletitle{{{CCured}}: Type-Safe Retrofitting of Legacy
  Software}.
\newblock \bibinfo{journal}{\emph{ACM Trans. Program. Lang. Syst.}}
  \bibinfo{volume}{27}, \bibinfo{number}{3} (\bibinfo{date}{May}
  \bibinfo{year}{2005}), \bibinfo{pages}{477--526}.
\newblock
\showISSN{0164-0925}
\href{https://doi.org/10.1145/1065887.1065892}{doi:\nolinkurl{10.1145/1065887.1065892}}


\bibitem[Nethercote and Seward(2007)]%
        {nethercoteValgrindFrameworkHeavyweight2007}
\bibfield{author}{\bibinfo{person}{Nicholas Nethercote} {and}
  \bibinfo{person}{Julian Seward}.} \bibinfo{year}{2007}\natexlab{}.
\newblock \showarticletitle{Valgrind: A Framework for Heavyweight Dynamic
  Binary Instrumentation}.
\newblock \bibinfo{journal}{\emph{SIGPLAN Not.}} \bibinfo{volume}{42},
  \bibinfo{number}{6} (\bibinfo{date}{June} \bibinfo{year}{2007}),
  \bibinfo{pages}{89--100}.
\newblock
\showISSN{0362-1340}
\href{https://doi.org/10.1145/1273442.1250746}{doi:\nolinkurl{10.1145/1273442.1250746}}


\bibitem[Noh et~al\mbox{.}(2026)]%
        {noh2026armmteperformancepractice}
\bibfield{author}{\bibinfo{person}{Taehyun Noh}, \bibinfo{person}{Yingchen
  Wang}, \bibinfo{person}{Tal Garfinkel}, \bibinfo{person}{Mahesh Madhav},
  \bibinfo{person}{Daniel Moghimi}, \bibinfo{person}{Mattan Erez}, {and}
  \bibinfo{person}{Shravan Narayan}.} \bibinfo{year}{2026}\natexlab{}.
\newblock \bibinfo{title}{ARM MTE Performance in Practice (Extended Version)}.
\newblock
\showeprint[arxiv]{2601.11786}~[cs.CR]
\urldef\tempurl%
\url{https://arxiv.org/abs/2601.11786}
\showURL{%
\tempurl}


\bibitem[Oleksenko et~al\mbox{.}(2018)]%
        {oleksenkoIntelMPXExplained2018b}
\bibfield{author}{\bibinfo{person}{Oleksii Oleksenko}, \bibinfo{person}{Dmitrii
  Kuvaiskii}, \bibinfo{person}{Pramod Bhatotia}, \bibinfo{person}{Pascal
  Felber}, {and} \bibinfo{person}{Christof Fetzer}.}
  \bibinfo{year}{2018}\natexlab{}.
\newblock \showarticletitle{Intel {{MPX Explained}}: {{A Cross-layer Analysis}}
  of the {{Intel MPX System Stack}}}.
\newblock \bibinfo{journal}{\emph{Proc. ACM Meas. Anal. Comput. Syst.}}
  \bibinfo{volume}{2}, \bibinfo{number}{2} (\bibinfo{date}{June}
  \bibinfo{year}{2018}), \bibinfo{pages}{28:1--28:30}.
\newblock
\href{https://doi.org/10.1145/3224423}{doi:\nolinkurl{10.1145/3224423}}


\bibitem[Oracle(0000)]%
        {oracleinc.MySQL}
\bibfield{author}{\bibinfo{person}{Inc. Oracle}.}
  \bibinfo{year}{[n.\,d.]}\natexlab{}.
\newblock \bibinfo{title}{{{MySQL}}}.
\newblock
\urldef\tempurl%
\url{https://www.mysql.com/}
\showURL{%
\tempurl}


\bibitem[{Oracle Corporation}(2026)]%
        {mysql}
\bibfield{author}{\bibinfo{person}{{Oracle Corporation}}.}
  \bibinfo{year}{2026}\natexlab{}.
\newblock \bibinfo{title}{{MySQL} Server}.
\newblock \bibinfo{howpublished}{\url{https://github.com/mysql/mysql-server}}.
\newblock


\bibitem[Park et~al\mbox{.}(2019)]%
        {parkLibmpkSoftwareAbstraction2019}
\bibfield{author}{\bibinfo{person}{Soyeon Park}, \bibinfo{person}{Sangho Lee},
  \bibinfo{person}{Wen Xu}, \bibinfo{person}{Hyungon Moon}, {and}
  \bibinfo{person}{Taesoo Kim}.} \bibinfo{year}{2019}\natexlab{}.
\newblock \showarticletitle{Libmpk: Software Abstraction for Intel Memory
  Protection Keys (Intel {{MPK}})}. In \bibinfo{booktitle}{\emph{Proceedings of
  the 2019 {{USENIX Conference}} on {{Usenix Annual Technical Conference}}}}
  \emph{(\bibinfo{series}{{{USENIX ATC}} '19})}. \bibinfo{publisher}{USENIX
  Association}, \bibinfo{address}{USA}, \bibinfo{pages}{241--254}.
\newblock
\showISBNx{978-1-939133-03-8}


\bibitem[Pavlo(2025)]%
        {pavloPytpcc2025}
\bibfield{author}{\bibinfo{person}{Andy Pavlo}.}
  \bibinfo{year}{2025}\natexlab{}.
\newblock \bibinfo{title}{Py-Tpcc}.
\newblock
\urldef\tempurl%
\url{https://github.com/apavlo/py-tpcc}
\showURL{%
\tempurl}


\bibitem[Project(0000)]%
        {thechromiumprojectMemorySafety}
\bibfield{author}{\bibinfo{person}{The~Chromium Project}.}
  \bibinfo{year}{[n.\,d.]}\natexlab{}.
\newblock \bibinfo{title}{Memory Safety}.
\newblock
\urldef\tempurl%
\url{https://www.chromium.org/Home/chromium-security/memory-safety/}
\showURL{%
\tempurl}


\bibitem[Raasveldt and M{\"u}hleisen(2019)]%
        {raasveldtDuckDBEmbeddableAnalytical2019}
\bibfield{author}{\bibinfo{person}{Mark Raasveldt} {and}
  \bibinfo{person}{Hannes M{\"u}hleisen}.} \bibinfo{year}{2019}\natexlab{}.
\newblock \showarticletitle{{{DuckDB}}: An {{Embeddable Analytical Database}}}.
  In \bibinfo{booktitle}{\emph{Proceedings of the 2019 {{International
  Conference}} on {{Management}} of {{Data}}}}
  \emph{(\bibinfo{series}{{{SIGMOD}} '19})}. \bibinfo{publisher}{Association
  for Computing Machinery}, \bibinfo{address}{New York, NY, USA},
  \bibinfo{pages}{1981--1984}.
\newblock
\showISBNx{978-1-4503-5643-5}
\href{https://doi.org/10.1145/3299869.3320212}{doi:\nolinkurl{10.1145/3299869.3320212}}


\bibitem[Redis(0000)]%
        {redisRedisRealtimeData}
\bibfield{author}{\bibinfo{person}{Redis}.}
  \bibinfo{year}{[n.\,d.]}\natexlab{}.
\newblock \bibinfo{title}{Redis - {{The Real-time Data Platform}}}.
\newblock
\urldef\tempurl%
\url{https://redis.io/}
\showURL{%
\tempurl}


\bibitem[{Redis Ltd.}(2026)]%
        {redis}
\bibfield{author}{\bibinfo{person}{{Redis Ltd.}}}
  \bibinfo{year}{2026}\natexlab{}.
\newblock \bibinfo{title}{{Redis}: In-Memory Data Structure Store}.
\newblock \bibinfo{howpublished}{\url{https://github.com/redis/redis}}.
\newblock


\bibitem[Rigger and Su(2020a)]%
        {optimization_bugs_in_dbs}
\bibfield{author}{\bibinfo{person}{Manuel Rigger} {and}
  \bibinfo{person}{Zhendong Su}.} \bibinfo{year}{2020}\natexlab{a}.
\newblock \showarticletitle{Detecting optimization bugs in database engines via
  non-optimizing reference engine construction}. In
  \bibinfo{booktitle}{\emph{Proceedings of the 28th ACM Joint Meeting on
  European Software Engineering Conference and Symposium on the Foundations of
  Software Engineering}} (Virtual Event, USA) \emph{(\bibinfo{series}{ESEC/FSE
  2020})}. \bibinfo{publisher}{Association for Computing Machinery},
  \bibinfo{address}{New York, NY, USA}, \bibinfo{pages}{1140–1152}.
\newblock
\showISBNx{9781450370431}
\href{https://doi.org/10.1145/3368089.3409710}{doi:\nolinkurl{10.1145/3368089.3409710}}


\bibitem[Rigger and Su(2020b)]%
        {finding_bugs_query_partitioning}
\bibfield{author}{\bibinfo{person}{Manuel Rigger} {and}
  \bibinfo{person}{Zhendong Su}.} \bibinfo{year}{2020}\natexlab{b}.
\newblock \showarticletitle{Finding bugs in database systems via query
  partitioning}.
\newblock \bibinfo{journal}{\emph{Proc. ACM Program. Lang.}}
  \bibinfo{volume}{4}, \bibinfo{number}{OOPSLA}, Article
  \bibinfo{articleno}{211} (\bibinfo{date}{Nov.} \bibinfo{year}{2020}),
  \bibinfo{numpages}{30}~pages.
\newblock
\href{https://doi.org/10.1145/3428279}{doi:\nolinkurl{10.1145/3428279}}


\bibitem[Rigger and Su(2020c)]%
        {testing_db_pivoted_query_synthesis}
\bibfield{author}{\bibinfo{person}{Manuel Rigger} {and}
  \bibinfo{person}{Zhendong Su}.} \bibinfo{year}{2020}\natexlab{c}.
\newblock \showarticletitle{Testing Database Engines via Pivoted Query
  Synthesis}. In \bibinfo{booktitle}{\emph{14th USENIX Symposium on Operating
  Systems Design and Implementation (OSDI 20)}}. \bibinfo{publisher}{USENIX
  Association}, \bibinfo{pages}{667--682}.
\newblock
\showISBNx{978-1-939133-19-9}
\urldef\tempurl%
\url{https://www.usenix.org/conference/osdi20/presentation/rigger}
\showURL{%
\tempurl}


\bibitem[Seltenreich(nd)]%
        {sqlsmith}
\bibfield{author}{\bibinfo{person}{Andreas Seltenreich}.}
  \bibinfo{year}{[n.\,d.]}\natexlab{}.
\newblock \bibinfo{title}{{SQLsmith}}.
\newblock
\urldef\tempurl%
\url{https://github.com/anse1/sqlsmith}
\showURL{%
Retrieved 2026-02-05 from \tempurl}


\bibitem[Serebryany et~al\mbox{.}(2012)]%
        {asan}
\bibfield{author}{\bibinfo{person}{Konstantin Serebryany},
  \bibinfo{person}{Derek Bruening}, \bibinfo{person}{Alexander Potapenko},
  {and} \bibinfo{person}{Dmitriy Vyukov}.} \bibinfo{year}{2012}\natexlab{}.
\newblock \showarticletitle{AddressSanitizer: A Fast Address Sanity Checker}.
  In \bibinfo{booktitle}{\emph{2012 {USENIX} Annual Technical Conference
  ({USENIX} {ATC} 12)}}. \bibinfo{publisher}{{USENIX} Association},
  \bibinfo{address}{Boston, MA}, \bibinfo{pages}{309--318}.
\newblock
\showISBNx{978-931971-93-5}
\urldef\tempurl%
\url{https://www.usenix.org/conference/atc12/technical-sessions/presentation/serebryany}
\showURL{%
\tempurl}


\bibitem[Serebryany et~al\mbox{.}(2024)]%
        {serebryanyGWPASanSamplingBasedDetection2024}
\bibfield{author}{\bibinfo{person}{Kostya Serebryany}, \bibinfo{person}{Chris
  Kennelly}, \bibinfo{person}{Mitch Phillips}, \bibinfo{person}{Matt Denton},
  \bibinfo{person}{Marco Elver}, \bibinfo{person}{Alexander Potapenko},
  \bibinfo{person}{Matt Morehouse}, \bibinfo{person}{Vlad Tsyrklevich},
  \bibinfo{person}{Christian Holler}, \bibinfo{person}{Julian Lettner},
  \bibinfo{person}{David Kilzer}, {and} \bibinfo{person}{Lander Brandt}.}
  \bibinfo{year}{2024}\natexlab{}.
\newblock \showarticletitle{{{GWP-ASan}}: {{Sampling-Based Detection}} of
  {{Memory-Safety Bugs}} in {{Production}}}. In
  \bibinfo{booktitle}{\emph{Proceedings of the 46th {{International
  Conference}} on {{Software Engineering}}: {{Software Engineering}} in
  {{Practice}}}} \emph{(\bibinfo{series}{{{ICSE-SEIP}} '24})}.
  \bibinfo{publisher}{Association for Computing Machinery},
  \bibinfo{address}{New York, NY, USA}, \bibinfo{pages}{168--177}.
\newblock
\showISBNx{979-8-4007-0501-4}
\href{https://doi.org/10.1145/3639477.3640328}{doi:\nolinkurl{10.1145/3639477.3640328}}


\bibitem[Serebryany et~al\mbox{.}(2018)]%
        {serebryanyMemoryTaggingHow2018}
\bibfield{author}{\bibinfo{person}{Kostya Serebryany}, \bibinfo{person}{Evgenii
  Stepanov}, \bibinfo{person}{Aleksey Shlyapnikov}, \bibinfo{person}{Vlad
  Tsyrklevich}, {and} \bibinfo{person}{Dmitry Vyukov}.}
  \bibinfo{year}{2018}\natexlab{}.
\newblock \bibinfo{title}{Memory {{Tagging}} and How It Improves {{C}}/{{C}}++
  Memory Safety}.
\newblock
\showeprint[arxiv]{1802.09517}~[cs]
\href{https://doi.org/10.48550/arXiv.1802.09517}{doi:\nolinkurl{10.48550/arXiv.1802.09517}}


\bibitem[Slutz(1998)]%
        {rags}
\bibfield{author}{\bibinfo{person}{Donald~R. Slutz}.}
  \bibinfo{year}{1998}\natexlab{}.
\newblock \showarticletitle{Massive Stochastic Testing of SQL}. In
  \bibinfo{booktitle}{\emph{Proceedings of the 24rd International Conference on
  Very Large Data Bases}} \emph{(\bibinfo{series}{VLDB '98})}.
  \bibinfo{publisher}{Morgan Kaufmann Publishers Inc.}, \bibinfo{address}{San
  Francisco, CA, USA}, \bibinfo{pages}{618–622}.
\newblock
\showISBNx{1558605665}


\bibitem[Stavrakakis et~al\mbox{.}(2024)]%
        {spp}
\bibfield{author}{\bibinfo{person}{Dimitrios Stavrakakis},
  \bibinfo{person}{Alexandrina Panfil}, \bibinfo{person}{MJin Nam}, {and}
  \bibinfo{person}{Pramod Bhatotia}.} \bibinfo{year}{2024}\natexlab{}.
\newblock \showarticletitle{SPP: Safe Persistent Pointers for Memory Safety}.
  In \bibinfo{booktitle}{\emph{2024 54th Annual IEEE/IFIP International
  Conference on Dependable Systems and Networks (DSN)}}.
  \bibinfo{pages}{37--52}.
\newblock
\href{https://doi.org/10.1109/DSN58291.2024.00019}{doi:\nolinkurl{10.1109/DSN58291.2024.00019}}


\bibitem[Sun et~al\mbox{.}(2025)]%
        {sunSweetSourCHERI2025}
\bibfield{author}{\bibinfo{person}{Xiaoyang Sun}, \bibinfo{person}{Jeremy
  Singer}, {and} \bibinfo{person}{Zheng Wang}.}
  \bibinfo{year}{2025}\natexlab{}.
\newblock \showarticletitle{Sweet or {{Sour CHERI}}: {{Performance
  Characterization}} of the {{Arm Morello Platform}}}. In
  \bibinfo{booktitle}{\emph{2025 {{IEEE International Symposium}} on {{Workload
  Characterization}} ({{IISWC}})}}. \bibinfo{publisher}{IEEE Computer Society},
  \bibinfo{pages}{423--438}.
\newblock
\showISBNx{979-8-3315-4917-6}
\href{https://doi.org/10.1109/IISWC66894.2025.00042}{doi:\nolinkurl{10.1109/IISWC66894.2025.00042}}


\bibitem[Szekeres et~al\mbox{.}(2013)]%
        {szekeres2013sok}
\bibfield{author}{\bibinfo{person}{Laszlo Szekeres}, \bibinfo{person}{Mathias
  Payer}, \bibinfo{person}{Tao Wei}, {and} \bibinfo{person}{Dawn Song}.}
  \bibinfo{year}{2013}\natexlab{}.
\newblock \showarticletitle{Sok: Eternal war in memory}. In
  \bibinfo{booktitle}{\emph{2013 IEEE Symposium on Security and Privacy}}.
  IEEE, \bibinfo{pages}{48--62}.
\newblock


\bibitem[Talibzade et~al\mbox{.}(2025)]%
        {talibzadeUsingLLMsVulnerability2025}
\bibfield{author}{\bibinfo{person}{Rustam Talibzade}, \bibinfo{person}{Idilio
  Drago}, {and} \bibinfo{person}{Francesco Bergadano}.}
  \bibinfo{year}{2025}\natexlab{}.
\newblock \showarticletitle{On {{Using LLMs}} for {{Vulnerability
  Classification}}}. In \bibinfo{booktitle}{\emph{Proceedings of the 2025
  {{Workshop}} on {{Large AI Systems}} and {{Models}} with {{Privacy}} and
  {{Security Analysis}}}} \emph{(\bibinfo{series}{{{LAMPS}} '25})}.
  \bibinfo{publisher}{Association for Computing Machinery},
  \bibinfo{address}{New York, NY, USA}, \bibinfo{pages}{68--71}.
\newblock
\showISBNx{979-8-4007-1896-0}
\href{https://doi.org/10.1145/3733800.3763267}{doi:\nolinkurl{10.1145/3733800.3763267}}


\bibitem[Tarek Ibn~Ziad et~al\mbox{.}(2023)]%
        {cucatch}
\bibfield{author}{\bibinfo{person}{Mohamed Tarek Ibn~Ziad},
  \bibinfo{person}{Sana Damani}, \bibinfo{person}{Aamer Jaleel},
  \bibinfo{person}{Stephen~W. Keckler}, {and} \bibinfo{person}{Mark
  Stephenson}.} \bibinfo{year}{2023}\natexlab{}.
\newblock \showarticletitle{cuCatch: A Debugging Tool for Efficiently Catching
  Memory Safety Violations in CUDA Applications}.
\newblock \bibinfo{journal}{\emph{Proc. ACM Program. Lang.}}
  \bibinfo{volume}{7}, \bibinfo{number}{PLDI}, Article \bibinfo{articleno}{111}
  (\bibinfo{date}{June} \bibinfo{year}{2023}), \bibinfo{numpages}{24}~pages.
\newblock
\href{https://doi.org/10.1145/3591225}{doi:\nolinkurl{10.1145/3591225}}


\bibitem[{The PostgreSQL Global Development Group}(2026)]%
        {postgresql}
\bibfield{author}{\bibinfo{person}{{The PostgreSQL Global Development Group}}.}
  \bibinfo{year}{2026}\natexlab{}.
\newblock \bibinfo{title}{{PostgreSQL}}.
\newblock \bibinfo{howpublished}{\url{https://www.postgresql.org/}}.
\newblock


\bibitem[Torkamani et~al\mbox{.}(2025)]%
        {torkamaniStreamliningSecurityVulnerability2025}
\bibfield{author}{\bibinfo{person}{Mohammad~Jalili Torkamani},
  \bibinfo{person}{Joey NG}, \bibinfo{person}{Nikita Mehrotra},
  \bibinfo{person}{Mahinthan Chandramohan}, \bibinfo{person}{Padmanabhan
  Krishnan}, {and} \bibinfo{person}{Rahul Purandare}.}
  \bibinfo{year}{2025}\natexlab{}.
\newblock \bibinfo{title}{Streamlining {{Security Vulnerability Triage}} with
  {{Large Language Models}}}.
\newblock
\showeprint[arxiv]{2501.18908}~[cs.SE]
\href{https://doi.org/10.48550/arXiv.2501.18908}{doi:\nolinkurl{10.48550/arXiv.2501.18908}}


\bibitem[Unterguggenberger et~al\mbox{.}(2023a)]%
        {crypt_mem_safety}
\bibfield{author}{\bibinfo{person}{Martin Unterguggenberger},
  \bibinfo{person}{David Schrammel}, \bibinfo{person}{Lukas Lamster},
  \bibinfo{person}{Pascal Nasahl}, {and} \bibinfo{person}{Stefan Mangard}.}
  \bibinfo{year}{2023}\natexlab{a}.
\newblock \showarticletitle{Cryptographically Enforced Memory Safety}. In
  \bibinfo{booktitle}{\emph{Proceedings of the 2023 ACM SIGSAC Conference on
  Computer and Communications Security}} (<conf-loc>, <city>Copenhagen</city>,
  <country>Denmark</country>, </conf-loc>) \emph{(\bibinfo{series}{CCS '23})}.
  \bibinfo{publisher}{Association for Computing Machinery},
  \bibinfo{address}{New York, NY, USA}, \bibinfo{pages}{889–903}.
\newblock
\showISBNx{9798400700507}
\href{https://doi.org/10.1145/3576915.3623138}{doi:\nolinkurl{10.1145/3576915.3623138}}


\bibitem[Unterguggenberger et~al\mbox{.}(2023b)]%
        {multitag}
\bibfield{author}{\bibinfo{person}{Martin Unterguggenberger},
  \bibinfo{person}{David Schrammel}, \bibinfo{person}{Pascal Nasahl},
  \bibinfo{person}{Robert Schilling}, \bibinfo{person}{Lukas Lamster}, {and}
  \bibinfo{person}{Stefan Mangard}.} \bibinfo{year}{2023}\natexlab{b}.
\newblock \showarticletitle{Multi-Tag: A Hardware-Software Co-Design for Memory
  Safety based on Multi-Granular Memory Tagging}. In
  \bibinfo{booktitle}{\emph{Proceedings of the 2023 ACM Asia Conference on
  Computer and Communications Security}} (Melbourne, VIC, Australia)
  \emph{(\bibinfo{series}{ASIA CCS '23})}. \bibinfo{publisher}{Association for
  Computing Machinery}, \bibinfo{address}{New York, NY, USA},
  \bibinfo{pages}{177–189}.
\newblock
\showISBNx{9798400700989}
\href{https://doi.org/10.1145/3579856.3590331}{doi:\nolinkurl{10.1145/3579856.3590331}}


\bibitem[van~der Veen et~al\mbox{.}(2012)]%
        {memory-errors}
\bibfield{author}{\bibinfo{person}{Victor van~der Veen},
  \bibinfo{person}{Nitish Dutt-Sharma}, \bibinfo{person}{Lorenzo Cavallaro},
  {and} \bibinfo{person}{Herbert Bos}.} \bibinfo{year}{2012}\natexlab{}.
\newblock \showarticletitle{Memory Errors: The Past, the Present, and the
  Future}, Vol.~\bibinfo{volume}{7462}.
\newblock
\showISBNx{978-3-642-33337-8}
\href{https://doi.org/10.1007/978-3-642-33338-5_5}{doi:\nolinkurl{10.1007/978-3-642-33338-5_5}}


\bibitem[Van~der Veen et~al\mbox{.}(2012)]%
        {van2012memory}
\bibfield{author}{\bibinfo{person}{Victor Van~der Veen},
  \bibinfo{person}{Nitish Dutt-Sharma}, \bibinfo{person}{Lorenzo Cavallaro},
  {and} \bibinfo{person}{Herbert Bos}.} \bibinfo{year}{2012}\natexlab{}.
\newblock \showarticletitle{Memory errors: The past, the present, and the
  future}. In \bibinfo{booktitle}{\emph{International Workshop on Recent
  Advances in Intrusion Detection}}. Springer, \bibinfo{pages}{86--106}.
\newblock


\bibitem[Vander~Stoep and Rebert(2024)]%
        {GoogleOnlineSecurity2024}
\bibfield{author}{\bibinfo{person}{Jeff Vander~Stoep} {and}
  \bibinfo{person}{Alex Rebert}.} \bibinfo{year}{2024}\natexlab{}.
\newblock \bibinfo{title}{Google {{Online Security Blog}}: {{Eliminating Memory
  Safety Vulnerabilities}} at the {{Source}}}.
\newblock
\urldef\tempurl%
\url{https://security.googleblog.com/2024/09/eliminating-memory-safety-vulnerabilities-Android.html}
\showURL{%
\tempurl}


\bibitem[Wahbe et~al\mbox{.}(1993)]%
        {wahbeEfficientSoftwarebasedFault1993}
\bibfield{author}{\bibinfo{person}{Robert Wahbe}, \bibinfo{person}{Steven
  Lucco}, \bibinfo{person}{Thomas~E. Anderson}, {and} \bibinfo{person}{Susan~L.
  Graham}.} \bibinfo{year}{1993}\natexlab{}.
\newblock \showarticletitle{Efficient Software-Based Fault Isolation}. In
  \bibinfo{booktitle}{\emph{Proceedings of the Fourteenth {{ACM}} Symposium on
  {{Operating}} Systems Principles}} \emph{(\bibinfo{series}{{{SOSP}} '93})}.
  \bibinfo{publisher}{Association for Computing Machinery},
  \bibinfo{address}{New York, NY, USA}, \bibinfo{pages}{203--216}.
\newblock
\showISBNx{978-0-89791-632-5}
\href{https://doi.org/10.1145/168619.168635}{doi:\nolinkurl{10.1145/168619.168635}}


\bibitem[Wang et~al\mbox{.}(2024a)]%
        {wangCherifyingLinuxPractical2024}
\bibfield{author}{\bibinfo{person}{Kui Wang}, \bibinfo{person}{Dmitry
  Kasatkin}, \bibinfo{person}{Vincent Ahlrichs}, \bibinfo{person}{Lukas Auer},
  \bibinfo{person}{Konrad Hohentanner}, \bibinfo{person}{Julian Horsch}, {and}
  \bibinfo{person}{Jan-Erik Ekberg}.} \bibinfo{year}{2024}\natexlab{a}.
\newblock \showarticletitle{Cherifying {{Linux}}: {{A Practical View}} on Using
  {{CHERI}}}. In \bibinfo{booktitle}{\emph{Proceedings of the 17th {{European
  Workshop}} on {{Systems Security}}}} \emph{(\bibinfo{series}{{{EuroSec}}
  '24})}. \bibinfo{publisher}{Association for Computing Machinery},
  \bibinfo{address}{New York, NY, USA}, \bibinfo{pages}{15--21}.
\newblock
\showISBNx{979-8-4007-0542-7}
\href{https://doi.org/10.1145/3642974.3652282}{doi:\nolinkurl{10.1145/3642974.3652282}}


\bibitem[Wang et~al\mbox{.}(2021)]%
        {ratel}
\bibfield{author}{\bibinfo{person}{Mingzhe Wang}, \bibinfo{person}{Zhiyong Wu},
  \bibinfo{person}{Xinyi Xu}, \bibinfo{person}{Jie Liang},
  \bibinfo{person}{Chijin Zhou}, \bibinfo{person}{Huafeng Zhang}, {and}
  \bibinfo{person}{Yu Jiang}.} \bibinfo{year}{2021}\natexlab{}.
\newblock \showarticletitle{Industry Practice of Coverage-Guided
  Enterprise-Level DBMS Fuzzing}. In \bibinfo{booktitle}{\emph{2021 IEEE/ACM
  43rd International Conference on Software Engineering: Software Engineering
  in Practice (ICSE-SEIP)}}. \bibinfo{pages}{328--337}.
\newblock
\href{https://doi.org/10.1109/ICSE-SEIP52600.2021.00042}{doi:\nolinkurl{10.1109/ICSE-SEIP52600.2021.00042}}


\bibitem[Wang et~al\mbox{.}(2024b)]%
        {cecsan}
\bibfield{author}{\bibinfo{person}{Xiaolei Wang}, \bibinfo{person}{Bin Zhang},
  \bibinfo{person}{Chaojing Tang}, {and} \bibinfo{person}{Long Zhang}.}
  \bibinfo{year}{2024}\natexlab{b}.
\newblock \showarticletitle{Highly Comprehensive and Efficient Memory Safety
  Enforcement with Pointer Tagging}. In \bibinfo{booktitle}{\emph{2024 54th
  Annual IEEE/IFIP International Conference on Dependable Systems and Networks
  Workshops (DSN-W)}}. \bibinfo{pages}{74--81}.
\newblock
\href{https://doi.org/10.1109/DSN-W60302.2024.00026}{doi:\nolinkurl{10.1109/DSN-W60302.2024.00026}}


\bibitem[Wesley~Filardo et~al\mbox{.}(2020)]%
        {wesleyfilardoCornucopiaTemporalSafety2020}
\bibfield{author}{\bibinfo{person}{Nathaniel Wesley~Filardo},
  \bibinfo{person}{Brett~F. Gutstein}, \bibinfo{person}{Jonathan Woodruff},
  \bibinfo{person}{Sam Ainsworth}, \bibinfo{person}{Lucian {Paul-Trifu}},
  \bibinfo{person}{Brooks Davis}, \bibinfo{person}{Hongyan Xia},
  \bibinfo{person}{Edward Tomasz~Napierala}, \bibinfo{person}{Alexander
  Richardson}, \bibinfo{person}{John Baldwin}, \bibinfo{person}{David
  Chisnall}, \bibinfo{person}{Jessica Clarke}, \bibinfo{person}{Khilan Gudka},
  \bibinfo{person}{Alexandre Joannou}, \bibinfo{person}{A. Theodore~Markettos},
  \bibinfo{person}{Alfredo Mazzinghi}, \bibinfo{person}{Robert~M. Norton},
  \bibinfo{person}{Michael Roe}, \bibinfo{person}{Peter Sewell},
  \bibinfo{person}{Stacey Son}, \bibinfo{person}{Timothy~M. Jones},
  \bibinfo{person}{Simon~W. Moore}, \bibinfo{person}{Peter~G. Neumann}, {and}
  \bibinfo{person}{Robert N.~M. Watson}.} \bibinfo{year}{2020}\natexlab{}.
\newblock \showarticletitle{Cornucopia: {{Temporal Safety}} for {{CHERI
  Heaps}}}. In \bibinfo{booktitle}{\emph{2020 {{IEEE Symposium}} on
  {{Security}} and {{Privacy}} ({{SP}})}}. \bibinfo{pages}{608--625}.
\newblock
\showISSN{2375-1207}
\href{https://doi.org/10.1109/SP40000.2020.00098}{doi:\nolinkurl{10.1109/SP40000.2020.00098}}


\bibitem[Woodruff et~al\mbox{.}(2014a)]%
        {cheri}
\bibfield{author}{\bibinfo{person}{Jonathan Woodruff},
  \bibinfo{person}{Robert~N.M. Watson}, \bibinfo{person}{David Chisnall},
  \bibinfo{person}{Simon~W. Moore}, \bibinfo{person}{Jonathan Anderson},
  \bibinfo{person}{Brooks Davis}, \bibinfo{person}{Ben Laurie},
  \bibinfo{person}{Peter~G. Neumann}, \bibinfo{person}{Robert Norton}, {and}
  \bibinfo{person}{Michael Roe}.} \bibinfo{year}{2014}\natexlab{a}.
\newblock \showarticletitle{The CHERI Capability Model: Revisiting RISC in an
  Age of Risk}.
\newblock \bibinfo{journal}{\emph{SIGARCH Comput. Archit. News}}
  \bibinfo{volume}{42}, \bibinfo{number}{3} (\bibinfo{date}{June}
  \bibinfo{year}{2014}), \bibinfo{pages}{457–468}.
\newblock
\showISSN{0163-5964}
\href{https://doi.org/10.1145/2678373.2665740}{doi:\nolinkurl{10.1145/2678373.2665740}}


\bibitem[Woodruff et~al\mbox{.}(2014b)]%
        {woodruffCHERICapabilityModel2014}
\bibfield{author}{\bibinfo{person}{Jonathan Woodruff},
  \bibinfo{person}{Robert~N.M. Watson}, \bibinfo{person}{David Chisnall},
  \bibinfo{person}{Simon~W. Moore}, \bibinfo{person}{Jonathan Anderson},
  \bibinfo{person}{Brooks Davis}, \bibinfo{person}{Ben Laurie},
  \bibinfo{person}{Peter~G. Neumann}, \bibinfo{person}{Robert Norton}, {and}
  \bibinfo{person}{Michael Roe}.} \bibinfo{year}{2014}\natexlab{b}.
\newblock \showarticletitle{The {{CHERI}} Capability Model: Revisiting {{RISC}}
  in an Age of Risk}.
\newblock \bibinfo{journal}{\emph{SIGARCH Comput. Archit. News}}
  \bibinfo{volume}{42}, \bibinfo{number}{3} (\bibinfo{date}{June}
  \bibinfo{year}{2014}), \bibinfo{pages}{457--468}.
\newblock
\showISSN{0163-5964}
\href{https://doi.org/10.1145/2678373.2665740}{doi:\nolinkurl{10.1145/2678373.2665740}}


\bibitem[Xia et~al\mbox{.}(2019)]%
        {xiaCHERIvokeCharacterisingPointer2019}
\bibfield{author}{\bibinfo{person}{Hongyan Xia}, \bibinfo{person}{Jonathan
  Woodruff}, \bibinfo{person}{Sam Ainsworth}, \bibinfo{person}{Nathaniel~W.
  Filardo}, \bibinfo{person}{Michael Roe}, \bibinfo{person}{Alexander
  Richardson}, \bibinfo{person}{Peter Rugg}, \bibinfo{person}{Peter~G.
  Neumann}, \bibinfo{person}{Simon~W. Moore}, \bibinfo{person}{Robert N.~M.
  Watson}, {and} \bibinfo{person}{Timothy~M. Jones}.}
  \bibinfo{year}{2019}\natexlab{}.
\newblock \showarticletitle{{{CHERIvoke}}: {{Characterising Pointer
  Revocation}} Using {{CHERI Capabilities}} for {{Temporal Memory Safety}}}. In
  \bibinfo{booktitle}{\emph{Proceedings of the 52nd {{Annual IEEE}}/{{ACM
  International Symposium}} on {{Microarchitecture}}}}
  \emph{(\bibinfo{series}{{{MICRO-52}}})}. \bibinfo{publisher}{Association for
  Computing Machinery}, \bibinfo{address}{New York, NY, USA},
  \bibinfo{pages}{545--557}.
\newblock
\showISBNx{978-1-4503-6938-1}
\href{https://doi.org/10.1145/3352460.3358288}{doi:\nolinkurl{10.1145/3352460.3358288}}


\bibitem[Xu et~al\mbox{.}(2023)]%
        {xuPANICPANassistedIntraprocess2023}
\bibfield{author}{\bibinfo{person}{Jiali Xu}, \bibinfo{person}{Mengyao Xie},
  \bibinfo{person}{Chenggang Wu}, \bibinfo{person}{Yinqian Zhang},
  \bibinfo{person}{Qijing Li}, \bibinfo{person}{Xuan Huang},
  \bibinfo{person}{Yuanming Lai}, \bibinfo{person}{Yan Kang},
  \bibinfo{person}{Wei Wang}, \bibinfo{person}{Qiang Wei}, {and}
  \bibinfo{person}{Zhe Wang}.} \bibinfo{year}{2023}\natexlab{}.
\newblock \showarticletitle{{{PANIC}}: {{PAN-assisted Intra-process Memory
  Isolation}} on {{ARM}}}. In \bibinfo{booktitle}{\emph{Proceedings of the 2023
  {{ACM SIGSAC Conference}} on {{Computer}} and {{Communications Security}}}}.
  \bibinfo{publisher}{ACM}, \bibinfo{address}{Copenhagen Denmark},
  \bibinfo{pages}{919--933}.
\newblock
\showISBNx{979-8-4007-0050-7}
\href{https://doi.org/10.1145/3576915.3623206}{doi:\nolinkurl{10.1145/3576915.3623206}}


\bibitem[Yang et~al\mbox{.}(2025)]%
        {dissecting_cgpu}
\bibfield{author}{\bibinfo{person}{Yang Yang}, \bibinfo{person}{Mohammad
  Sonji}, {and} \bibinfo{person}{Adwait Jog}.} \bibinfo{year}{2025}\natexlab{}.
\newblock \showarticletitle{Dissecting Performance Overheads of Confidential
  Computing on GPU-based Systems}. In \bibinfo{booktitle}{\emph{2025 IEEE
  International Symposium on Performance Analysis of Systems and Software
  (ISPASS)}}. \bibinfo{pages}{1--16}.
\newblock
\href{https://doi.org/10.1109/ISPASS64960.2025.00020}{doi:\nolinkurl{10.1109/ISPASS64960.2025.00020}}


\bibitem[Ye et~al\mbox{.}(2014)]%
        {wpbound}
\bibfield{author}{\bibinfo{person}{Ding Ye}, \bibinfo{person}{Yu Su},
  \bibinfo{person}{Yulei Sui}, {and} \bibinfo{person}{Jingling Xue}.}
  \bibinfo{year}{2014}\natexlab{}.
\newblock \showarticletitle{WPBOUND: Enforcing Spatial Memory Safety
  Efficiently at Runtime with Weakest Preconditions}. In
  \bibinfo{booktitle}{\emph{2014 IEEE 25th International Symposium on Software
  Reliability Engineering}}. \bibinfo{pages}{88--99}.
\newblock
\href{https://doi.org/10.1109/ISSRE.2014.20}{doi:\nolinkurl{10.1109/ISSRE.2014.20}}


\bibitem[Yedidia(2024)]%
        {yedidiaLightweightFaultIsolation2024}
\bibfield{author}{\bibinfo{person}{Zachary Yedidia}.}
  \bibinfo{year}{2024}\natexlab{}.
\newblock \showarticletitle{Lightweight {{Fault Isolation}}: {{Practical}},
  {{Efficient}}, and {{Secure Software Sandboxing}}}. In
  \bibinfo{booktitle}{\emph{Proceedings of the 29th {{ACM International
  Conference}} on {{Architectural Support}} for {{Programming Languages}} and
  {{Operating Systems}}, {{Volume}} 2}}. \bibinfo{publisher}{ACM},
  \bibinfo{address}{La Jolla CA USA}, \bibinfo{pages}{649--665}.
\newblock
\showISBNx{979-8-4007-0385-0}
\href{https://doi.org/10.1145/3620665.3640408}{doi:\nolinkurl{10.1145/3620665.3640408}}


\bibitem[Yee et~al\mbox{.}(2010)]%
        {yeeNativeClientSandbox2010}
\bibfield{author}{\bibinfo{person}{Bennet Yee}, \bibinfo{person}{David Sehr},
  \bibinfo{person}{Gregory Dardyk}, \bibinfo{person}{J.~Bradley Chen},
  \bibinfo{person}{Robert Muth}, \bibinfo{person}{Tavis Ormandy},
  \bibinfo{person}{Shiki Okasaka}, \bibinfo{person}{Neha Narula}, {and}
  \bibinfo{person}{Nicholas Fullagar}.} \bibinfo{year}{2010}\natexlab{}.
\newblock \showarticletitle{Native {{Client}}: A Sandbox for Portable,
  Untrusted X86 Native Code}.
\newblock \bibinfo{journal}{\emph{Commun. ACM}} \bibinfo{volume}{53},
  \bibinfo{number}{1} (\bibinfo{date}{Jan.} \bibinfo{year}{2010}),
  \bibinfo{pages}{91--99}.
\newblock
\showISSN{0001-0782, 1557-7317}
\href{https://doi.org/10.1145/1629175.1629203}{doi:\nolinkurl{10.1145/1629175.1629203}}


\bibitem[Yu et~al\mbox{.}(2023)]%
        {capstone}
\bibfield{author}{\bibinfo{person}{Jason~Zhijingcheng Yu},
  \bibinfo{person}{Conrad Watt}, \bibinfo{person}{Aditya Badole},
  \bibinfo{person}{Trevor~E. Carlson}, {and} \bibinfo{person}{Prateek Saxena}.}
  \bibinfo{year}{2023}\natexlab{}.
\newblock \showarticletitle{Capstone: A Capability-based Foundation for
  Trustless Secure Memory Access}. In \bibinfo{booktitle}{\emph{32nd USENIX
  Security Symposium (USENIX Security 23)}}. \bibinfo{publisher}{USENIX
  Association}, \bibinfo{address}{Anaheim, CA}, \bibinfo{pages}{787--804}.
\newblock
\showISBNx{978-1-939133-37-3}
\urldef\tempurl%
\url{https://www.usenix.org/conference/usenixsecurity23/presentation/yu-jason}
\showURL{%
\tempurl}


\bibitem[Zhang et~al\mbox{.}(2023)]%
        {catamaran}
\bibfield{author}{\bibinfo{person}{Yiyu Zhang}, \bibinfo{person}{Tianyi Liu},
  \bibinfo{person}{Zewen Sun}, \bibinfo{person}{Zhe Chen},
  \bibinfo{person}{Xuandong Li}, {and} \bibinfo{person}{Zhiqiang Zuo}.}
  \bibinfo{year}{2023}\natexlab{}.
\newblock \showarticletitle{Catamaran: Low-Overhead Memory Safety Enforcement
  via Parallel Acceleration}. In \bibinfo{booktitle}{\emph{Proceedings of the
  32nd ACM SIGSOFT International Symposium on Software Testing and Analysis}}
  (Seattle, WA, USA) \emph{(\bibinfo{series}{ISSTA 2023})}.
  \bibinfo{publisher}{Association for Computing Machinery},
  \bibinfo{address}{New York, NY, USA}, \bibinfo{pages}{816–828}.
\newblock
\showISBNx{9798400702211}
\href{https://doi.org/10.1145/3597926.3598098}{doi:\nolinkurl{10.1145/3597926.3598098}}


\bibitem[Zhao et~al\mbox{.}(2024)]%
        {Survey_of_Hardware_Improvement_to_Secure_Program_Execution}
\bibfield{author}{\bibinfo{person}{Lianying Zhao}, \bibinfo{person}{He Shuang},
  \bibinfo{person}{Shengjie Xu}, \bibinfo{person}{Wei Huang},
  \bibinfo{person}{Rongzhen Cui}, \bibinfo{person}{Pushkar Bettadpur}, {and}
  \bibinfo{person}{David Lie}.} \bibinfo{year}{2024}\natexlab{}.
\newblock \showarticletitle{A Survey of Hardware Improvements to Secure Program
  Execution}.
\newblock \bibinfo{journal}{\emph{ACM Comput. Surv.}} \bibinfo{volume}{56},
  \bibinfo{number}{12}, Article \bibinfo{articleno}{311} (\bibinfo{date}{Oct.}
  \bibinfo{year}{2024}), \bibinfo{numpages}{37}~pages.
\newblock
\showISSN{0360-0300}
\href{https://doi.org/10.1145/3672392}{doi:\nolinkurl{10.1145/3672392}}


\bibitem[Zhong et~al\mbox{.}(2020)]%
        {squirrel}
\bibfield{author}{\bibinfo{person}{Rui Zhong}, \bibinfo{person}{Yongheng Chen},
  \bibinfo{person}{Hong Hu}, \bibinfo{person}{Hangfan Zhang},
  \bibinfo{person}{Wenke Lee}, {and} \bibinfo{person}{Dinghao Wu}.}
  \bibinfo{year}{2020}\natexlab{}.
\newblock \showarticletitle{SQUIRREL: Testing Database Management Systems with
  Language Validity and Coverage Feedback}. In
  \bibinfo{booktitle}{\emph{Proceedings of the 2020 ACM SIGSAC Conference on
  Computer and Communications Security}} (Virtual Event, USA)
  \emph{(\bibinfo{series}{CCS '20})}. \bibinfo{publisher}{Association for
  Computing Machinery}, \bibinfo{address}{New York, NY, USA},
  \bibinfo{pages}{955–970}.
\newblock
\showISBNx{9781450370899}
\href{https://doi.org/10.1145/3372297.3417260}{doi:\nolinkurl{10.1145/3372297.3417260}}


\bibitem[Zhong and Rigger(2026)]%
        {sqlancer_plus_plus}
\bibfield{author}{\bibinfo{person}{Suyang Zhong} {and} \bibinfo{person}{Manuel
  Rigger}.} \bibinfo{year}{2026}\natexlab{}.
\newblock \bibinfo{title}{Scaling Automated Database System Testing}.
\newblock
\showeprint[arxiv]{2503.21424}~[cs.SE]
\urldef\tempurl%
\url{https://arxiv.org/abs/2503.21424}
\showURL{%
\tempurl}


\end{thebibliography}

\clearpage
\appendix

\begin{table}[t]
    \footnotesize
    \centering
    \caption{Experimental platform specifications.}
    \label{tab:platform-specs}
    \begin{tabular}{lccc}
        \toprule
        & \textbf{Morello} & \textbf{Ampere 1a} & \textbf{Xeon 6438Y+} \\
        \midrule
        Extension & CHERI & MTE & MPK \\
        ISA & Armv8.2 & Armv8.6 & x86-64 \\
        Microarch. & Neoverse N1 & AmpereOne & Golden Cove \\
        Sockets $\times$ Cores & $1 \times 4$ & $1 \times 192$ & $2 \times 32$ \\
        Frequency & \qty{2.5}{GHz} & \qty{3.2}{GHz} & \qty{2.0}{GHz} \\
        L1 I-Cache & \qty{64}{KiB} & \qty{16}{KiB} & \qty{32}{KiB} \\
        L1 D-Cache & \qty{64}{KiB} & \qty{64}{KiB} & \qty{48}{KiB} \\
        L2 Cache & \qty{1}{MiB} & \qty{2}{MiB} & \qty{2}{MiB} \\
        L3 Cache & \qty{1}{MiB} (shared/2 cores) & --- & \qty{60}{MiB} (shared/socket) \\
        RAM & \qty{16}{GiB} & \qty{512}{GiB} & \qty{2}{TiB} \\
        OS & Debian 13.2 & NixOS 25.11 & NixOS 25.11 \\
        Kernel & Morello Linux 6.7.0 & Linux 6.17.13 & Linux 6.0.0 \\
        \bottomrule
        \multicolumn{4}{l}{\footnotesize All cache sizes are per core unless noted otherwise.}
    \end{tabular}
\end{table}

\section{Additional Architectural Analysis}
\label{appendix:arch_analysis}

\subsection{MPK}
\label{subsec:hw-mpk}

\begin{figure}[t]
    \centering
    \includegraphics[width=\columnwidth]{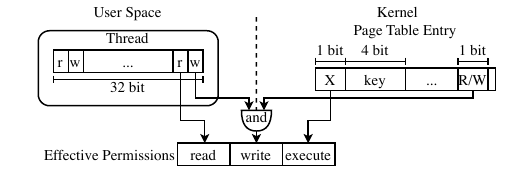}
    \caption{MPK access check mechanism.}
    \label{fig:arch-mpk-overview}
\end{figure}

\myparagraph{Experimental Setup}
We evaluate \ac{MPK} on a dual-socket Intel Xeon Gold 6438Y+ CPU with 32 cores per socket and 2 threads per core, featuring \qty{2}{TiB} of memory, running NixOS 25.05 and Linux 6.0.0.
Each core features \qty{32}{KiB} of L1i cache, \qty{48}{KiB} of L1d cache, \qty{2}{MiB} of L2 cache, while each core additionally features \qty{60}{MiB} of L3 cache.

\myparagraph{Mechanism}
\ac{MPK} introduces a per-thread isolation mechanism on a page-level granularity.
Four unused bits in the \ac{PTE} specify one of 16 distinct protection keys.
Each thread then specifies its access rights in the thread-local PKRU register, which holds two bits for each key, specifying read and write permissions.
This allows for very fast switching of access rights, compared to \textit{mprotect} system calls.
Code using \ac{MPK} allocates a new protection key by calling \textit{pkey\_alloc}.
\textit{pkey\_mprotect} associates a protection key with a memory region, while \textit{pkey\_free} marks the key as free again.
To mirror \textit{mprotect}'s model, user-space code must synchronize the PKRU register between threads~\cite{parkLibmpkSoftwareAbstraction2019}.
\ac{MPK} \emph{complements} \textit{mprotect}'s permissions:
With \textit{mprotect}, users can select between five permissions: inaccessible, read-only, read-execute, read-write, and read-write-execute.
Execute-only is not possible.
Conversely, \ac{MPK} allows to enable read or write permissions.
The effective permission is that of the intersection of the page table entry and the PKRU register.
Using \ac{MPK} allows to set a page to execute-only, by disallowing reading through \ac{MPK}.

However, \ac{MPK} imposes some drawbacks:
As the number of keys is limited to 16, allocating new keys may fail.
Freeing keys does not invalidate existing keys, allowing for key use-after-free scenarios where pages using the freed are accessible using the reallocated key.
Both of these problems can be solved by virtualizing keys in software, e.g., through libmpk~\cite{parkLibmpkSoftwareAbstraction2019}.
Since the protection key is set per page, protection is more coarse-grained than CHERI or MTE: Providing fine-grained memory safety by protecting individual objects is not feasible.
Instead, \ac{MPK} is used to divide sensitive and non-sensitive data (e.g., shadow stacks), isolate different compartments (e.g., JIT-compiled functions), or allow for execute-only permissions.

\myparagraph{Microarchitectural Analysis}
We evaluate the overheads of MPK, in particular the cost of reading, writing, and updating the PKRU register, as well as the cost of the newly introduced system calls in \cref{tab:mpk-insts}. %
As accessing the PKRU register is thread-local and done in user space, it performs very well.
Updating the permissions completes in under \qty{19}{cycles}.
The system calls are much more expensive, taking around \qty{270}{cycles} to allocate or free a protection key and around \qty{450}{cycles} to change the protection key of a memory page.

\begin{table}
    \footnotesize
    \caption{MPK Instructions}
    \label{tab:mpk-insts}
    \begin{tabularx}{\columnwidth}{lXc}
        \toprule
        \textbf{Inst} & \textbf{Description} & \textbf{Cycles ($\downarrow$)} \\
        \midrule
        rdpkru & Read protection key rights. & 4.02 \\
        wrpkru & Writes protection key rights. & 17.78 \\
        pkru update & Read, change and write protection key rights. & 18.66 \\
        \midrule
        \textbf{System Call} & \textbf{Description} & \textbf{Cycles ($\downarrow$)} \\
        \midrule
        pkey\_alloc & Allocate a new protection key. & 272.84 \\
        pkey\_free & Free a protection key. & 268.36 \\
        pkey\_mprotect & Change protection key for region. & 447.75 \\
        \bottomrule
        \multicolumn{3}{l}{$\downarrow$: lower is better}
    \end{tabularx}
    
\end{table}

\section{Memory Safety Bug Study}
\label{sec:appendix-bug-study}

This section describes the full methodology used to collect, classify, and analyze
memory safety bugs across the database systems studied in \cref{subsec:bug_study}.

\myparagraph{Scope and Time Window}
The study covers two complementary datasets with overlapping system coverage.\\

\noindent Our \textit{bug report study} covers nine database systems:
\begin{multicols}{3}
\begin{itemize}[leftmargin=*]
    \item DuckDB~\cite{duckdb}
    \item Redis~\cite{redis}
    \item RocksDB~\cite{rocksdb}
    \item ClickHouse~\cite{clickhouse}
    \item MariaDB~\cite{mariadb}
    \item MySQL~\cite{mysql}
    \item LevelDB~\cite{leveldb}
\end{itemize}
\end{multicols}

\noindent Our \textit{CVE study} covers thirteen database systems:
\begin{multicols}{3}
\begin{itemize}[leftmargin=*]
    \item MySQL~\cite{mysql}
    \item SQLite~\cite{sqlite}
    \item MariaDB~\cite{mariadb}
    \item Redis~\cite{redis}
    \item LevelDB~\cite{leveldb}
    \item RocksDB~\cite{rocksdb}
    \item DuckDB~\cite{duckdb}
    \item LadyBug~\cite{ladybugdbLadybugDB}
    \item PostgreSQL~\cite{postgresql}
    \item MongoDB~\cite{mongodb}
    \item Memcached~\cite{memcached}
    \item InfluxDB~\cite{influxdb}
    \item Redshift~\cite{redshift}
\end{itemize}
\end{multicols}

\noindent All data collection is restricted to the ten-year window \textit{2015--2025} to ensure
coverage of modern vulnerability trends.

\subsection{CVE Collection}
\label{subsec:cve-collection}

We download the full \textit{CVEProject/cvelistV5}
snapshot~\cite{cvelistv5}\footnote{\url{https://github.com/CVEProject/cvelistV5}}, which provides every CVE in structured JSON format following the CVE~5.0 schema.
To isolate the relevant CVEs for the systems considered in this study, we scan the \textit{containers.cna.affected} fields (\textit{product} and \textit{packageName}) for case-insensitive matches against the database name.
We did not include the CVE \textit{description} in our pattern-matching search because we observed that it returned various CVEs unrelated to the database system itself.
CVE records whose filenames encode a year prior to 2015 are skipped at the filename level before parsing.
Results are filtered by the \textit{cveMetadata.datePublished} field to
enforce the 2015-2025 window.
The output is a per-database list of matching CVE files that are fed into the classification step.

\subsection{Bug Report Collection}
\label{subsec:bug-collection}

Since CVEs capture only publicly disclosed, security-relevant vulnerabilities,
we expand our study to include a dataset with bug reports scraped from each project's official issue tracker.
All scrapers impose a \SI{0.5}{\second} inter-request delay to respect rate limits of the bug tracker hosts.
We save each bug as a self-contained JSON file.

\myparagraph{GitHub Issues}
For \textit{DuckDB, RocksDB, Redis, LevelDB, and ClickHouse},
we query the GitHub REST~API (\textit{/issues}) with \textit{state=closed},
sorted ascending by \textit{updated\_at}, and pre-filtered with the API's
\textit{since} parameter set to \textit{2015-01-01T00:00:00Z}.
To work around GitHub's 10,000-result depth limit per sorted query, the scraper
slides a \textit{since} cursor forward to the latest \textit{updated\_at} timestamp seen whenever the page counter approaches the depth limit (page~$\geq 90$), effectively re-anchoring the query window.
Pull requests are excluded, and results are post-filtered to enforce the 2015-2025 window.

\myparagraph{Jira}
For MariaDB\footnote{\url{https://jira.mariadb.org}},
we query each project's public Jira REST~API (\textit{/rest/api/2/search}) using
JQL date constraints (\textit{created >= "2015-01-01" AND created <= "2025-12-31"}), and paginate through all results in batches of 50.
Issue comments are concatenated to the description to provide a full context to
the classifier.

\myparagraph{MySQL Bug Tracker}
For MySQL, we use the CSV export endpoint of the official MySQL bug tracker\footnote{\url{https://bugs.mysql.com/search-csv.php}}.
We set the \textit{bug\_age} parameter to the number of days elapsed since \textit{2015-01-01}, pre-filtering server-side, and post-filter individual rows against the target window using the \textit{Entered} timestamp in the CSV.
For each bug ID, the full HTML report page (\textit{bugs.mysql.com/bug.php?id=\{ID\}}) is fetched and parsed to extract the title, status, and comment text.

\subsection{Keyword Filtering}
\label{subsec:keyword-filter}
Before invoking the LLM classifier, we apply a case-insensitive regular-expression
search over each bug's full text (title, description, and comments) using the
following list of memory-safety terms:

\begin{center}
\small
\setlength{\fboxsep}{6pt}
\fbox{\parbox{0.9\columnwidth}{%
buffer overflow/overrun/underflow/underrun,
out-of-bounds (OOB, OOBR, OOBW), array bounds, off-by-one,
stack overflow, heap overflow, stack smashing,
integer overflow/underflow,
null pointer dereference,
double free, multiple free, free-after-free,
use-after-free (UAF), stack/heap use-after,
uninitialized memory/variable,
wild pointer, dangling pointer/reference,
memory safety/corruption, heap/stack corruption,
invalid/illegal memory access, invalid read/write,
segmentation fault (SIGSEGV), address sanitizer, ASLR
}}
\end{center}

Bugs that match at least one term form a \emph{keyword set} that serves as a baseline against which LLM results are evaluated.

\subsection{LLM Classification}
\label{subsec:llm-classification}

After performing the keyword-based filtering and analysis, we realized that the produced set contained a large amount of misclassified bugs/CVEs.
Therefore, for each bug or CVE we construct a structured prompt and submit it to a locally hosted \textit{gpt-oss-120b} model\footnote{\url{https://huggingface.co/unsloth/gpt-oss-120b-GGUF}} (Q4\_K\_M quantization) served by \textit{llama-server}~\cite{llama_cpp} on an H100 GPU with continuous batching enabled and 16 parallel completion slots.
Descriptions are truncated to 12,000 characters to stay within the model's context budget. 
Up to 16 classification requests are dispatched concurrently per run.
The system and user prompts for bug reports and CVEs are shown below.

\begin{tcolorbox}[colback=blue!5!white, colframe=blue!50!black,
                  title=Bug Report Classification Prompt,
                  fonttitle=\bfseries\small, fontupper=\small]
\textbf{System:} You are an expert software security researcher specializing
in database systems.

\textbf{User:}
Analyze the following bug report.\\
\underline{First}, classify the issue into one of the following types:
1. "\textit{bug}": Actual bugs (flaws in the system implementation).
2. "\textit{programmer\_error}": Problems on the side of the programmer (misuse).
3. "\textit{other}": Requests or other issues (feature requests, questions, etc.).

\underline{Second}, IF the issue is a "\textit{bug}", determine if it describes a **Memory Safety Issue**.
If it is not a "\textit{bug}", "\textit{is\_memory\_safety}" must be false.

\underline{Third}, IF the issue is a "bug" AND you can determine the primary operational implication
with HIGH confidence, classify it into exactly one of the following:
\begin{itemize}[leftmargin=*]
\item \textit{"data\_loss}": Data is permanently lost or unrecoverable.
\item \textit{"data\_corruption}": Stored data becomes physically or logically inconsistent.
\item \textit{"incorrect\_query\_results}": Queries silently return wrong data (no crash).
\item \textit{"system\_crash}": The database process crashes, aborts, or panics.
\item \textit{"resource\_exhaustion}": Memory leak, disk/CPU saturation, connection pool depletion.
\item \textit{"concurrency\_hazard}": Deadlocks, race conditions, or isolation violations under concurrency.
\item \textit{"transaction\_violation}": ACID guarantees are broken (lost updates, dirty reads, etc.).
\item \textit{"durability\_violation}": Committed transactions are lost after a crash or restart.
\item \textit{"replication\_inconsistency}": Primary and replica data diverge or replication fails.
\item \textit{"denial\_of\_service}": Server becomes unresponsive or hangs without crashing.
\item \textit{"none}": No clear operational implication can be determined with high confidence, or the issue is not a "bug".
\end{itemize}

\underline{Bug Report}:\\
Title: {[title\_placeholder]}\\
Description: {[description\_placeholder]}\\
\underline{Task}:\\
Classify this bug. Answer with JSON format:
\begin{verbatim}
{"issue_type": "...",
 "is_memory_safety": true/false,
 "implication": "...",
 "confidence": "...",
 "reason": "..."}
\end{verbatim}
Only provide the JSON."
\end{tcolorbox}

\begin{tcolorbox}[colback=blue!5!white, colframe=blue!50!black,
                  title=CVE Classification Prompt,
                  fonttitle=\bfseries\small, fontupper=\small]
\textbf{System:} You are an expert software security researcher specializing
in database systems and CVE analysis.

\textbf{User:} Analyze the following CVE (Common Vulnerabilities and Exposures) report.\\
\underline{First}, classify the issue into one of the following types:
1. "\textit{bug}": Actual bugs (flaws in the system implementation).
2. "\textit{programmer\_error}": Problems on the side of the programmer (misuse).
3. "\textit{other}": Requests or other issues (feature requests, questions, etc.).\\
\underline{Second}, IF the issue is a "\textit{bug}", determine if it describes a **Memory Safety Issue**.
If it is not a "\textit{bug}", "\textit{is\_memory\_safety}" must be false.
Memory safety issues INCLUDE:
\begin{itemize}[leftmargin=*]
\item Buffer overflow, heap overflow, stack overflow
\item Use-after-free, double-free, dangling pointer
\item Null pointer dereference
\item Out-of-bounds read/write
\item Uninitialized memory access
\item Integer overflow ONLY if it leads to memory corruption
\end{itemize}

Memory safety issues DO NOT INCLUDE:
\begin{itemize}[leftmargin=*]
\item SQL injection, XSS, CSRF (these are injection/logic bugs)
\item Authentication bypass, privilege escalation (unless memory-related)
\item Denial of service through resource exhaustion (unless memory corruption)
\item Logic errors, race conditions (unless causing memory corruption)
\end{itemize}

\underline{Third}, IF the issue is a "bug" AND you can determine the primary operational implication, with HIGH confidence, classify it into exactly one of the following:
\begin{itemize}[leftmargin=*]
\item \textit{"data\_loss}": Data is permanently lost or unrecoverable.
\item \textit{"data\_corruption}": Stored data becomes physically or logically inconsistent.
\item \textit{"incorrect\_query\_results}": Queries silently return wrong data (no crash).
\item \textit{"system\_crash}": The database process crashes, aborts, or panics.
\item \textit{"resource\_exhaustion}": Memory leak, disk/CPU saturation, connection pool depletion.
\item \textit{"concurrency\_hazard}": Deadlocks, race conditions, or isolation violations under concurrency.
\item \textit{"transaction\_violation}": ACID guarantees are broken (lost updates, dirty reads, etc.).
\item \textit{"durability\_violation}": Committed transactions are lost after a crash or restart.
\item \textit{"replication\_inconsistency}": Primary and replica data diverge or replication fails.
\item \textit{"denial\_of\_service}": Server becomes unresponsive or hangs without crashing.
\item \textit{"none}": No clear operational implication can be determined with high confidence, or the issue is not a "bug".
\end{itemize}

\underline{CVE Report}:\\
CVE ID: [CVE\_ID placeholder]\\
CWE: [CWE\_IDs placeholder]\\
Affected: [Vendor/Product placeholder]\\
Description: [Description placeholder]\\
\underline{TASK}:\\
Classify this CVE. Answer with JSON format:
\begin{verbatim}
{"issue_type": "...",
 "is_memory_safety": true/false,
 "implication": "...",
 "confidence": "...",
 "reason": "..."}
\end{verbatim}
Only provide the JSON."
\end{tcolorbox}

\noindent The model produces a JSON object with five fields:
\begin{enumerate}
  \item \textbf{issue\_type}: one of \textit{bug}, \textit{programmer\_error},
    or \textit{other}.
  \item \textbf{is\_memory\_safety}: a boolean, forced to \textit{false}
    if the \textit{issue\_type} is not a \textit{bug}.
  \item \textbf{implication}: the primary operational consequence, constrained
    to one of: \textit{data\_loss},
    \textit{data\_corruption}, \textit{system\_crash}, \textit{incorrect\_query\_results},
    \textit{resource\_exhaustion},
    \textit{concurrency\_\linebreak hazard}, \textit{transaction\_violation}, \textit{replication\_inconsistency},
    \textit{durability\_violation}, \textit{denial\_of\_service};
    \textit{none} if no implication can be determined with high confidence
    or the entry is not classified as a \textit{bug}.
  \item \textbf{confidence}: one of \textit{high}, \textit{medium}, or \textit{low}.
  \item \textbf{reason}: a brief free-text explanation of the classification.
\end{enumerate}

To reduce sensitivity to non-deterministic model outputs, each
classification run is repeated \emph{three times independently},
producing three result sets.
In our analysis, a bug is counted as a \textit{memory safety bug} only when \textbf{all} three
repetitions agree (i.e., the intersection of the three LLM sets).

\subsection{Result Analysis}
\label{subsec:result-analysis}

Our analysis scripts produce a set of complementary statistics that we collect
to gain confidence in the results presented in \cref{subsec:bug_study}.
The full outputs are retained for validation and manual inspection.
More precisely, our analysis produces two types of output:

\myparagraph{Aggregate statistics}
For each database and each repetition, we extract: the total number of entries
processed; the count classified as \textit{bug}, \textit{programmer error},
or \textit{other}; and the subset flagged as a memory safety bug.
The keyword set size is gathered in parallel as an independent baseline.
Tracking these numbers across all three repetitions allows us to assess the
stability of the classifier and identify any significant variation between runs.

\myparagraph{Intersection analysis \& result validation}
We compute all non-empty intersections across the four signal
sets (Keywords, LLM repetition~1, LLM repetition~2, LLM repetition~3),
to observe the \emph{exclusive} contribution of each subset combination
(i.e., entries that appear in exactly that combination and no other).
This reveals the degree of agreement between the keyword heuristic and the
LLM classifier, and quantifies how consistently each entry is flagged across
repeated classifications.
Note that the keyword set serves here as a comparative signal only; the
\textbf{confirmed} memory safety count is derived exclusively from the
three-way LLM intersection, as this produced more accurate results.
Finally, to further validate the automated classification, we conducted
a manual inspection of a sample of flagged entries, cross-checking titles,
descriptions, and LLM outputs against the expected classification.

\rev[rev1]{R1.O8,R1.O9,R1.O16,R1.O17}{
\myparagraph{Limitations}
This pipeline is a motivating analysis, not a validated vulnerability classifier, and its numbers should be read as estimates.
We use a quantized (Q4\_K\_M) \textit{gpt-oss-120b} model rather than a full-precision frontier model, and we do not report precision, recall, or inter-annotator agreement against a manually labeled ground-truth set; dedicated works such as DUALLM~\cite{liWhatTheyFix2026} provide such validated baselines.
Our three-way agreement rule---counting a report as a memory-safety bug only when all three runs concur---has two distinct effects.
The first concerns \emph{coverage}: requiring agreement filters out reports that only one or two runs flagged, i.e., the cases the model classifies inconsistently.
These borderline cases are dropped from the counted set, so the rule trades away some coverage in exchange for a set of entries that is more likely to be correct.
The second concerns \emph{confidence}: one might hope that three runs act like three independent judgments, so that a mistake in one run is caught by the others.
This does not hold here, because we sample the same model three times.
When such a model is uncertain about a case, it tends to make the \emph{same} error on every run rather than a different one each time, so the runs are correlated rather than independent~\cite{byerlySelfConsistencyFallsShort2026}.
Agreement across runs therefore reflects the model's consistency, and provides less assurance than three genuinely independent judgments would.
We also classify issue-tracker reports and \acp{CVE} only, and exclude fixes that land as pull requests without an associated issue (\cref{subsec:bug-collection}); issue-text-only classification is known to be weaker than classification that also ingests commit messages and diffs~\cite{liWhatTheyFix2026}.
Consequently, our claims concern memory-safety \emph{reports} rather than all memory-safety bugs.
Finally, MySQL's bug tracker redacts security-sensitive issues, so memory-safety bugs for MySQL are systematically under-reported in the source we scrape, which partly explains its low memory-safety share in \cref{subsec:bug_study}.
}

\end{document}